\documentclass[twocolumn]{aastex6}
\usepackage{color}
\usepackage{float}
\usepackage{mathtools}
\usepackage{animate}
\shorttitle{Linearly polarized structures near SGP}
\newcommand\rah{\mbox{$^{\mathrm h}$}}%
\newcommand\ram{\mbox{$^{\mathrm m}$}}%
\renewcommand{\vec}[1]{\mathbf{#1}}

\turnoffedit
\begin{document}

% ***********************************************************
\title{Low frequency observations of linearly polarized structures in the interstellar medium near the south Galactic pole}

% Include current builders list.
\author{E.~Lenc\altaffilmark{1,2},
B.~M.~Gaensler\altaffilmark{1,2,3},
X.~H.~Sun\altaffilmark{1},
E.~M.~Sadler\altaffilmark{1,2},
A.~G.~Willis\altaffilmark{4},
N.~Barry\altaffilmark{5},
A.~P.~Beardsley\altaffilmark{5,6},
M.~E.~Bell\altaffilmark{7,2},
G.~Bernardi\altaffilmark{8},
J.~D.~Bowman\altaffilmark{6},
F.~Briggs\altaffilmark{9,2},
J.~R.~Callingham\altaffilmark{1,2,7},
R.~J.~Cappallo\altaffilmark{10},
P.~Carroll\altaffilmark{5},
B.~E.~Corey\altaffilmark{10},
A.~de~Oliveira-Costa\altaffilmark{11},
A.~A.~Deshpande\altaffilmark{12},
J.~S.~Dillon\altaffilmark{11,13},
K.~S.~Dwarkanath\altaffilmark{12},
D.~Emrich\altaffilmark{14},
A.~Ewall-Wice\altaffilmark{11},
L.~Feng\altaffilmark{11},
B.-Q.~For\altaffilmark{15},
R.~Goeke\altaffilmark{11},
L.~J.~Greenhill\altaffilmark{16},
P.~Hancock\altaffilmark{14,2},
B.~J.~Hazelton\altaffilmark{5,17},
J.~N.~Hewitt\altaffilmark{11},
L.~Hindson\altaffilmark{18},
N.~Hurley-Walker\altaffilmark{14},
M.~Johnston-Hollitt\altaffilmark{18},
D.~C.~Jacobs\altaffilmark{6},
A.~D.~Kapi\'{n}ska\altaffilmark{15,2},
D.~L.~Kaplan\altaffilmark{19},
J.~C.~Kasper\altaffilmark{20,16},
H.-S.~Kim\altaffilmark{21,2},
E.~Kratzenberg\altaffilmark{10},
J.~Line\altaffilmark{21,2},
A.~Loeb\altaffilmark{16},
C.~J.~Lonsdale\altaffilmark{10},
M.~J.~Lynch\altaffilmark{14},
B.~McKinley\altaffilmark{21},
S.~R.~McWhirter\altaffilmark{10},
D.~A.~Mitchell\altaffilmark{7,2},
M.~F.~Morales\altaffilmark{5},
E.~Morgan\altaffilmark{11},
J.~Morgan\altaffilmark{14,2},
T.~Murphy\altaffilmark{1,2},
A.~R.~Neben\altaffilmark{11},
D.~Oberoi\altaffilmark{22},
A.~R.~Offringa\altaffilmark{23},
S.~M.~Ord\altaffilmark{7,2},
S.~Paul\altaffilmark{12},
B.~Pindor\altaffilmark{21,2},
J.~C.~Pober\altaffilmark{24},
T.~Prabu\altaffilmark{12},
P.~Procopio\altaffilmark{21,2},
J.~Riding\altaffilmark{21,2},
A.~E.~E.~Rogers\altaffilmark{10},
A.~Roshi\altaffilmark{25},
N.~Udaya~Shankar\altaffilmark{12},
S.~K.~Sethi\altaffilmark{12},
K.~S.~Srivani\altaffilmark{12},
L.~Staveley-Smith\altaffilmark{15,2},
R.~Subrahmanyan\altaffilmark{12,2},
I.~S.~Sullivan\altaffilmark{5},
M.~Tegmark\altaffilmark{11},
Nithyanandan~Thyagarajan\altaffilmark{6},
S.~J.~Tingay\altaffilmark{14,2,26},
C.~Trott\altaffilmark{14,2},
M.~Waterson\altaffilmark{14,9},
R.~B.~Wayth\altaffilmark{14,2},
R.~L.~Webster\altaffilmark{21,2},
A.~R.~Whitney\altaffilmark{10},
A.~Williams\altaffilmark{14},
C.~L.~Williams\altaffilmark{11},
C.~Wu\altaffilmark{15},
J.~S.~B.~Wyithe\altaffilmark{21,2} and
Q.~Zheng\altaffilmark{18}}

\altaffiltext{1}{Sydney Institute for Astronomy, School of Physics, The University of Sydney, NSW 2006, Australia}
\altaffiltext{2}{ARC Centre of Excellence for All-sky  Astrophysics (CAASTRO)}
\altaffiltext{3}{Dunlap Institute for Astronomy and Astrophysics, University of Toronto, 50 St.\ George Street, Toronto, ON M5S 3H4, Canada}
\altaffiltext{4}{National Research Council of Canada, Herzberg Astronomy and  Astrophysics , P.O. Box 248, Penticton, BC, Canada}
\altaffiltext{5}{Department of Physics, University of Washington, Seattle, WA 98195, USA}
\altaffiltext{6}{School of Earth and Space Exploration, Arizona  State University, Tempe, AZ 85287, USA}
\altaffiltext{7}{CSIRO Astronomy and Space Science (CASS), PO Box  76, Epping, NSW 1710, Australia}
\altaffiltext{8}{Department of Physics and Electronics, Rhodes University, PO Box 94, Grahamstown 6140, South Africa}
\altaffiltext{9}{Research School of Astronomy and Astrophysics, Australian National University, Canberra, ACT 2611, Australia}
\altaffiltext{10}{MIT Haystack Observatory, Westford, MA 01886, USA}
\altaffiltext{11}{MIT Kavli Institute for Astrophysics and Space  Research, Massachusetts Institute of Technology, Cambridge, MA  02139, USA}
\altaffiltext{12}{Raman Research Institute, Bangalore 560080, India}
\altaffiltext{13}{Department of Astronomy, University of California, Berkeley, CA 94720, USA}
\altaffiltext{14}{International Centre for Radio Astronomy Research, Curtin University, Bentley, WA 6102, Australia}
\altaffiltext{15}{ICRAR University of Western Australia, Crawley, WA 6009, Australia}
\altaffiltext{16}{Harvard-Smithsonian Center for Astrophysics, Cambridge, MA 02138, USA}
\altaffiltext{17}{University of Washington, eScience Institute, Seattle, WA 98195, USA}
\altaffiltext{18}{School of Chemical \& Physical Sciences,  Victoria University of Wellington, Wellington 6140, New Zealand}
\altaffiltext{19}{Department of Physics, University of Wisconsin--Milwaukee, Milwaukee, WI 53201, USA}
\altaffiltext{20}{University of Michigan, Department of Atmospheric, Oceanic and Space Sciences, Ann Arbor, MI 48109, USA}
\altaffiltext{21}{School of Physics, The University of Melbourne, Parkville, VIC 3010, Australia}
\altaffiltext{22}{National Centre for Radio Astrophysics, Tata Institute for Fundamental Research, Pune 411007, India}
\altaffiltext{23}{ASTRON, The Netherlands Institute for Radio Astronomy, Postbus 2, 7990 AA, Dwingeloo, The Netherlands}
\altaffiltext{24}{Brown University, Department of Physics, Providence, RI 02912, USA}
\altaffiltext{25}{National Radio Astronomy Observatory, Charlottesville and Greenbank, USA}
\altaffiltext{26}{Osservatorio di Radio Astronomia, Istituto Nazionale di Astrofisica, Bologna, Italy, 40123}
%\altaffil{ARC Centre of Excellence for All-sky Astrophysics (CAASTRO)}

% ***********************************************************
\begin{abstract}

We present deep polarimetric observations at 154\,MHz with the Murchison Widefield Array (MWA), covering 625\,deg$^{2}$ centered on $\alpha=0\rah$, $\delta=-27\arcdeg$. The sensitivity available in our deep observations allows \edit1{an in-band,} frequency-dependent analysis of polarized structure for the first time at long wavelengths. Our analysis suggests that the polarized structures are dominated by intrinsic emission but may also have a foreground Faraday screen component. At these wavelengths, the compactness of the MWA baseline distribution provides excellent snapshot sensitivity to large-scale structure. The observations are sensitive to diffuse polarized emission at $\sim54\arcmin$ resolution with a sensitivity of 5.9\,mJy\,beam$^{-1}$ and compact polarized sources at $\sim2.4\arcmin$ resolution with a sensitivity of 2.3\,mJy\,beam$^{-1}$ for a subset (400\,deg$^{2}$) of this field. The sensitivity allows the effect of ionospheric Faraday rotation to be spatially and temporally measured directly from the diffuse polarized background. Our observations reveal large-scale structures ($\sim1\arcdeg$--$8\arcdeg$ in extent) in linear polarization clearly detectable in $\sim2$ minute snapshots, which would remain undetectable by interferometers with minimum baseline lengths $>110\,m$ at 154 MHz. The brightness temperature of these structures is on average 4\,K in polarized intensity, peaking at 11\,K. Rotation measure synthesis reveals that the structures have Faraday depths ranging from $-2$\,rad\,m$^{-2}$ to 10\,rad\,m$^{-2}$ with a large fraction peaking at $\sim+1$\,rad\,m$^{-2}$. We estimate a distance of $51\pm 20$\,pc to the polarized emission based on measurements of the in-field pulsar J2330$-$2005. We detect four extragalactic linearly polarized point sources within the field in our compact source survey. Based on the known polarized source population at 1.4\,GHz and non-detections at 154\,MHz, we estimate an upper limit on the depolarization ratio of 0.08 from 1.4\,GHz to 154\,MHz.

\end{abstract}


% ***********************************************************
\section{Introduction}
\label{sec:introduction}
\setcounter{footnote}{0}

The interstellar medium (ISM) of the Milky Way hosts a variety of physical mechanisms that define the structure and evolution of the Galaxy. It is a multi-phase medium composed of a tenuous plasma that is permeated by a large-scale magnetic field and is highly turbulent \citep{McKee:2007v45p565, Haverkorn:2015}. Despite advances in theory and simulation \citep{Burkhart:2012v749p145}, our understanding of the properties of the ISM has been limited by the dearth of observational data against which to test.

The local ISM, particularly within the local bubble \citep{Lallement:2003v411p447}, has been very poorly studied. Studies using multi-wavelength observations of diffuse emission \citep{Puspitarini:2014v566p13} show that the local bubble appears to be open-ended towards the south Galactic pole. Polarimetry from stars can be a useful probe \citep{Berdyugin2001v368p635, Berdyugin2004v424p873, Berdyugin:2014v561p24}, however these are sparsely sampled for stars within the local bubble region (a few tens of parsec to $\sim$$100$\,pc). Observations of pulsars can also be used to probe conditions in the line of sight to the source \citep{Mao:2010v717p1170} however the density of such sources is low, even more so if only nearby sources are considered and for directions at mid or high Galactic latitudes.

Radio observations of diffuse polarized emission have become a valuable tool for understanding the structure and properties of the ISM. At $350$\,MHz, it has been demonstrated that diffuse polarization could result from gradients in rotation measure and that they could be used to study the structure of the diffuse ionized gas \citep{Wieringa:1993v268p215, Haverkorn:2000v356p13, Haverkorn:2004v421p1011}. \citet{Gaensler:2011v478p214} observed features at 1.4\,GHz associated with the turbulent ISM using polarization gradient maps. Such features have also been observed as part of the Canadian Galactic Place Survey at 1.4\,GHz \citep{Taylor:2003v125p314} carried out at the Dominion Radio Astrophysical Observatory, the S-band Polarization All Sky Survey (S-PASS) at 2.3\,GHz with the Parkes radio telescope \citep{Carretti:2010v438p276, Iacobelli:2014v566p5} and at 4.8\,GHz at Urumqi as part of the Sino-German $\lambda$6\,cm Polarization Survey of the Galactic Plane \citep{Han:2013v23p82,Sun:2011v527p74, Sun:2014v437p2936}. These centimeter-wavelength observations are significantly less affected by depth depolarization than longer wavelength ones and can probe the ISM out to kilo-parsec distances. However, as they are also sensitive to the local ISM, they cannot distinguish between nearby structures and more distant ones. Longer wavelength observations provide a means to do so; depth depolarization is so significant at these wavelengths that only local regions of the ISM can be seen. As such, they provide a valuable tool for probing the local ISM.

Long wavelength polarimetric observations are particularly sensitive to small changes in Faraday rotation, as a result of fluctuations in the magnetized plasma, which are difficult to detect at shorter wavelengths. Several such studies have been performed with synthesis telescopes at long wavelengths, e.g. WSRT between $325$ and $375$\,MHz \citep{Wieringa:1993v268p215, Haverkorn:2000v356p13, Haverkorn:2003v403p1031, Haverkorn:2003v403p1045, Haverkorn:2003v404p233}, WSRT at 150\,MHz \citep{Bernardi:2009v500p965, Bernardi:2010v522p67, Iacobelli:2013v549p56}; LOFAR at 150\,MHz \citep{Jelic:2014v1407p2093}, and at 189\,MHz with an MWA prototype \citep{Bernardi:2013v771p105}, but none of these were sensitive to structures larger than $\sim$$1\arcdeg$. \edit1{LOFAR observations of the 3C196 field at 150\,MHz \citep{Jelic:2015v583p137} achieved sensitivity to spatial scales up to $\sim$$5\arcdeg$ by utilising a dual-inner-HBA mode \citep{vanHaarlem:2013v556p2}. However, only a limited number of short-baselines are available in this mode and sensitivity is compromised to provide them.} Single dish polarimetric observations at long wavelengths provide access to large-scale structure but so far there has only been one such observation \citep{Mathewson:1965v18p635} and it suffered from poor sensitivity and spatial sampling. Furthermore, single dish observations below 300\,MHz also lack resolution.

The Murchison Widefield Array (MWA) can help to bridge the gap that exists between existing single-dish and interferometric observations at long wavelengths. The MWA is a low frequency ($72$--$300$\,MHz) interferometer located in Western Australia \citep{Tingay:2013v30p7}, with four key science themes: 1) searching for emission from the epoch of reionization (EoR); 2) Galactic and extragalactic surveys; 3) transient science; and 4) solar, heliospheric, and ionospheric science and space weather \citep{Bowman:2013v30p31B}. The array has a very wide field-of-view (over 600 deg$^{2}$ at 154\,MHz) and the dense compact distribution of baselines provides excellent sensitivity to structure on scales up to $14\arcdeg$ in extent at 154\,MHz. Most importantly for this project, the high sensitivity observations can, for the first time, enable a frequency-dependent analysis of large-scale polarized structure. The large number of baselines provide high sensitivity ($\sim$$100$\,mJy rms for a 1 s integration) and dense $(u,v)$-coverage for snapshot imaging. Visibilities can be generated with a spectral resolution of 10\,kHz  and with cadences as low as 0.5 s with the current MWA correlator \citep{Ord:2015v32p6O}, however, typical imaging is performed on $>112$\,s time-scales. 

In this paper, we present results from the first deep MWA survey of diffuse polarization and polarized point sources, for an EoR field situated just west of the South Galactic Pole (SGP). The primary aims of the survey are to study polarized structures in the local ISM, localize them, and gain insights into the processes that generate them. Secondary aims include a study of the polarized point source population at long wavelengths and also an overall evaluation of the polarimetric capabilities of the MWA. 

In Section \ref{sec:mwaobs} we describe the MWA observations and data reduction. In Section \ref{sec:results} we present our diffuse total intensity and polarization maps, apply rotation measure synthesis, analyze the effects of the ionosphere on the observed Faraday rotation, create both continuum and frequency-dependent polarization gradient maps, and search for polarized point sources. In Section \ref{sec:discussion} we explore the nature of the diffuse polarization, estimate the distance to the observed polarized features, study the linearly polarized point source population, discuss possible causes for the polarized structures based on frequency-dependent observations, perform a structure function analysis, and study the observed Faraday depth spectra. A summary and conclusion is provided in Section \ref{sec:conclusion}.

% ***********************************************************
\section{Observations and data reduction}
\label{sec:mwaobs}

All observations were carried out with the 128 tile MWA, located at the Murchison Radio Observatory in Western Australia. Each tile consists of a regular $4\times4$ grid of dual-polarization dipoles. The dipole signals are combined in an analog beamformer, using a set of switchable delay lines, to form a tile beam.

More specifically, data for this investigation were obtained from observations associated with MWA proposals G0008 GLEAM (A Galactic and Extragalactic All-Sky MWA Survey) and G0009 EoR (Epoch of Reionization)\footnote{See \url{http://www.mwatelescope.org/astronomers} for a list of currently active observing proposals.}. The two projects utilize two different observing strategies; GLEAM \citep[][and Hurley-Walker et al., in prep.]{Wayth:2015v32p25} uses a drift-scan observing mode, i.e., the tiles always point to the meridian, whereas the EoR observations track the field over $\sim$$4$ hours with quantized beamformer settings that are separated by about 7 degrees \citep{Trott:2014v31p26, Paul:2014v793p28, Jacobs:2016}. The EoR observations enable deep scans of individual fields whereas the GLEAM observations minimize instrumental systematics by maintaining a consistent observing set up. While the GLEAM observations are not as deep as the EoR observations, they are observed in multiple \edit1{30.72\,MHz} frequency bands and thus enable frequency-dependent polarization characteristics to be explored \edit1{over a wider range of wavelengths}.

While a vast quantity of EoR and GLEAM data has already been collected, our investigation here primarily focusses on the MWA EoR-0 field which is centered on $\alpha=0\rah$, $\delta=-27\arcdeg$, approximately 10 degrees west of the South Galactic Pole ($b=-90\arcdeg$). Only a small subset of the available data has been used in this initial study of the characteristics of linearly polarized diffuse emission in this region. Specifically, two epochs of 154\,MHz EoR data (so-called ``low-band'' by the MWA EoR community) and one epoch of multi-band GLEAM data (centered on 154\,MHz, 185\,MHz and 216\,MHz) which contains the EoR-0 region have been selected. A summary of parameters associated with the 3 epochs of observations used in this investigation can be found in Table \ref{table:obsSummary}. The epoch 1 EoR data corresponds to a quiet period in the ionosphere whereas epoch 3 coincides with the arrival of a coronal mass ejection that propagated from the Sun \citep{Kaplan:2015v809p12} and interacted with the ionosphere. For polarimetric studies, our interest is primarily in the 154\,MHz data (low-band EoR data) as this band is less prone to polarization leakage than at higher frequencies where inaccuracies in the MWA beam model become significant \citep{Sutinjo:2015v50p52S}.

\begin{table*}[t]
\centering
\caption{Details of MWA polarization observations in the EoR-0 field.}
\label{table:obsSummary}
\begin{tabular}{l c c c c c c c c c}
\hline\hline
Epoch & Project & RA & Dec & Obs. Date  & Start Time & End Time & $N_{obs}$\textsuperscript{a} & Band  & $t_{int}$\textsuperscript{b} \\
      &         &    &     &            & (UTC)      & (UTC)    &                              & (MHz) & (s) \\ [0.5ex]
\hline
1  & EoR     & 0$\rah$00$\ram$00$\fs$00 & -27$\arcdeg$00$\arcmin$00$\farcs$0 & 2013-08-26 & 15:04:08 & 18:27:28 & 44 & 138.88-169.60 & 0.5 \\
2(a) & GLEAM   & 0$\rah$03$\ram$16$\fs$01 & -26$\arcdeg$46$\arcmin$49$\farcs$1 & 2013-11-25 & 11:58:56 & 12:00:48 & 1 & 138.88-169.60 & 0.5 \\
2(b) & GLEAM   & 0$\rah$05$\ram$16$\fs$41 & -26$\arcdeg$46$\arcmin$48$\farcs$7 & 2013-11-25 & 12:00:56 & 12:02:48 & 1 & 169.60-200.32 & 0.5 \\
2(c) & GLEAM   & 0$\rah$07$\ram$16$\fs$81 & -26$\arcdeg$46$\arcmin$48$\farcs$7 & 2013-11-25 & 12:02:56 & 12:04:48 & 1 & 200.32-231.04 & 0.5 \\
3  & EoR     & 0$\rah$00$\ram$00$\fs$00 & -27$\arcdeg$00$\arcmin$00$\farcs$0 & 2014-11-06 & 12:56:32 & 14:09:44 & 36 & 138.88-169.60 & 2.0 \\ [1ex]
\hline
\multicolumn{10}{l}{\textsuperscript{a}\footnotesize{Total number of 112 s snapshots used from observation.}} \\
\multicolumn{10}{l}{\textsuperscript{b}\footnotesize{Visibility integration time.}} \\
\end{tabular}
\end{table*}

For the EoR and GLEAM observations, the MWA correlator was configured to generate visibilities in 24 coarse channels each with 32 $\times$ 40\,kHz fine channels, providing a total bandwidth of 30.72\,MHz. Nine fine channels per coarse channel are always flagged, one central channel and four edge channels on either side to remove aliasing introduced by the poly-phase filter bank \citep{Ord:2015v32p6O}. Observations are typically recorded in 112\,s ``snapshots'' with either 0.5\,s or\,2.0 s integration times. For GLEAM, observations cycle through five frequency bands on a per-snapshot basis, this investigation only considers the three upper frequency bands. For EoR observations, the band is centered on 154\,MHz but the beam-former pointing is regularly adjusted to ensure that the EoR field remains near the center of the field-of-view.

\subsection{Primary Beam, Flux Density and Bandpass Calibration}
\label{sec:mwacal}

The visibility data in each snapshot was flagged for radio frequency interference (RFI) using \textsc{aoflagger} \citep{Offringa:2012v539p95O}. A benefit of the radio quiet environment within the MRO is that less than 1\% of data is typically flagged as a result of RFI \citep{Offringa:2015v32p8O}.

Calibration was carried out using the real-time calibration and imaging system, referred to as the \textsc{rts} \citep{Mitchell:2008v2p707, Ord:2010v122p1353}, but utilized in an off-line mode to perform additional polarimetric analysis. For all observations, a pointed scan of 3C444 was used to calibrate the bandpass and to set the absolute flux scale.  The flux \edit1{density} of 3C444 at 154\,MHz is 81\,Jy with a spectral index\footnote{Where $S\propto\nu^{\alpha}$.} of $\alpha=-0.88$ \citep{Slee:1977v43p1, Slee:1995v48p143}, tied to the \cite{Baars:1977v61p99B} flux scale. The uncertainty on the absolute calibration is estimated to be better than 10\% (Hurley-Walker et al., in prep.).

For each observing epoch and frequency band, targeted observations of 3C444 were used to measure the direction independent bandpass gains with the \textsc{rts}, and a polynomial fit was determined for each of the 24 coarse channels. After the bandpass was applied, complex Jones matrices were fitted and the overall solution derived was applied to all visibility data associated with the same observing session, using the calibration scheme described in Section 2.1 of \citet{Bernardi:2013v771p105}. Independent solutions were obtained for each observing session and frequency band. Previous experience has shown that bandpass solutions are stable over an entire night of observing, and so it was assumed that the solutions were not time variable \citep{Bernardi:2013v771p105, Hurley-Walker:2014v31p45}. The relative phase between the instrumental polarizations, i.e. the XY-phase, was not constrained during calibration due to the lack of a bright polarized calibrator in the field. This will result in an excess of leakage from Stokes U into V\citep{Sault:1996v117p149}, however, based on observations with a 32-tile prototype of the MWA \citep{Bernardi:2013v771p105}, we estimate this will result in no more than $20-30\%$ leakage.

The \textsc{rts} uses a simple short-dipole analytic model to determine the tile beam used for calibration and imaging. While this is an over-simplification of the true tile beam over the entire frequency range and field-of-view available to the MWA, it has been shown \citep{Bernardi:2013v771p105} that this is sufficient for polarimetric observations below 200\,MHz and restricted to fields passing close to or through zenith. To minimize polarization leakage as a result of deviations of the primary beam model from the true beam, only near-zenith observations of the EoR-0 field have been included in this study. Similarly, our investigations primarily use data from the EoR low-band (154\,MHz), where the model and true beam match well \citep{Sutinjo:2015v50p52S}. Based on observational tests \citep{Sutinjo:2015v50p52S}, we estimate polarization leakage (primarily Stokes I into Stokes Q) of approximately 1\% near zenith and a few percent towards the edge of a typical 25\arcdeg$\times$25\arcdeg field.

\subsection{Imaging}
\label{sec:mwaimg}

\begin{figure*}[t]
\centering
\epsscale{1.15}
\plotone{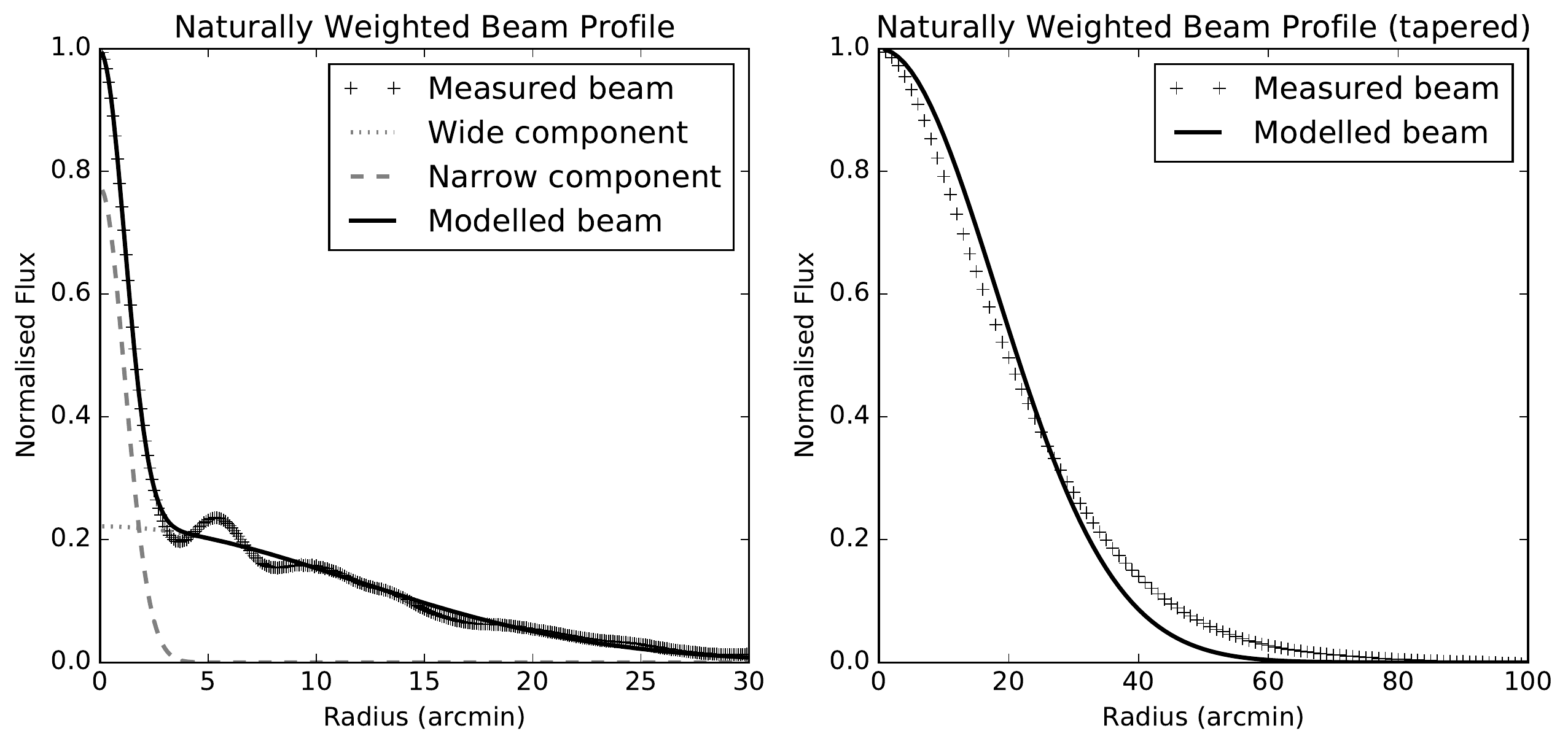}
\caption{Comparison of naturally weighted synthesized beam without (left) and with (right) $(u,v)$-tapering. The measured synthesized beam in both instances shows the radially averaged profile of the beam. The untapered beam (left) was modelled with a two-component Gaussian; one corresponding to the narrow component and another corresponding to the wide component. The tapered beam (right) was modelled with a single Gaussian component.}
\label{fig:beam}
\end{figure*}

Using all available baselines to generate a naturally weighted image results in a point spread function (PSF) with a narrow $6\arcmin$ full width at half maximum (FWHM) Gaussian-like peak associated with the longest baselines and a broad $\sim$$50\arcmin$ FWHM Gaussian-like component associated with the dense inner core of the MWA. Figure \ref{fig:beam} shows a cut of the naturally-weighted MWA beam profile and its decomposition into narrow and broad Gaussian-like components. The \textsc{rts} does not perform image deconvolution on extended emission during the imaging process and the structured naturally-weighted PSF complicates flux scale measurements of diffuse features. To ensure a near-Gaussian beam and to improve imaging of large-scale features, $(u,v)$ visibilities were tapered with an 82\,$\lambda$ Gaussian taper and baselines above $300\lambda$ were excluded. The effect of the tapering can be seen in Figure \ref{fig:beam}; the resulting naturally-weighted PSF is now a single-component near-Gaussian beam with 54$\arcmin\times$47$\arcmin$ FWHM at a position angle of -1.8$\arcdeg$. This corresponds to a conversion factor of 1\,Jy\,beam$^{-1}$ = 5.6\,K at 154\,MHz \citep{Wrobel:1999v180p171}.

For a 112\,s snapshot image using the $(u,v)$-tapered visibilities, the PSF response in the image plane exhibits two weak ($3\%$ level) negative point-like sidelobes $\sim$$4\arcdeg$ from the peak and a two slightly stronger ($10\%$ level) positive point-like sidelobes $\sim$$8\arcdeg$ from the PSF peak. The sidelobe levels reduce with longer integration times. By avoiding deconvolution we can minimize processing requirements while only incurring image flux density errors, as a result of non-Gaussian PSF structure, of the order of a few percent. To verify the fidelity of the diffuse structure in the dirty maps, a single EoR snapshot was calibrated and deconvolved using \textsc{miriad} \citep{Sault:1995v77p433}. The deconvolution process did not greatly affect the diffuse structures in the image and the resulting image was found to be consistent, to within a few percent, with the dirty images generated by the \textsc{rts} in the zenith region. It should be noted that \textsc{miriad} does not have the capability to calibrate nor correct MWA data for wide-field polarimetric effects and so the results are only valid near zenith. To validate the wider field polarimetric results from the \textsc{rts} a second independent processing pipeline based on \textsc{wsclean} \citep{Offringa:2014v444p606O} was used to compare results against. This pipeline also has internal knowledge of the MWA beam and can apply the appropriate corrections for wide-field polarimetry. The output dirty maps from the \textsc{wsclean} pipeline were found to be consistent with the \textsc{rts} maps over the available field-of-view and for all four Stokes parameters. Subtle edge differences were noted, at a level less than $1\%$, owing to slightly different implementations of the MWA beam model.

% Beam with MWA 128T

Using the \textsc{rts}, calibrated 25$\arcdeg$ wide full Stokes (I, Q, U, and V) dirty image cubes were generated for each snapshot, with 160\,kHz frequency channels across the 30.72\,MHz band. The images \edit1{are} corrected for dipole projection effects and wide-field effects across the entire field-of-view \edit1{during the resampling stage \citep{Ord:2010v122p1353}}. A sampling of 3 pixels across the naturally-weighted synthesized beam is used in the final imaging. Assuming a receiver temperature of 50\,K and a sky temperature of 350\,K \citep{Tingay:2013v30p7} and taking into consideration the flagging, weighting and baselines used for imaging, we estimate a theoretical sensitivity of 35\,mJy\,beam$^{-1}$ (1$\sigma$) per snapshot over the entire 30.72\,MHz band at 154\,MHz. When combined with all 44 snapshots in our deepest field (epoch 1), this results in a theoretical sensitivity of 5.3\,mJy\,beam$^{-1}$. Using the continuum Stokes V image of the deepest field as a guide, we measure an actual image rms of 5.9\,mJy\,beam$^{-1}$. Table \ref{table:obsnoise} summarizes the measured continuum image noise and the synthesized beam parameters for all epochs processed in ``diffuse'' imaging mode. For total intensity, image rms is dominated by classical confusion and sidelobe confusion \citep{Wayth:2015v32p25, Franzen:2015}; this is also true for the point-source and pulsar imaging presented below. Similarly, for linear polarization, image rms is limited by diffuse polarized structure within the observed field.

\begin{table*}[t]
\centering
\caption{Summary of measured image noise and synthesized beam characteristics for all epochs and imaging modes. The measured image noise for Stokes I, Q, U and V are listed under columns $\sigma_{i}$, $\sigma_{q}$, $\sigma_{u}$ and $\sigma_{v}$, respectively. In all cases, $\sigma_{i}$ is dominated by classical confusion and sidelobe confusion. For diffuse imaging modes, $\sigma_{q}$ and $\sigma_{u}$ are dominated by diffuse polarized structure in the field. $\theta_{\text{maj}}$ and $\theta_{\text{min}}$ are the major and minor axis of the synthesized beam (FWHM), respectively. PA is the position angle of the synthesized beam measured from north to east.}
\label{table:obsnoise}
\begin{tabular}{l c c c c c c c c}
\hline\hline
Epoch & Mode    & $\sigma_{i}$   & $\sigma_{q}$   & $\sigma_{u}$   & $\sigma_{v}$   & $\theta_{\text{maj}}$         & $\theta_{\text{min}}$         & PA \\
      &         & (Jy\,beam$^{-1}$) & (Jy\,beam$^{-1}$) & (Jy\,beam$^{-1}$) & (Jy\,beam$^{-1}$)   &  (arcmin)                     & (arcmin)                  & (deg) \\ [0.5ex]
\hline
1     & Diffuse & 630            & 80             & 102            & 5.9            & $54$  & $47$  & $-1.8$ \\
1     & Point   & 9.0            & 3.1            & 2.4            & 2.3            & $2.4$ & $2.2$ & $-47$ \\
1     & Pulsar  & 24             & 1.6            & 1.5            & 1.1            & $4.6$ & $3.8$ & $86$ \\
2(a)  & Diffuse & 690            & 88             & 115            & 39             & $54$  & $47$  & $-1.8$ \\
2(b)  & Diffuse & 510            & 240            & 104            & 60             & $48$  & $41$  & $11$ \\
2(c)  & Diffuse & 406            & 254            & 140            & 120            & $43$  & $37$  & $11$ \\
3     & Diffuse & 600            & 87             & 112            & 6.2            & $54$  & $47$  & $-1.8$ \\
3     & Point   & 11.3           & 3.0            & 2.4            & 2.3            & $2.4$ & $2.2$ & $-47$ \\
3     & Pulsar  & 32             & 1.4            & 1.5            & 1.0            & $4.1$ & $3.7$ & $-89$ \\ [1ex]
\hline
\end{tabular}
\end{table*}

Full Stokes dirty image cubes of the inner 400 square degree region of the field were also produced using uniformly weighted images (the restricted field-of-view was due to memory limitations encountered when processing a field at increased resolution). The image cubes are considered ``dirty'' as no deconvolution was performed. All baselines shorter than 50 $\lambda$ were excluded and no $(u,v)$-tapering was applied. The resulting cubes were better suited for searches of polarized point sources as large-scale emission was effectively filtered out. Table \ref{table:obsnoise} summarizes the measured continuum image noise and the synthesized beam parameters for the two epochs processed in this imaging mode (designated as ``point'' mode).

Additional targeted imaging was performed in an attempt to detect a known field pulsar, PSR J2330$-$2005 (PSR B2327$-$20), to aid in localizing linearly polarized features. While not ideal, owing to increased sidelobe confusion and PSF structure, natural weighting was used to improve sensitivity. All available baselines were utilized except those below $100\lambda$; these were excluded to limit confusion and contamination from diffuse emission. Full Stokes dirty image cubes of a 16 square degree region centered on the pulsar ($\alpha=23\rah30\ram26\fs885$, $\delta=-20\arcdeg05\arcmin29\farcs63$) were produced for epochs 1 and 3. Table \ref{table:obsnoise} summarizes the measured continuum image noise and the synthesized beam parameters for the two epochs processed in this imaging mode (designated as ``pulsar'' mode).

% ***********************************************************
\section{Results}
\label{sec:results}

\subsection{Total Intensity Continuum Maps}

\begin{figure*}[t]
\epsscale{1.17}
\plotone{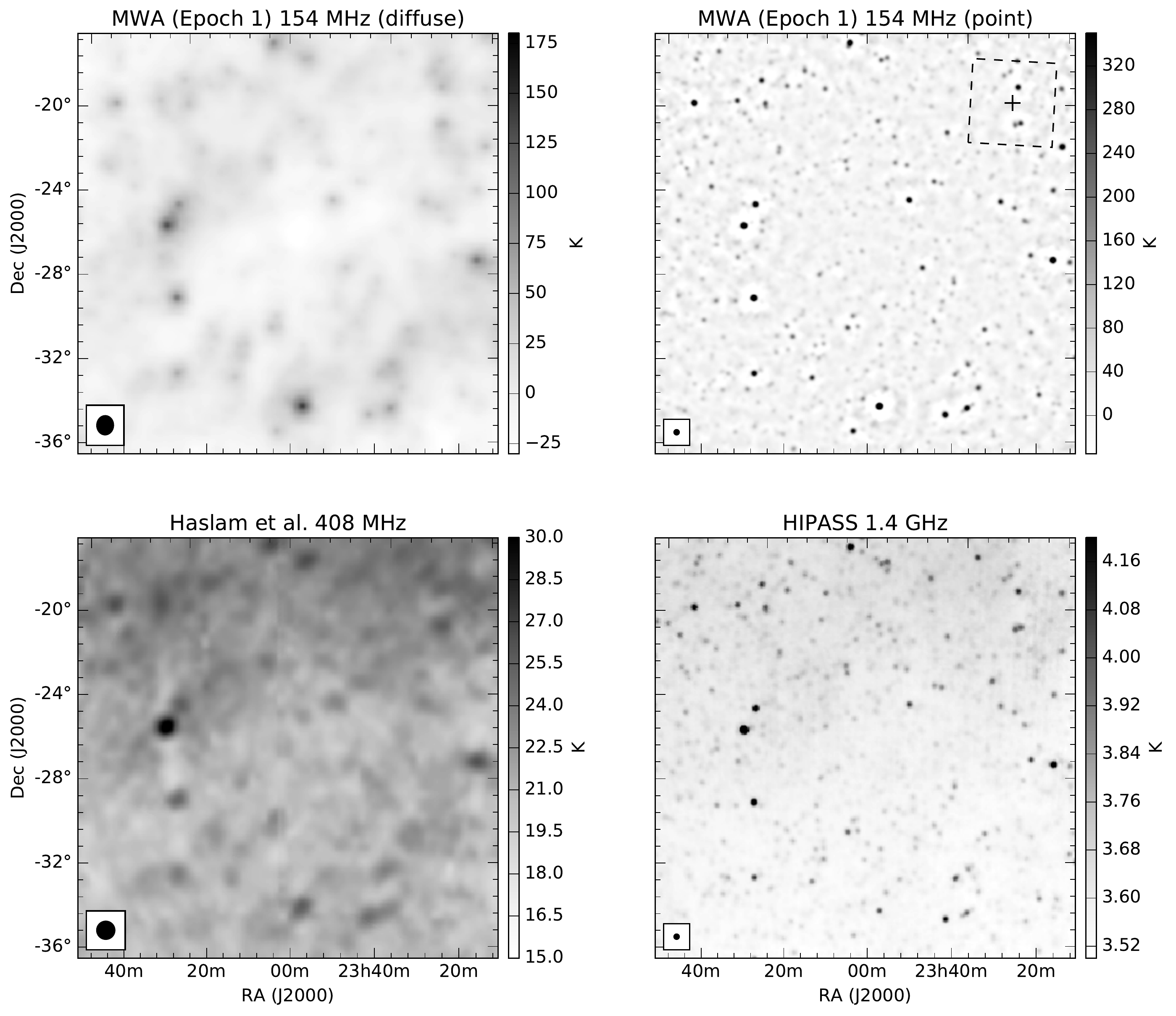}
\caption{Total intensity maps showing 20$\arcdeg\times 20\arcdeg$ portion of the EoR-0 field. The synthesized beam is shown by the ellipse in the bottom-left corner of each map. [Top Left] Dirty MWA map at 154\,MHz optimized for diffuse imaging (naturally weighted and $(u,v)$-tapered). Synthesized beam size is 54$\arcmin\times$47$\arcmin$ FWHM at a position angle of -1.8$\arcdeg$. [Top Right] Dirty MWA map at 154\,MHz optimized for point source imaging (uniformly weighted with short baseline cut-off applied). The map has been convolved with a $14.4\arcmin$ FWHM beam to highlight sources. The dashed inset box marks the region imaged in a targeted analysis of PSR J2330$-$2005; the cross marks the location of the pulsar. [Bottom Left] A reprocessed \citet{Haslam:1982v47p1} 408\,MHz map \citep{Remazeilles:2015v451p4311}. Beam FWHM is $51\arcmin$. [Bottom Right] A reprocessed 1.4\,GHz HIPASS map \citep{Calabretta:2014v31p7}. Beam FWHM is $14.4\arcmin$.}
\label{fig:eorallI}
\end{figure*}

The band-averaged epoch 1 total intensity (Stokes I) images optimized for diffuse emission and for point-source analysis are shown in Figure \ref{fig:eorallI}. While neither of these images have been deconvolved for the PSF, the beam has been shown to be near-Gaussian and does not significantly degrade the images. To demonstrate that these dirty maps accurately recover diffuse structures, reprocessed 408\,MHz \citet{Haslam:1982v47p1} \citep{Remazeilles:2015v451p4311} and 1.4\,GHz HIPASS \citep{Calabretta:2014v31p7} images of the same region have been included in Figure \ref{fig:eorallI} for comparison. The low level diffuse emission observed in the MWA diffuse map correlates well with diffuse emission seen in the \citet{Haslam:1982v47p1} 408\,MHz data and while these features are weaker at 1.4\,GHz they are also present in the HIPASS 1.4\,GHz data.

\subsection{Full Stokes Diffuse Maps}
\label{sec:stokesmaps}

\begin{figure*}
\epsscale{1.17}
\plotone{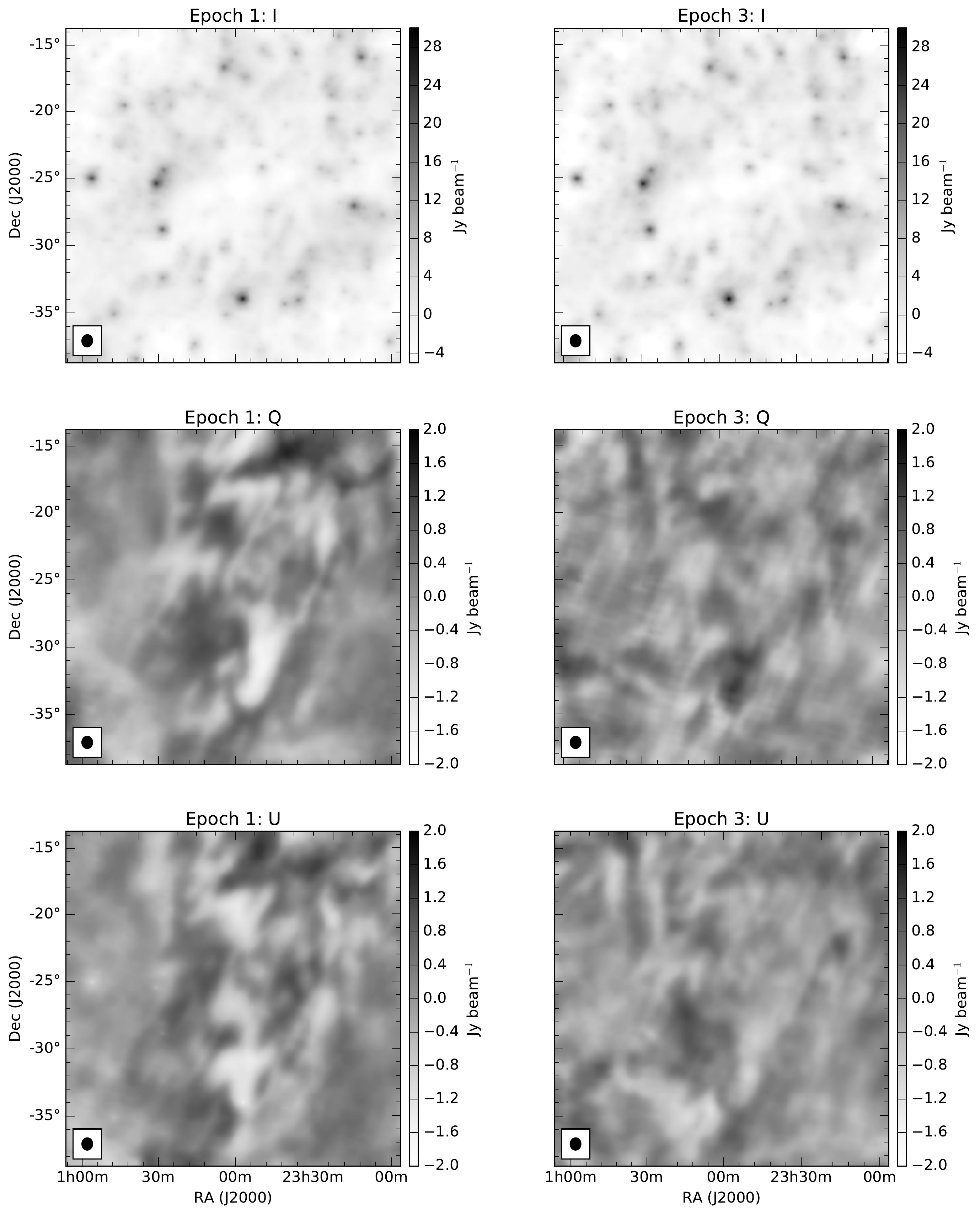}
\caption{$25\arcdeg\times25\arcdeg$ Stokes I, Q and U images from deep epoch 1 and epoch 3 data centered on the EoR-0 field. All images are naturally weighted ($(u,v)$-tapered), band-averaged and dirty (not deconvolved). The synthesized beam size is 54$\arcmin\times$47$\arcmin$ FWHM at a position angle of -1.8$\arcdeg$ and is shown by the ellipse in the bottom-left corner of each map. The Stokes I images are highly consistent between epochs but the linear polarization (Stokes Q and Stokes U) are not; this is due to differing ionospheric conditions between the two epochs.}
\label{fig:eoriqu}
\end{figure*}

The resulting band-averaged total intensity (Stokes I) and linear polarization (Stokes Q and U) dirty images for the epoch 1 and 3 observations of the EoR-0 field are shown in Figure \ref{fig:eoriqu}. As no point-source subtraction or peeling (point-source subtraction with direction-dependent calibration) was performed, the Stokes I images are confusion limited and dominated by point sources within the field. Despite the presence of sources with peak brightnesses exceeding 25\,Jy\,beam$^{-1}$ in total intensity, the linear polarization maps contain mostly smooth features and these are uncorrelated with features in Stokes I. A few of the brightest sources are just perceptible in the polarization maps at about the $1\%$ level but do not affect the overall structure of the diffuse emission seen in those maps. The Q and U maps are mostly dominated by smooth extended structures ranging from $1\arcdeg$ to $8\arcdeg$ in extent and filament-like features, a number of which are approximately aligned in a north-west direction. Note that while the Stokes I maps are virtually identical in epoch 1 and 3, the Stokes Q and U maps are quite different. In particular, the epoch 3 U image appears to exhibit features found in the epoch 1 Q image and the epoch 3 Q image appears to have inverted features from the epoch 1 U image. The changes observed between the epochs appear consistent with a rotation in the Q$-$U plane. As will be shown in Section \ref{sec:ionfr}, these changes are a result of ionospheric Faraday rotation.

\begin{figure*}[t]
\epsscale{1.17}
\plotone{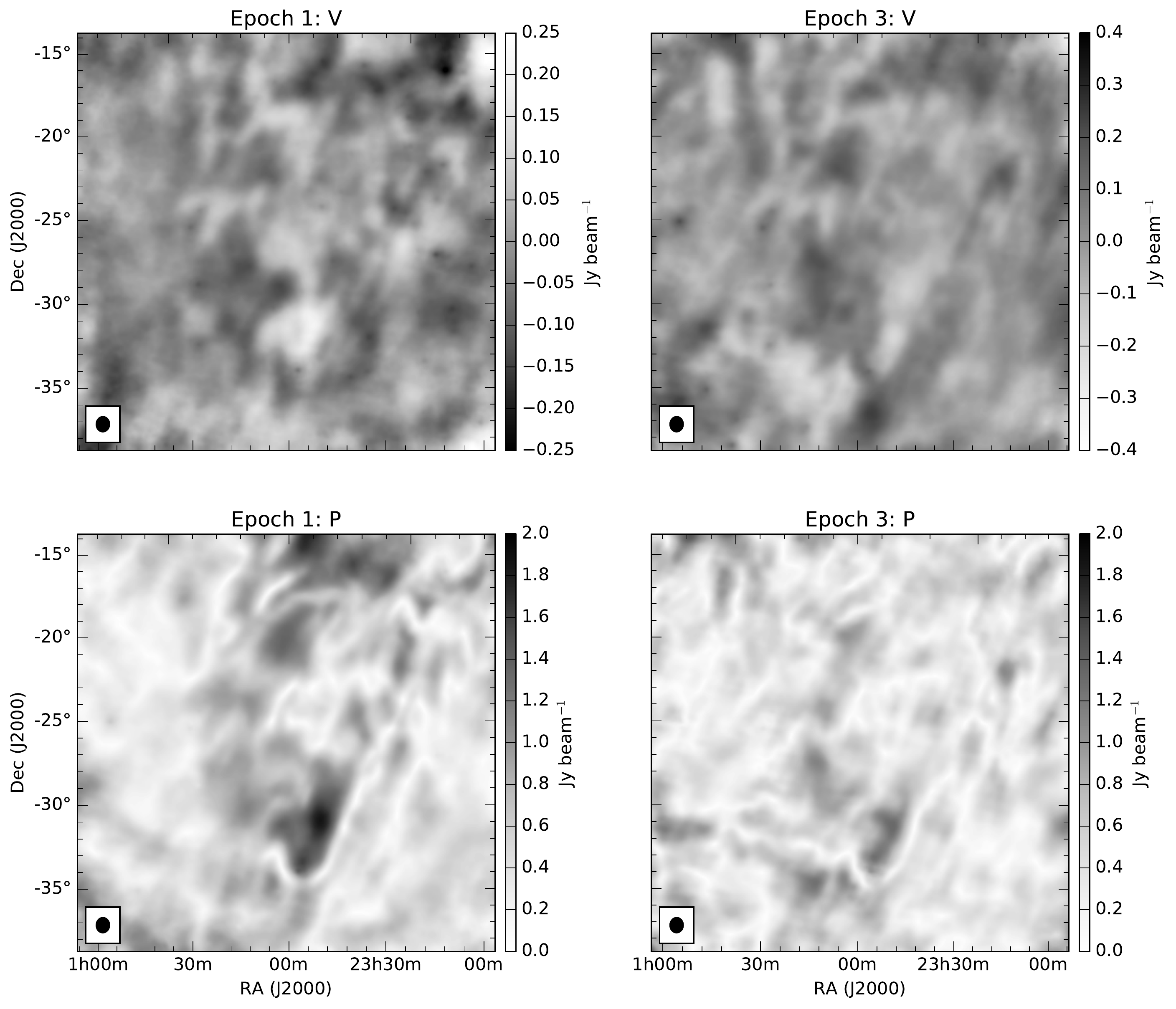}
\caption{Circular polarization (Stokes V) and polarized intensity (P) images. Image details are as for Figure \ref{fig:eoriqu}, but the intensity scales of the Stokes V images have been adjusted to highlight the leakage seen from Stokes U. The polarized intensity images do not show consistent structure as a result of differing ionospheric conditions between the two epochs.}
\label{fig:eorvp}
\end{figure*}

For comparison and diagnostic purposes, the band-averaged circular polarization (Stokes V) \edit1{images} and polarized intensity (P) images \edit1{(formed from the band-averaged Q and U images such that $P=\sqrt{Q^{2}+U^{2}}$)} are shown for epochs 1 and 3 in Figure \ref{fig:eorvp}. The circular polarization maps clearly exhibit leakage from Stokes U into Stokes V at the $\sim$$10\%$ level in epoch 1 and $\sim$$20\%$ level in epoch 3; due to a combination of frequency-dependent XY phase errors that have not been accounted for and uncertainties in the beam model. We note that this is relatively high but even if corrected for, the improvement in Stokes U would only be at a level that is already dominated by existing errors associated with the PSF and sidelobe confusion. We also note that the leakage from Stokes U to Stokes V is not prominent in the uniformly weighted image used for point source analysis; because the point sources are significantly weaker in Stokes U compared to the diffuse emission and so any leakage would be below the Stokes V noise level.

Comparing the polarized intensity images between epoch 1 and 3, one would expect that the images should remain constant between epochs. However, there are clear differences between the two. The epoch 1 image has significantly brighter structures whereas the epoch 3 image does not. These differences are caused by significantly different ionospheric conditions between the two epochs resulting in different levels of depolarization in the band-averaged Q and U images used to form the polarized intensity images. The polarized intensity image from epoch 1 is dominated by a large ($\sim$$4\arcdeg\times 7\arcdeg$) and bright feature (peaking at $\sim$$1.8$\,Jy\,beam$^{-1}$) centered around $\alpha=23\rah 50\ram$, $\delta=-31\arcdeg$. Depolarization canals, unresolved regions with little or no emission in linear polarization, are further clearly visible; many of them laying preferentially in a north-west orientation. The most prominent depolarization canals appear to be associated with the bright extended feature. In particular, one curves around the lower extent of the feature (starting around $\alpha=0\rah$, $\delta=-35\arcdeg$) and then extends linearly towards the north-west edge of the field (through $\alpha=23\rah30\ram$, $\delta=-25\arcdeg$).

\subsection{Rotation Measure Synthesis}
\label{sec:rmsynthesis}

When propagating through a magnetized plasma, a linearly polarized signal undergoes Faraday rotation. The effect is particularly pronounced at long wavelengths as the magnitude of rotation is proportional to the wavelength squared:

\begin{equation}
\label{eq:rm}
\chi(\lambda^{2}) = \chi_{0} + \phi\lambda^{2}
\end{equation}

where $\chi(\lambda^{2})$ is the measured linear polarization angle (rad) at wavelength $\lambda$ (m), $\chi_{0}$ is the intrinsic polarization angle (rad) and the overall strength of the effect is characterized by the Faraday depth $\phi$ (rad\,m$^{-2}$). The Faraday depth along the sightline to a source is defined as \citep{Burn:1966v133p67B}:

\begin{equation}
\label{eq:fd}
\phi(d) = 0.81 \int_{d}^{0}n_{e}B_{\parallel}\cdot dl\,,
\end{equation}

where $n_{e}$ is the electron density (cm$^{-3}$) and $B_{\parallel}$ is the magnetic field component parallel to the line of sight ($\mu$G). The integral is performed along the line of sight (of which $dl$ is the differential element) from the observer to a distance $d$ (pc). If Faraday rotation is not taken into consideration at long wavelengths, sources at any appreciable Faraday depth will depolarize over the available observing band, an effect known as bandwidth depolarization.

Rotation Measure (RM) synthesis \citep{Brentjens:2005v441p1217} is a technique that takes advantage of the Fourier relationship between the complex polarized intensity as a function of wavelength squared, $P(\lambda^{2})=Q(\lambda^{2})+iU(\lambda^{2})$, and the Faraday dispersion function (FDF) $F(\phi)$ which is the polarized intensity as a function of Faraday depth \citep{Burn:1966v133p67B}, i.e.

\begin{equation}
\label{eq:rms}
P(\lambda^{2})=W(\lambda^{2})\int_{-\infty}^{+\infty}F(\phi)e^{2i\phi\lambda^2}\text{d}\phi
\end{equation}

where $W(\lambda^{2})$ is a weighting function and $\phi$ is the Faraday depth. RM synthesis reconstructs the Faraday dispersion function $F(\phi)$ from an irregularly sampled $P(\lambda^{2})$. The rotation measure spread function (RMSF), which is the Faraday depth equivalent of the point spread function, is the Fourier transform of the weighting function and depends on bandwidth, channel weighting and wavelength.

In general, frequency channels may be weighted by $W(\lambda^{2})$ to account for varying sensitivity across the band. However, measuring the Q and U image noise in the presence of large-scale structures that vary dramatically as a function of frequency is problematic. To simplify processing, we have weighted all frequency channels in the image cubes equally, i.e. $W(\lambda^{2})=1$. We anticipate only a slight loss in overall sensitivity resulting from this choice of weighting scheme as the observed sensitivity across the band is relatively smooth when measured in uniformly weighted images. Using definitions from \citet{Brentjens:2005v441p1217}, for the 154\,MHz band with 160\,kHz channels, the resulting RMSF provides a resolution of $\delta\phi$=2.3 rad\,m$^{-2}$, maximum-scale size sensitivity of 1.0 rad\,m$^{-2}$ and Faraday depth range of $\lvert\phi_{max}\rvert$=160 rad\,m$^{-2}$. As the maximum-scale size is smaller than the resolution $\delta\phi$, these observations cannot resolve Faraday thick structures.

The incomplete sampling available in $\lambda^{2}$ results in side-lobes at about the $10\%$ level in the RMSF. These have been accounted for by using the RM clean algorithm, as described by \citet{Heald:2009v259p591}. In summary, the RM clean algorithm deconvolves peaks in Faraday space with the RMSF to determine the location and amplitude of clean components. The resulting clean components are then convolved with a Gaussian restoring function that has a FWHM equivalent to the RMSF resolution (i.e. $\delta\phi$=2.3 rad\,m$^{-2}$).

\begin{figure*}[t]
\epsscale{1.17}
\plotone{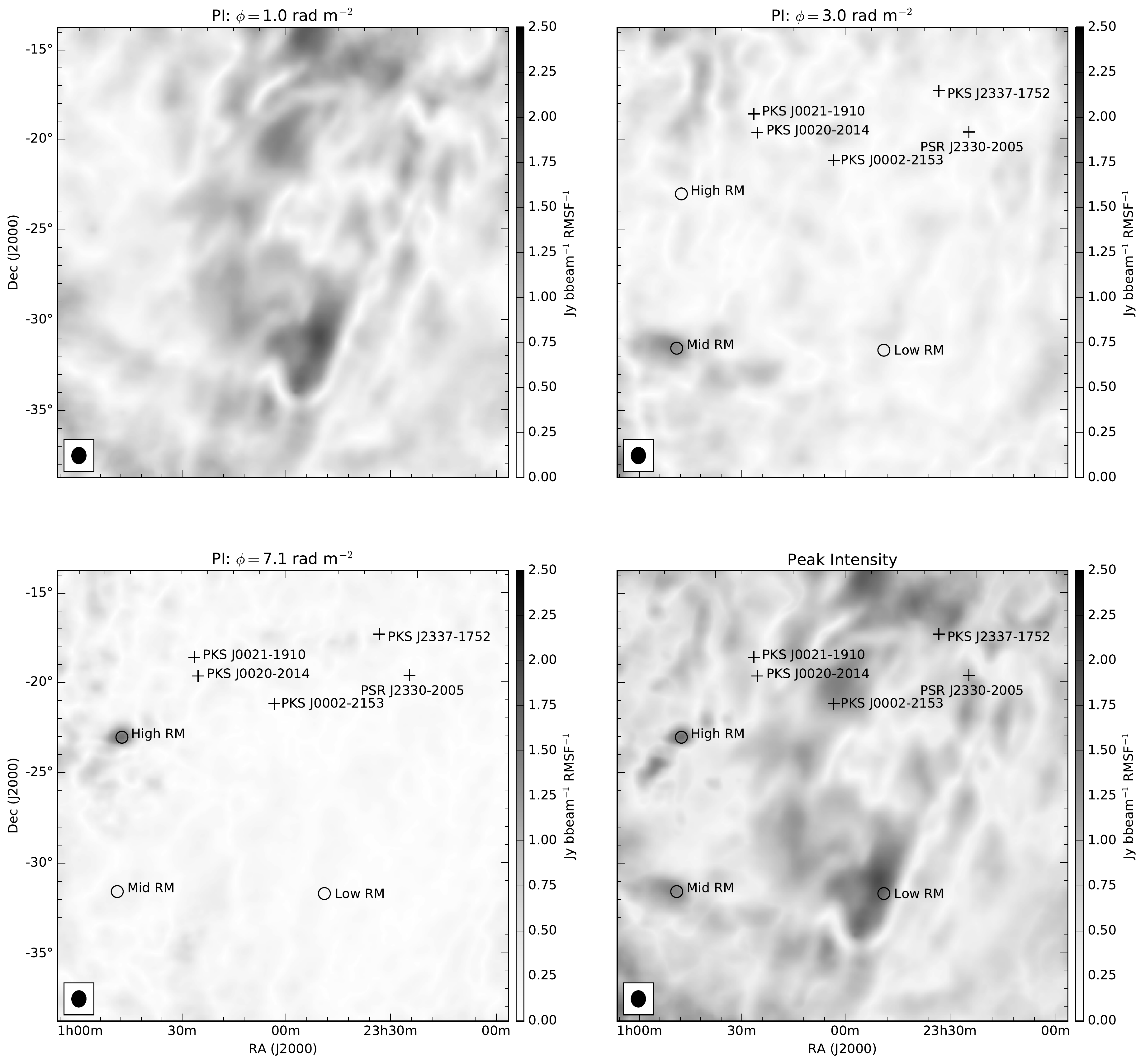}
\caption{Polarized features observed at three Faraday depths in epoch 1 data taken from the RM cleaned cube (corrected for ionospheric Faraday rotation). The RMSF has a FWHM of 2.3\,rad\,m$^{-2}$. [Top Left] $\phi=+1.0$\,rad\,m$^{-2}$. [Top Right] $\phi=+3.0$\,rad\,m$^{-2}$. [Bottom Left] $\phi=+7.1$\,rad\,m$^{-2}$. [Bottom Right] Peak intensities in the Faraday depth spectra at each spatial pixel. The flux scale is in Jy\,beam$^{-1}$\,RMSF$^{-1}$. The synthesized beam, shown as a filled ellipse, is 54$\arcmin\times$47$\arcmin$ FWHM at a position angle of -1.8$\arcdeg$. Circles mark the locations of diffuse features referred to in the text. Crosses mark locations of polarized point sources detected in high resolution imaging; these sources are not visible in the low resolution images as they are dominated by the presence of large-scale structure.}
\label{fig:rmfeatures}
\end{figure*}

\begin{figure*}[t]
\epsscale{1.17}
\plotone{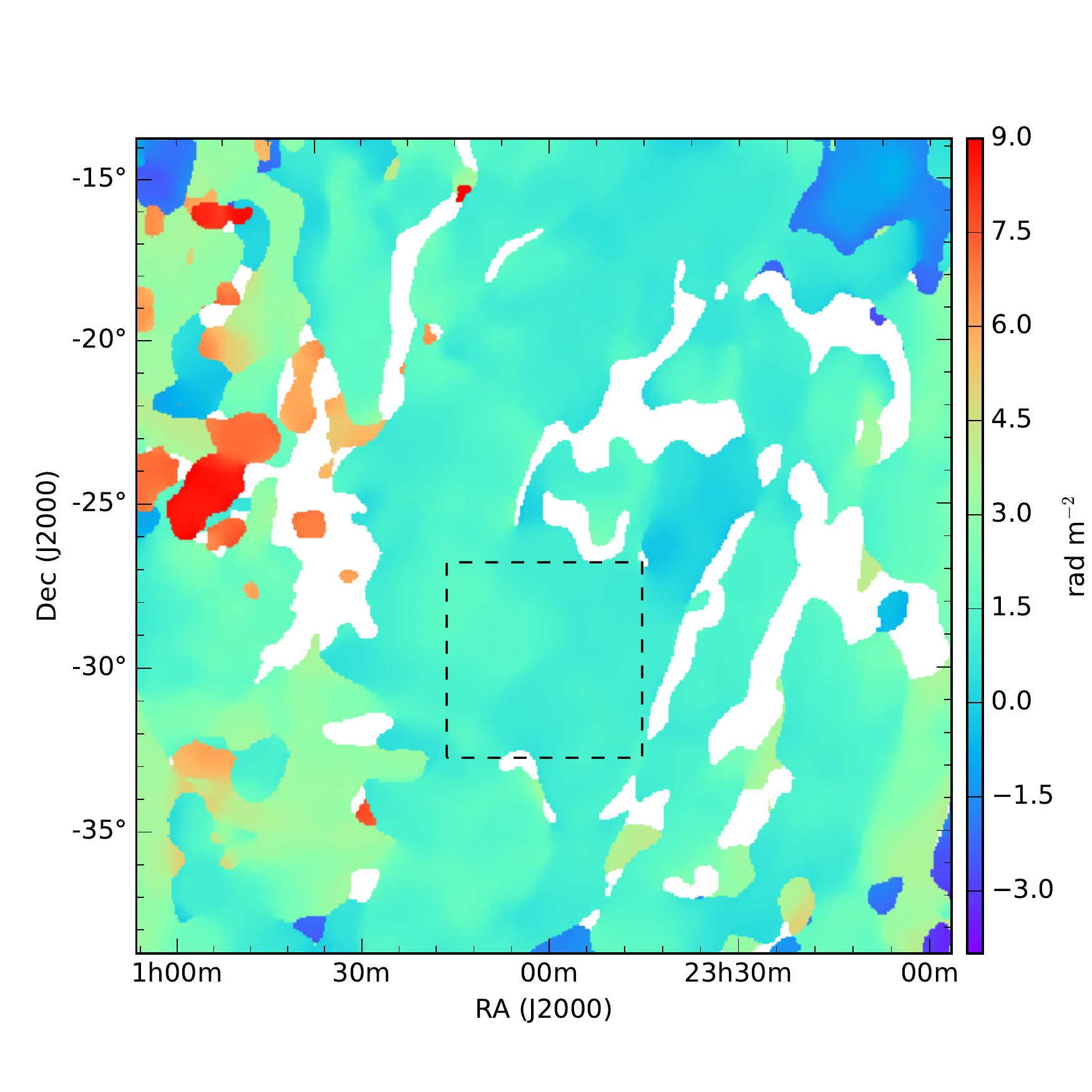}
\caption{Faraday depth measure at peak polarized intensity for each pixel in the Faraday depth cube (corrected for ionospheric Faraday rotation). Regions with low signal to noise have been blanked out. The Faraday depth is measured in rad\,m$^{-2}$. The $6\arcdeg\times 6\arcdeg$ dashed square shows the extent of a region containing both smooth Faraday depth variations and high levels of polarized intensity.}
\label{fig:rmmap}
\end{figure*}

Figure \ref{fig:rmfeatures} highlights features observed in Faraday space at three different Faraday depths ($+1.0$ rad\,m$^{-2}$, $+3.0$ rad\,m$^{-2}$ and $+7.1$ rad\,m$^{-2}$) in the epoch 1 data. The vast majority of the diffuse structure appears at low Faraday depths ($\sim$$1$ rad\,m$^{-2}$) and is dominated by a bright extended feature (labelled ``Low RM'') that was noted in the epoch 1 polarized intensity map (see Figure \ref{fig:eorvp}). Depolarization canals also dominate the entire field-of-view at this Faraday depth with several of the more significant canals oriented in an approximately SE$-$NW alignment. At a Faraday depth of $+3.0$ rad\,m$^{-2}$ the bulk of the features seen at $+1.0$ rad\,m$^{-2}$ are gone and are replaced by a number of $\sim$$2\arcdeg-4\arcdeg$ wide structures in the SE corner of the EoR-0 field (the brightest of which is labelled ``Mid RM''). At $+7.1$ rad\,m$^{-2}$, a small, barely resolved, feature is seen to the east (labelled as ``High RM''). Just SE of this source is a slightly more extended component that peaks at $\sim$$+9$ rad\,m$^{-2}$. The mid-to-high RM features may be associated with the increased level of diffuse polarized emission observed by \cite{Bernardi:2013v771p105} towards the south Galactic pole at similar Faraday depths. Figure \ref{fig:rmmap} shows the Faraday depth at peak emission in the Faraday depth cube for each line of sight in the field. The figures show that the majority of the EoR-0 field is dominated by features at low Faraday depths and that these features vary quite smoothly across the field. Apart from the small number of sources already described at higher Faraday depths, there are a number of weak features at negative Faraday depths near the edge of the field; these are not likely to be associated with real features and are caused by a combination of decreased sensitivity at the edge of the field and sidelobe structures contaminating the field at low signal to noise.

\begin{figure*}[t]
\epsscale{1.2}
\plotone{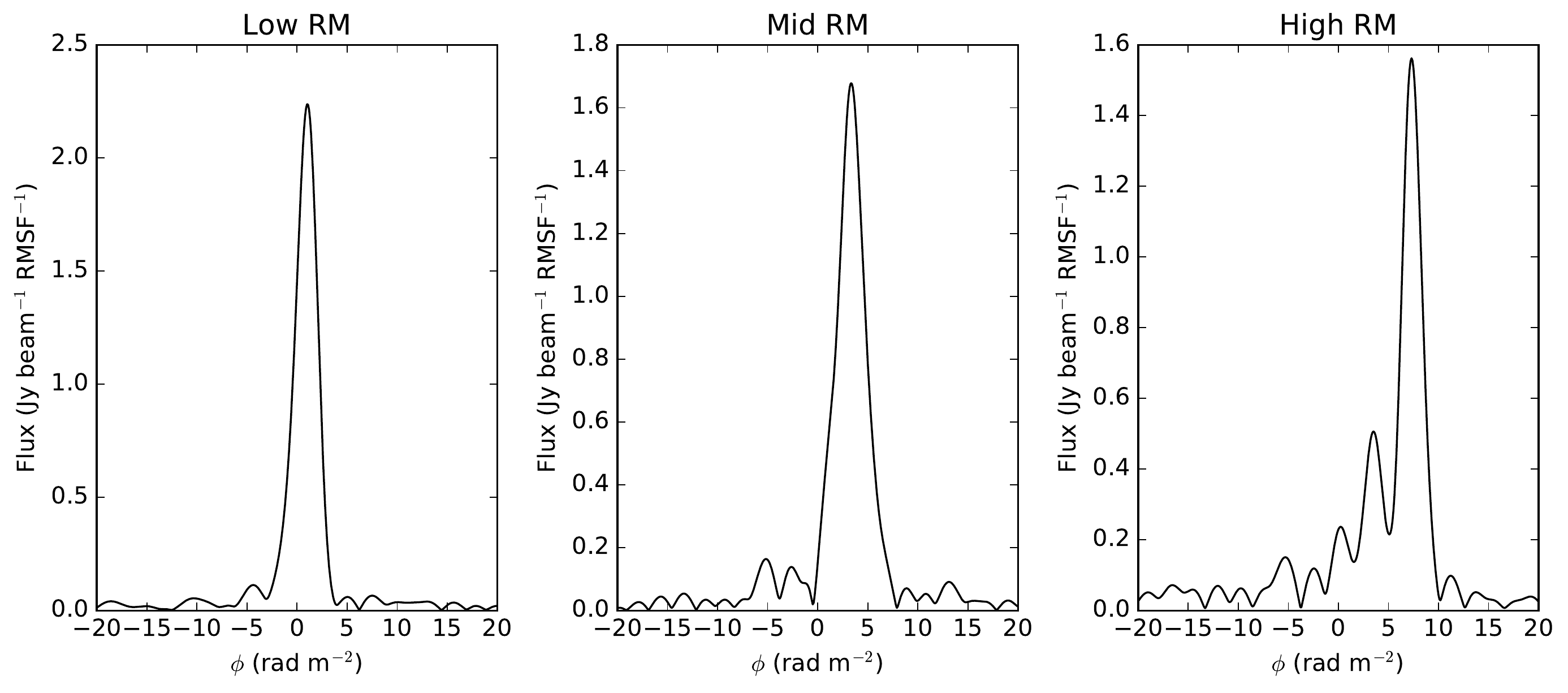}
\caption{Sample Faraday dispersion functions (corrected for ionospheric Faraday rotation) shown for the low ($\phi=+1.0$ rad\,m$^{-2}$), medium ($\phi=+3.0$ rad\,m$^{-2}$) and high ($\phi=+7.1$ rad\,m$^{-2}$) RM features highlighted in Figure \ref{fig:rmfeatures}. The three Faraday dispersion functions shown here correspond to the three peaks visible in the RM distribution plot shown in Figure \ref{fig:rmdist}. No features above the noise floor are seen outside of the Faraday range shown here. The Faraday dispersion functions were deconvolved with \textsc{rmclean} \citep{Heald:2009v259p591}.}
\label{fig:fdfsample}
\end{figure*}

The deconvolved Faraday dispersion functions for the samples taken in the low and high RM regions are shown in Figure \ref{fig:fdfsample}. The residual rms is 16\,mJy\,beam$^{-1}$\,RMSF$^{-1}$, 42\,mJy\,beam$^{-1}$\,RMSF$^{-1}$ and 29\,mJy\,beam$^{-1}$\,RMSF$^{-1}$ for the low, mid and high RM sources, respectively. The high RM FDF appears to contain more than one peak: a main peak at $\phi=+7.2$ rad\,m$^{-2}$, an intermediate 530\,mJy\,beam$^{-1}$\,RMSF$^{-1}$ peak at $\phi=+3.5$ rad\,m$^{-2}$ and a 230\,mJy\,beam$^{-1}$\,RMSF$^{-1}$ peak at $\phi=+0.7$ rad\,m$^{-2}$. The additional minor peaks in the high RM FDF fluctuate between epoch 1 and 3 and are due to side lobe contamination which introduces frequency dependent structure into the Faraday spectra that is also time dependent.

\subsection{RM distribution and Ionospheric Faraday Rotation}
\label{sec:ionfr}

The ionosphere affects observations through positional shifts of background sources and through Faraday rotation of linearly polarized signals. A gradient in total electron content (TEC) of the ionosphere across the field-of-view will result in positional shifts of sources; this has been observed previously with the MWA and studied in detail \citep{Loi:2015v42p3707,Loi:2015v50p574,Loi:2015v453p2731}. For diffuse polarization the effect is an order of magnitude smaller than the size of the features being studied and can safely be ignored.

\begin{figure*}[t]
\epsscale{1.2}
\plotone{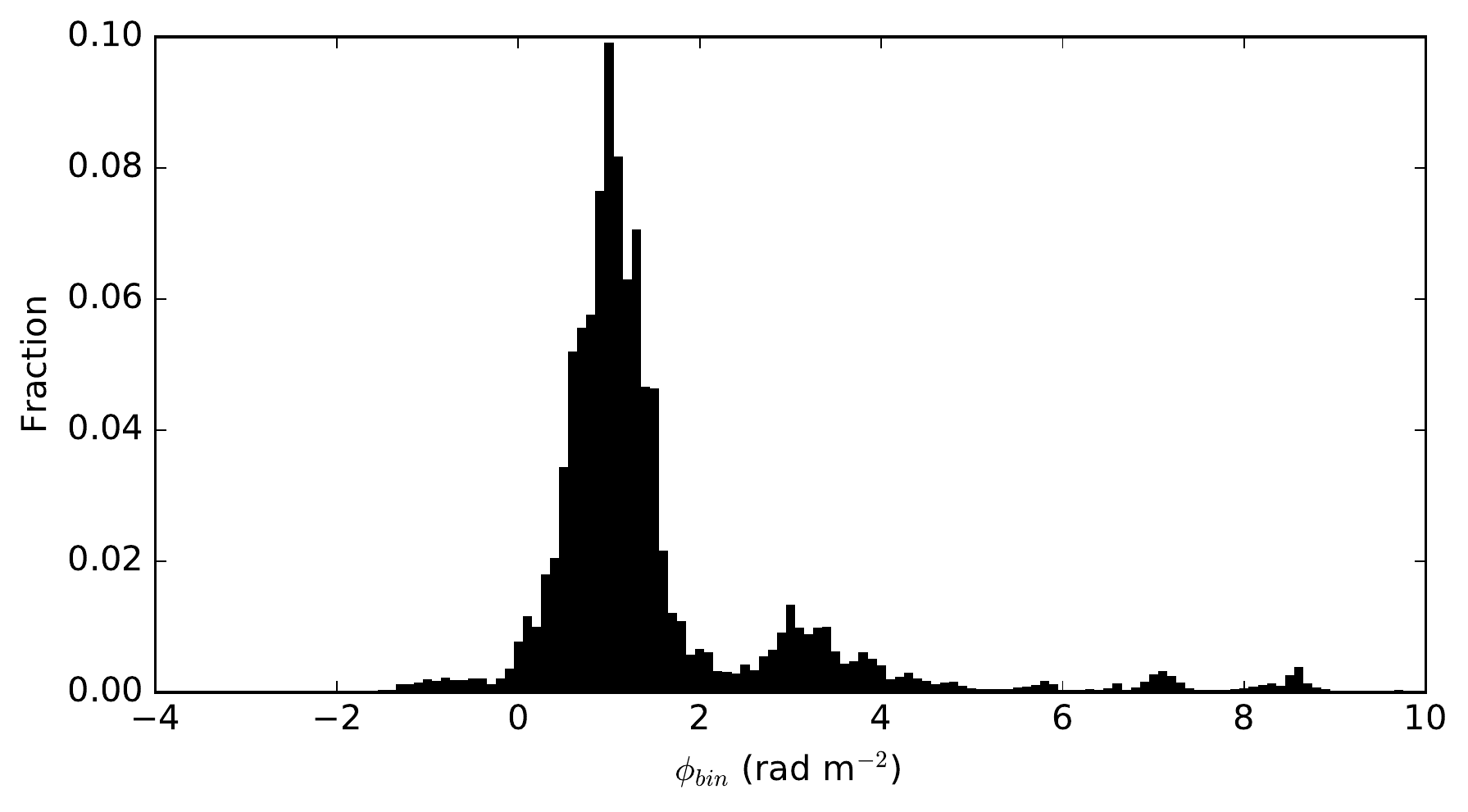}
\caption{Distribution of rotation measures in the EoR-0 field after ionospheric correction and performing \textsc{rmclean} to remove sidelobes in Faraday space. The binning width is 0.1 rad\,m$^{-2}$. The main peak of the distribution is at 1.0 rad\,m$^{-2}$. Sub-peaks in the distribution are due to features present at higher Faraday depths.}
\label{fig:rmdist}
\end{figure*}

The absolute TEC, in combination with the magnetic field in the direction being observed, can measurably contribute to the observed RM of a background source. The TEC at a given time, from a given location on the Earth, observed towards a particular line of sight, can be estimated based on Global Positioning System (GPS) models \citep{Arora:2015v32p29}. Using the TEC estimated by these models in combination with terrestrial magnetic field models, the predicted ionospheric component of Faraday rotation may be determined. An implementation of these models can be found in the \textsc{albus} \citep{Willis:2016} package. We used \textsc{albus} to determine the mean Faraday rotation introduced as a result of the ionosphere during the course of the epoch 1 observation and estimated it to be $-0.7\pm0.2$ rad\,m$^{-2}$. When corrected for ionospheric Faraday rotation, the distribution of RM in the EoR-0 field peaks at 1.0 rad\,m$^{-2}$ with a standard deviation of $\sigma_{\phi}=0.34$ rad\,m$^{-2}$; see Figure \ref{fig:rmdist}. A few further sub-peaks are seen within this distribution but the vast majority of features are contained within $-2<\phi<+10$ rad\,m$^{-2}$.

\begin{figure*}[t]
\epsscale{1.1}
\plotone{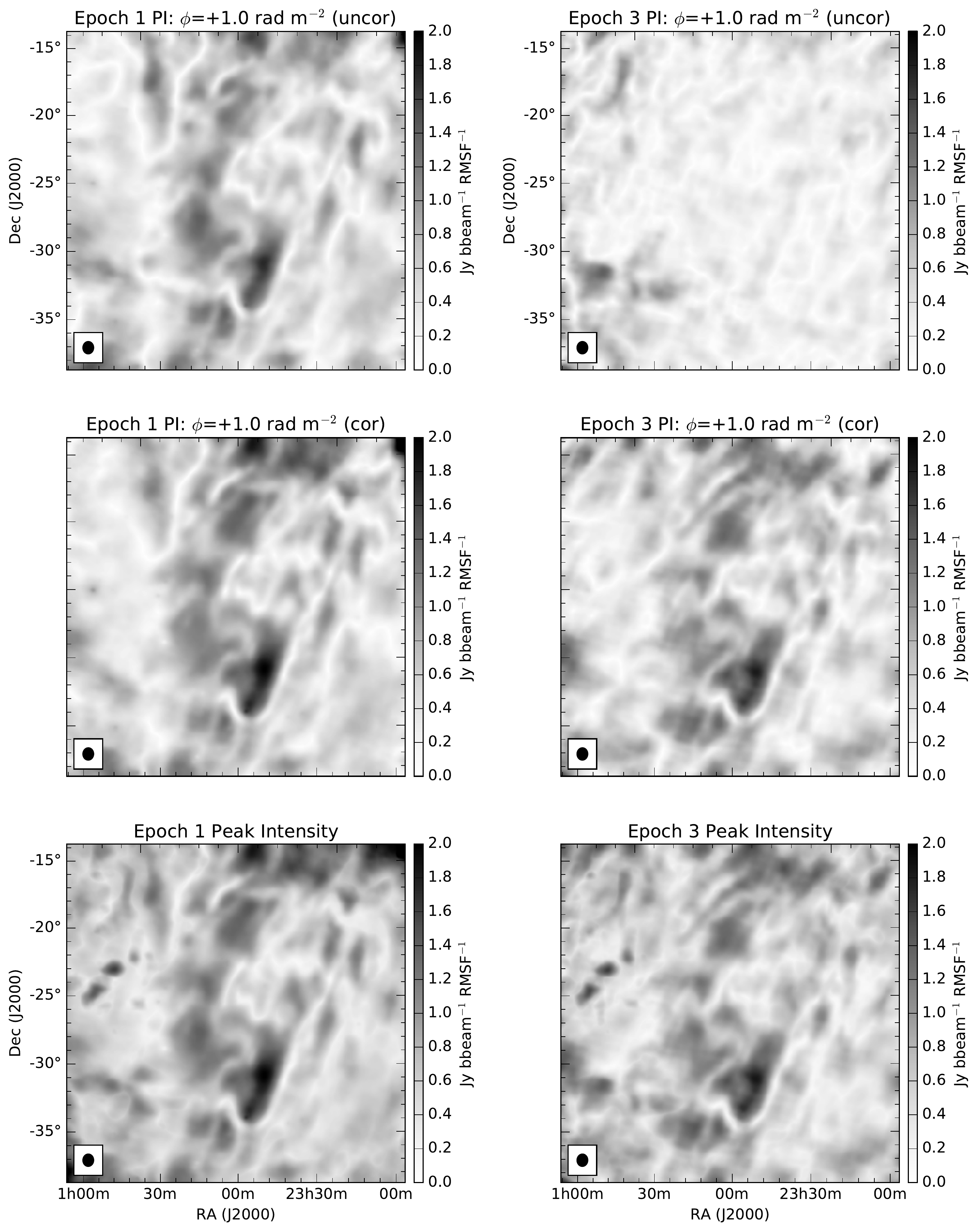}
\caption{Effect of ionospheric Faraday rotation on diffuse polarization as observed in epochs 1 and 3. Top: Polarized intensity at $\phi=+1.0$ rad\,m$^{-2}$ before correcting for ionospheric Faraday rotation. Center: Polarized intensity at $\phi=+1.0$\,rad\,m$^{-2}$ after correcting for ionospheric Faraday rotation. Bottom: Peak intensity in the Faraday spectra at each spatial pixel. \textsc{albus} reports an ionospheric Faraday rotation of $-0.7$\,rad\,m$^{-2}$ in epoch 1 and $-2.2$\,rad\,m$^{-2}$ in epoch 3. Differences observed near the edge of the field between the two epochs after ionospheric correction are due to primary beam errors and sidelobe confusion.}
\label{fig:eoriono}
\end{figure*}

When images are compared across the three epochs at 154\,MHz, the total intensity maps remain unchanged. However, there are clear differences in the linear polarization maps, particularly between epoch 1 and the two subsequent epochs. Using RM synthesis and searching for peak emission in Faraday depth for each line of sight revealed that all of the observed structures were consistent between the epochs but that they had been shifted in Faraday space. This is shown in Figure \ref{fig:eoriono} for the epoch 1 and epoch 3 data; here the peak intensity maps are consistent but the peak emission occurs at $+0.3$\,rad\,m$^{-2}$ in epoch 1 and at $\phi=-1.2$\,rad\,m$^{-2}$ in epoch 3. Both \textsc{albus} \citep{Willis:2016} and \textsc{ionfr} \citep{SotomayorBeltran:2013v552p58S} predict a shift in the ionospheric component of $\sim$$-1.5$\,rad\,m$^{-2}$ in Faraday rotation from epoch 1 to epoch 3 (ionospheric RM of $-0.7$\,rad\,m$^{-2}$ for epoch 1 and $-2.2$\,rad\,m$^{-2}$ for epoch 3). When these shifts are applied to the Faraday depth cubes, the peak emission for both epochs occurs at $\sim$$1.0$\,rad\,m$^{-2}$, verifying that the shift is associated with the ionosphere and not caused by variability or the instrument. \edit1{The significantly higher Faraday rotation induced by the ionosphere in epoch 3, most likely as a result of a known Coronal Mass Ejection event \citep{Kaplan:2015v809p12}, also explains the structural differences observed between the two epochs in the band-averaged Q and U images shown in Figure \ref{fig:eoriqu}.} This highlights the need for a correction to mitigate the effects of ionospheric Faraday rotation. However, it also demonstrates that the ionosphere is quite stable as a function of time and direction over the MWA field-of-view and that, to first order, a single shift in Faraday depth (as opposed to a grid of multiple direction dependent shifts) is sufficient to correct for these ionospheric effects. 

\begin{figure*}[t]
\epsscale{1.2}
\plotone{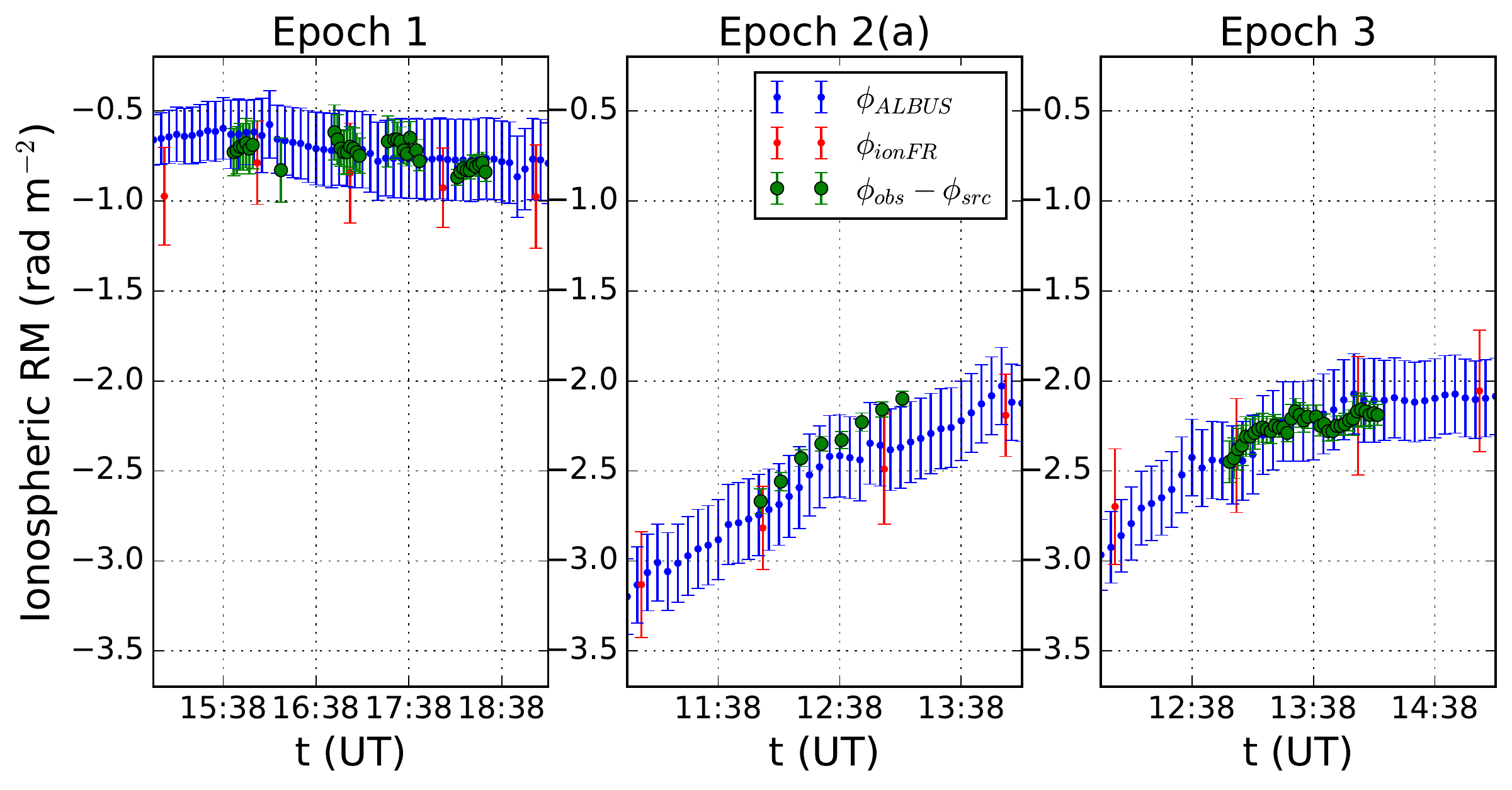}
\caption{Measurements of ionospheric Faraday rotation ($\phi_{obs}$ is the observed Faraday rotation of a source and $\phi_{src}$ is the previously determined ionosphere-corrected Faraday rotation of the source) at the position of the high RM feature shown in Figure \ref{fig:rmfeatures} for each observed snapshot and epoch (green points). The blue and red points are the predicted components of ionospheric Faraday rotation from \textsc{albus} ($\phi_{ALBUS}$) and \textsc{ionfr} ($\phi_{ionFR}$), respectively. Note that an additional seven 2 minute snapshots from epoch 2(a) were used here to track the high RM feature as it drifted through the zenith-pointed beam of that observation, however those snapshots were not used in subsequent processing as the bulk of the EoR-0 field had already drifted out of the field-of-view.}
\label{fig:rmvst2}
\end{figure*}

An interesting possibility with the MWA, given the high sensitivity to diffuse structures that the instrument provides, is to use the diffuse polarized background to track and measure the influence of ionospheric and heliospheric Faraday rotation. An estimate of the ionosphere-corrected Faraday rotation towards a source of bright diffuse emission ($\phi_{src}$) can be determined by performing a least-squares fit that minimizes $\phi_{src} - (\phi_{obs} - \phi_{ALBUS})$ over multiple observing snapshots and/or epochs, where $\phi_{obs}$ is the observed Faraday rotation towards the source and $\phi_{ALBUS}$ is the estimated ionospheric Faraday rotation at the time of the observation. By observing over multiple epochs, to overcome the relatively large errors associated with individual predictions of $\phi_{ALBUS}$, the overall error in $\phi_{src}$ can be minimized. Once an estimate of the ionosphere-corrected Faraday rotation is established, it can be used to estimate the ionospheric component of Faraday rotation at that source location at any epoch. Using this approach for all available snapshots in each of the three epochs, the fit to $\phi_{src}$ of the high RM source (see Figure \ref{fig:rmfeatures}) was determined to be $\phi_{src}=7.20\pm0.01$ rad\,m$^{-2}$. Figure \ref{fig:rmvst2} plots the ionospheric component ($\phi_{obs}-\phi_{src}$) at the high RM source location for each snapshot and epoch. The measured component tracks both predictive models quite well from epoch to epoch and even from snapshot to snapshot. This demonstrates that observations of diffuse polarization may allow the effects of the ionosphere to effectively be calibrated in fields where the RM structure has been previously determined, without the need to resort to predictive models. The technique also has the potential to aid ionospheric studies by mapping ionospheric changes both temporally and spatially over a wide field-of-view. In combination with predictive models, this technique may also provide a means to detect and track the propagation of space weather events, as caused by coronal mass ejections or solar flares, by observing the shift they impart on the RM signature of the diffuse polarized background \citep{Oberoi:2004v52p1415}.

\subsection{Gradient maps}
\label{sec:gmaps}

To examine filamentary magnetized structures believed to result from turbulence in the local interstellar medium \citet{Burkhart:2012v749p145}, polarization gradient maps of the Stokes vector (Q and U) were formed using the method described by \citet{Gaensler:2011v478p214}. The polarization gradient function is defined as:

\begin{equation}
\label{eq:pgf}
\lvert \vec {\nabla P} \rvert = \sqrt{\left(\frac{\partial Q}{\partial x}\right)^{2} + \left(\frac{\partial U}{\partial x}\right)^{2} + \left(\frac{\partial Q}{\partial y}\right)^{2} + \left(\frac{\partial U}{\partial y}\right)^{2}}\,.
\end{equation}

\begin{figure*}[t]
\epsscale{1.2}
\plotone{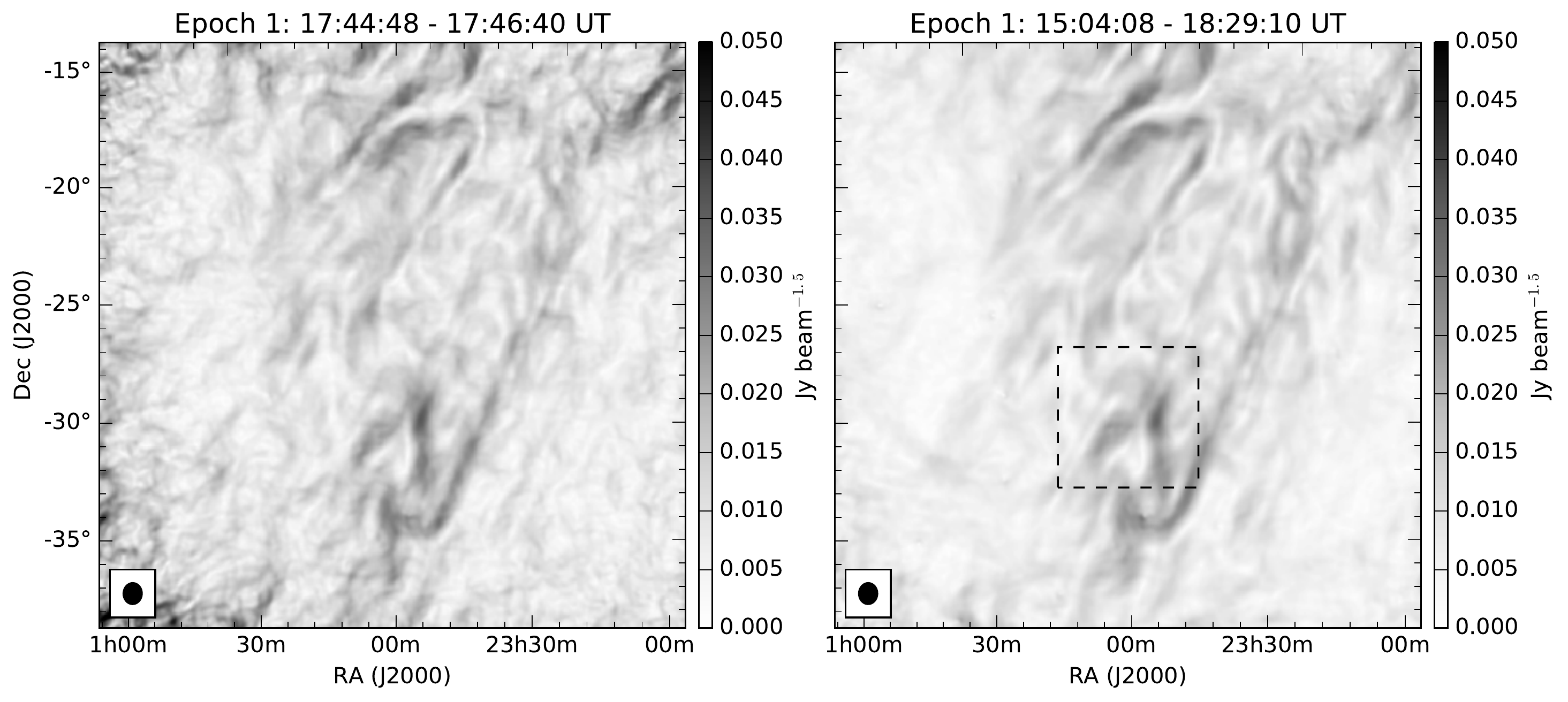}
\caption{Left: An ionosphere-corrected polarization gradient map from a single 112 s snapshot of epoch 1 data. Right: The ionosphere-corrected polarization gradient map derived from time-averaged  epoch 1 data ($44\times112$ s snapshots). Units are in Jy\,beam$^{-1.5}$. The $6\arcdeg\times 6\arcdeg$ dashed square shows the extent of a region containing both smooth Faraday depth variations and high levels of polarized intensity.}
\label{fig:eorgt}
\end{figure*}

Time-dependent effects were examined by comparing changes in the gradient map from snapshot to snapshot and against the time-averaged data set. Figure \ref{fig:eorgt} shows gradient maps from the epoch 1 observation using a single 112 s snapshot compared against a gradient map using all of the available data from that epoch. When corrected for ionospheric effects, the structures seen in these gradient maps are centered at a Faraday depth of $+1.0$ rad\,m$^{-2}$ as this is where the bulk of the polarized emission exists (see Figure \ref{fig:rmdist}). 

The most prominent features in Figure \ref{fig:eorgt} remain stable as a function of time. Minor changes can be seen in the fainter structures and these are primarily due to a combination of image noise; sidelobe confusion; and errors associated with incomplete $(u,v)$ sampling. We note that the first few snapshots of epoch 1 are detrimentally affected by the presence of the Galactic plane within a far sidelobe as this epoch included low elevation beam-former pointings that were not used in the epoch 2 or epoch 3 data. The projected baselines of the MWA are severely foreshortened for sources at low elevation and the array is particularly sensitive to bright and extended sources in those locations \citep{Thyagarajan:2015v804p14}. Nonetheless, when integrating data over a wide range of hour angles, the source sidelobe effects are diluted and the exclusion of the affected snapshots has a minimal impact on the final integrated gradient map. Gradient maps were also produced for the remaining epochs at 154\,MHz but no significant changes were observed once the maps were corrected for ionospheric Faraday rotation. 

\begin{figure*}[t]
\epsscale{1.18}
\plotone{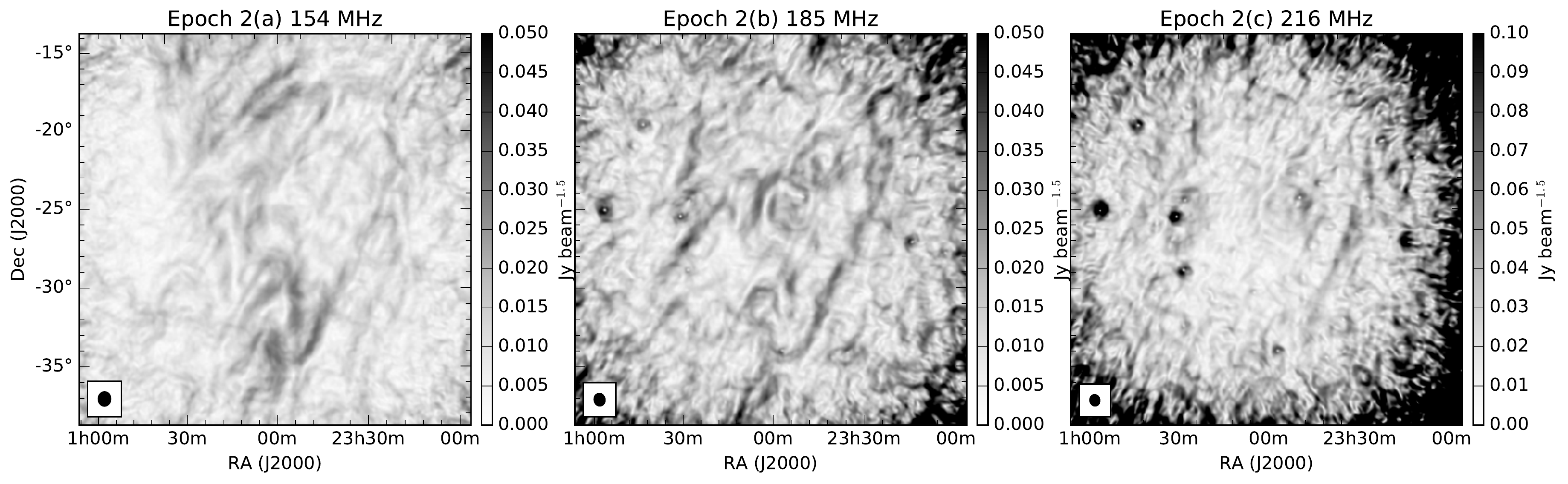}
\caption{Gradient maps from ionosphere-corrected epoch 2 data of the EoR-0 field in three available frequency bands: 154\,MHz (left), 185\,MHz (center) and 216\,MHz (right). Clear evidence of polarization leakage, primarily contamination from Stokes I point sources, is seen in the higher frequency bands where beam model errors are more apparent. The available field-of-view also decreases as a function of increasing frequency and so the higher frequency maps exhibit increased levels of noise at the edge of the field. }
\label{fig:gleamgv}
\end{figure*}

To examine the evolution of gradient map features as a function of observing frequency, gradient maps were created for the three GLEAM bands, i.e. epochs 2(a), 2(b) and 2(c). The EoR-0 field was only fully visible in the first snapshot of the epoch 2 data as the Sun was in the process of setting just as the field was passing through zenith. As such, the sensitivity is limited to that of a single snapshot in these data. Figure \ref{fig:gleamgv} shows the resulting gradient maps for each band of epoch 2 data, with channels across each 30.72\,MHz band averaged to increase signal to noise. The point-like sources that begin to appear in epoch 2(b) and dominate in epoch 2(c), result from apparent polarization leakage from bright Stokes I sources. The level of leakage increases with frequency and angular distance from zenith. For the most part, this leakage is due to errors in the primary beam model and will be reduced once improved beam models are implemented into the imaging pipeline \citep{Sutinjo:2015v50p52S}. The leakage in epoch 2(a) is minimal as this is where the beam model and instrument were designed to perform optimally; thus its behaviour is well-defined. Increased noise levels are also evident at the edge of the field in the higher frequency bands as a result of the decreased field-of-view available in those bands.

Comparing the gradient maps in the three different bands, the dominant features are stationary with respect to spatial coordinates in the epoch 2(a) and 2(b) images. Some features, such as the linear feature that runs from NW to SE near the western edge, remain persistent over all three of the bands. The weaker structures are more difficult to trace, particularly in epoch 2(c), as a result of the poor sensitivity available in a single snapshot and systematic issues associated with the beam, noise and available field-of-view.  

\begin{figure*}[t]
\animategraphics[controls,width=\textwidth]{25}{f13_}{00}{23}
\caption{Epoch 1 EoR-0 gradient maps for 24 $\times$ 1.28\,MHz channels across the 154\,MHz band. Note that \textsc{Adobe Acrobat Reader} is required to view the animated gradient map cube shown here.}
\label{fig:eorgv2}
\end{figure*}

An alternative view of the frequency dependent behaviour of the polarized gradients in the field can be obtained from the deep epoch 1 EoR-0 observations, albeit over the limited range of frequencies available in that observation ($138.88$--$169.60$\,MHz). The epoch 1 data provide sufficient sensitivity, as a result of the longer tracking observation available, to study the polarization gradient evolution on a per-160\,kHz channel basis.  Figure \ref{fig:eorgv2} presents an animation of the gradient map as a function of frequency across the 30.72\,MHz band of the 154\,MHz epoch 1 observations; in this animation, subsets of $6\times160$\,kHz channels have been frequency-averaged to form a smaller number of 1.28\,MHz channels. The polarized gradients are now more prominent in each of the frequency channels and can be seen to vary smoothly in intensity as a function of frequency but show no significant spatial movement. In general, once features appear at lower frequencies, they continue to persist up to the higher frequency gradient maps. For example, the gradient around the bright polarized feature labelled "Low RM" in Figure \ref{fig:rmfeatures} persists over the entire band. However, an SE to NW gradient appears towards the east and west only in the upper portion of the band. Similarly, features in the northern part of the image, some forming loop-like structures, also only appear in the upper end of the band.

\subsection{Polarized point sources}
\label{sec:pointsources}

\begin{figure}[t]
\epsscale{1.1}
\plotone{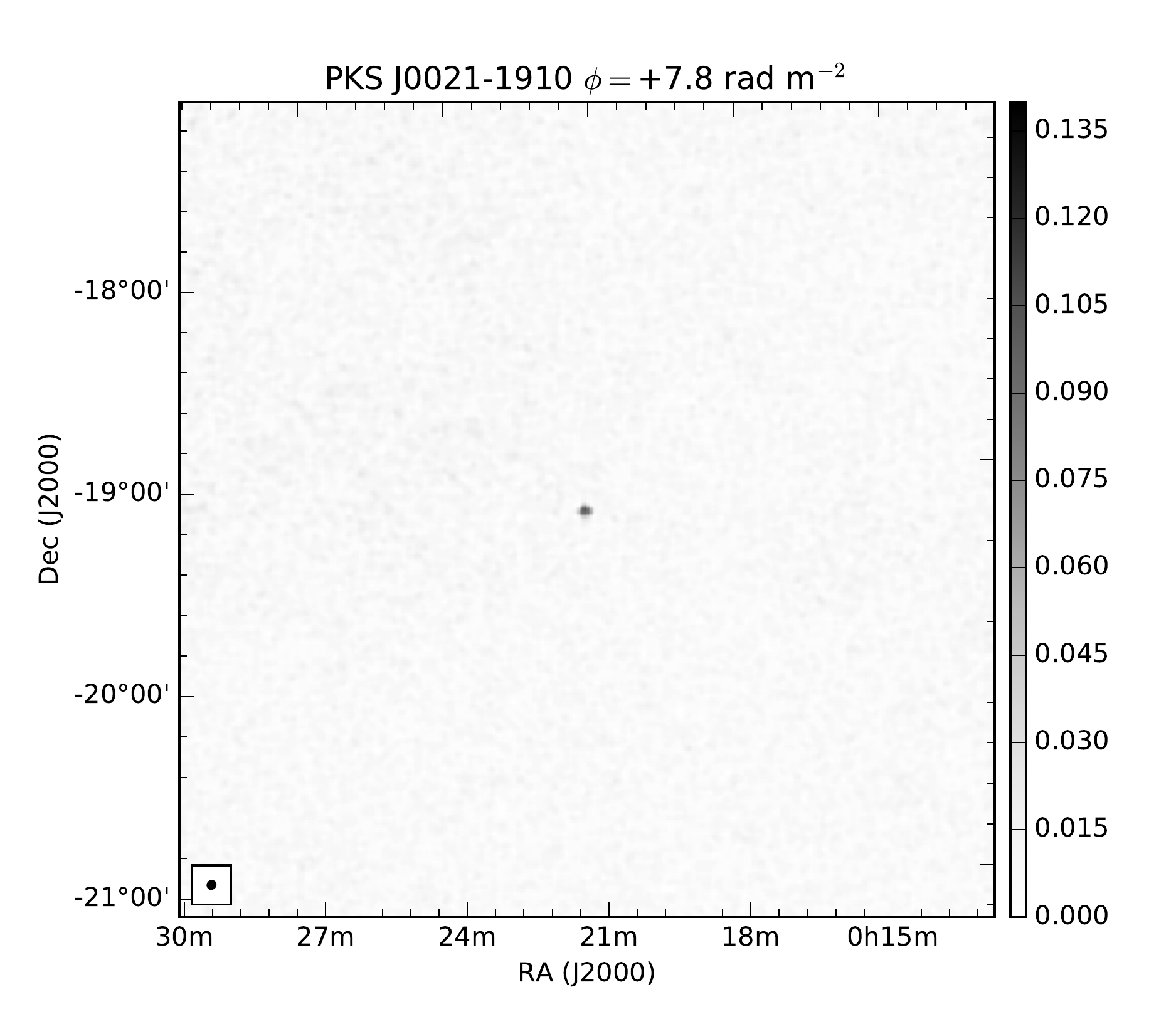}
\caption{Cut-out showing the polarized source PKS J0021$-$1910. The synthesized beam is 2.2\arcmin$\times$2.4\arcmin FWHM with a position angle of -47\arcdeg.}
\label{fig:pksimage}
\end{figure}

\begin{figure}[t]
\epsscale{1.0}
\plotone{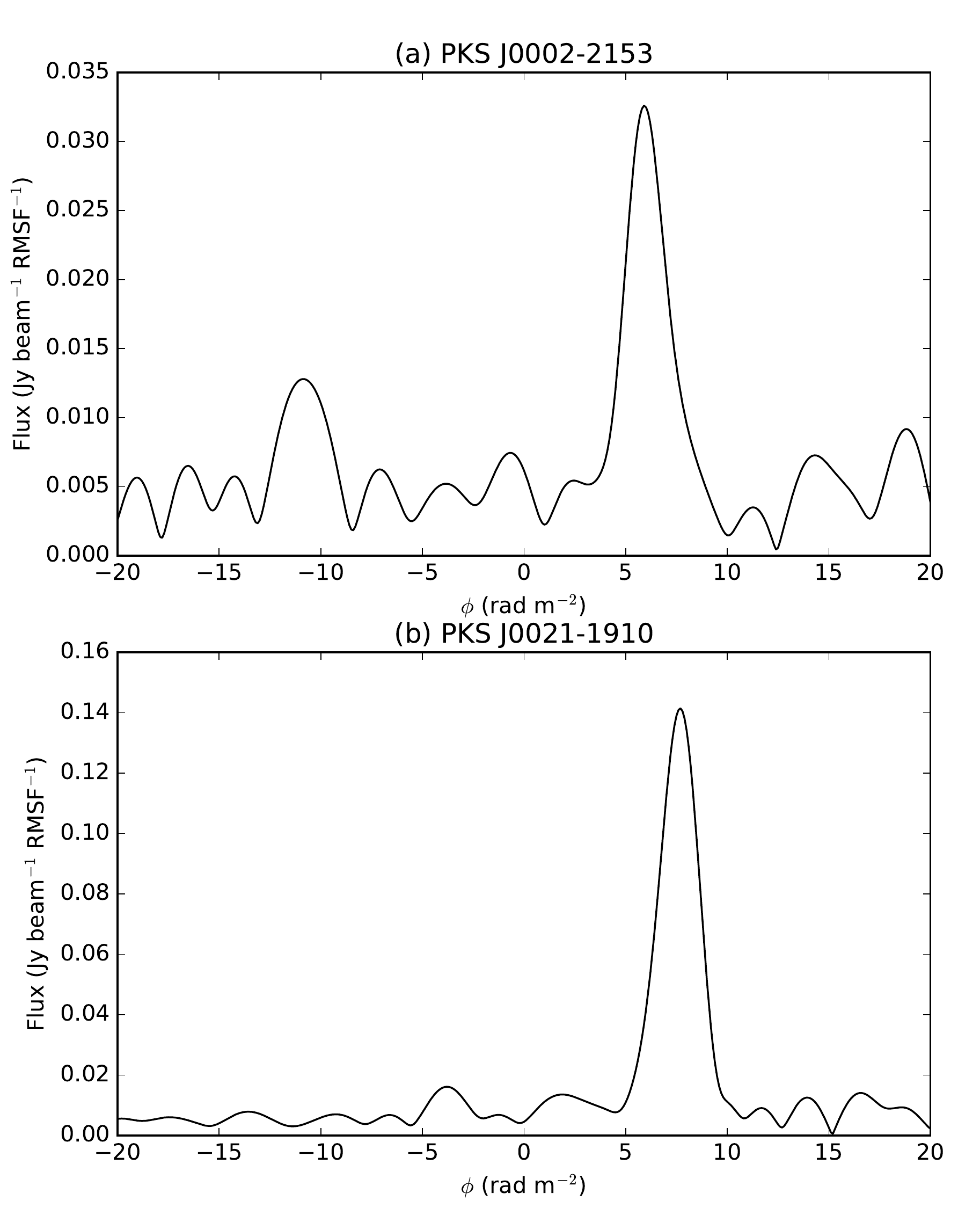}
\caption{Sources detected with high-resolution imaging. (a) Faraday dispersion function for PKS J0002$-$2153. RM$=+5.7$ rad\,m$^{-2}$. Noise is 2.5\,mJy\,beam$^{-1}$ RMSF$^{-1}$ rms. (b) Faraday dispersion function for PKS J0021$-$1910. RM$=+7.8$ rad\,m$^{-2}$. Noise is 3.1\,mJy\,beam$^{-1}$ RMSF$^{-1}$ rms.}
\label{fig:fdfpks1}
\end{figure}

The uniformly weighted polarization image cubes generated using all of the longest MWA baselines are well suited for the detection of polarized point sources. An initial inspection of the RM cube, created using the uniformly weighted Stokes Q and U cubes, reveals a clear detection of the extragalactic source PKS J0021$-$1910 (PKS B0018$-$194). A cut-out image for the source and its associated Faraday dispersion function are shown in Figure \ref{fig:pksimage} and Figure \ref{fig:fdfpks1}, respectively. In the uncorrected epoch 3 data the source has an RM of $+5.6$ rad\,m$^{-2}$ and is detected with a signal to noise of 10 in each snapshot. When corrected for ionospheric effects using \textsc{albus} predictions of the ionospheric Faraday rotation, a fit of $\phi=+7.8\pm0.1$ rad\,m$^{-2}$ is obtained for this source. With a total intensity of 4.7\,Jy\,beam$^{-1}$ and a polarized intensity of 140\,mJy\,beam$^{-1}$, the source is 3.0\% polarized.  The polarimetric characteristics of the source are consistent with a measurement made at 1.4\,GHz, RM $=3.6\pm 5.2$\,rad\,m$^{-2}$ and 3.67\% polarization \citep{Taylor:2009v702v1230}.

A subsequent search was performed concentrating on the locations of known polarized sources, using the \citet{Taylor:2009v702v1230} catalogue as a reference. The catalogue contains 399 polarized sources within the 400 sq. degree region imaged around the EoR-0 field. We use a conservative 14$\sigma$ cut-off in the time-averaged data cube to ensure that spurious detections are not made as a result of polarization leakage from bright Stokes I sources and the associated sidelobe structure this leakage introduces into the Faraday spectra. Any sources with $\lvert\text{RM}\rvert<1$ rad\,m$^{-2}$ were also filtered out; as these would most likely trigger false-positives as a result of polarization leakage. In all, two sources were detected: PKS 0002$-$2153 (PKS B2359$-$221) and PKS J0021$-$1910 (PKS B0018$-$194). The Faraday dispersion functions for these sources are shown in Figure \ref{fig:fdfpks1}.

\begin{figure*}[t]
\epsscale{1.2}
\plotone{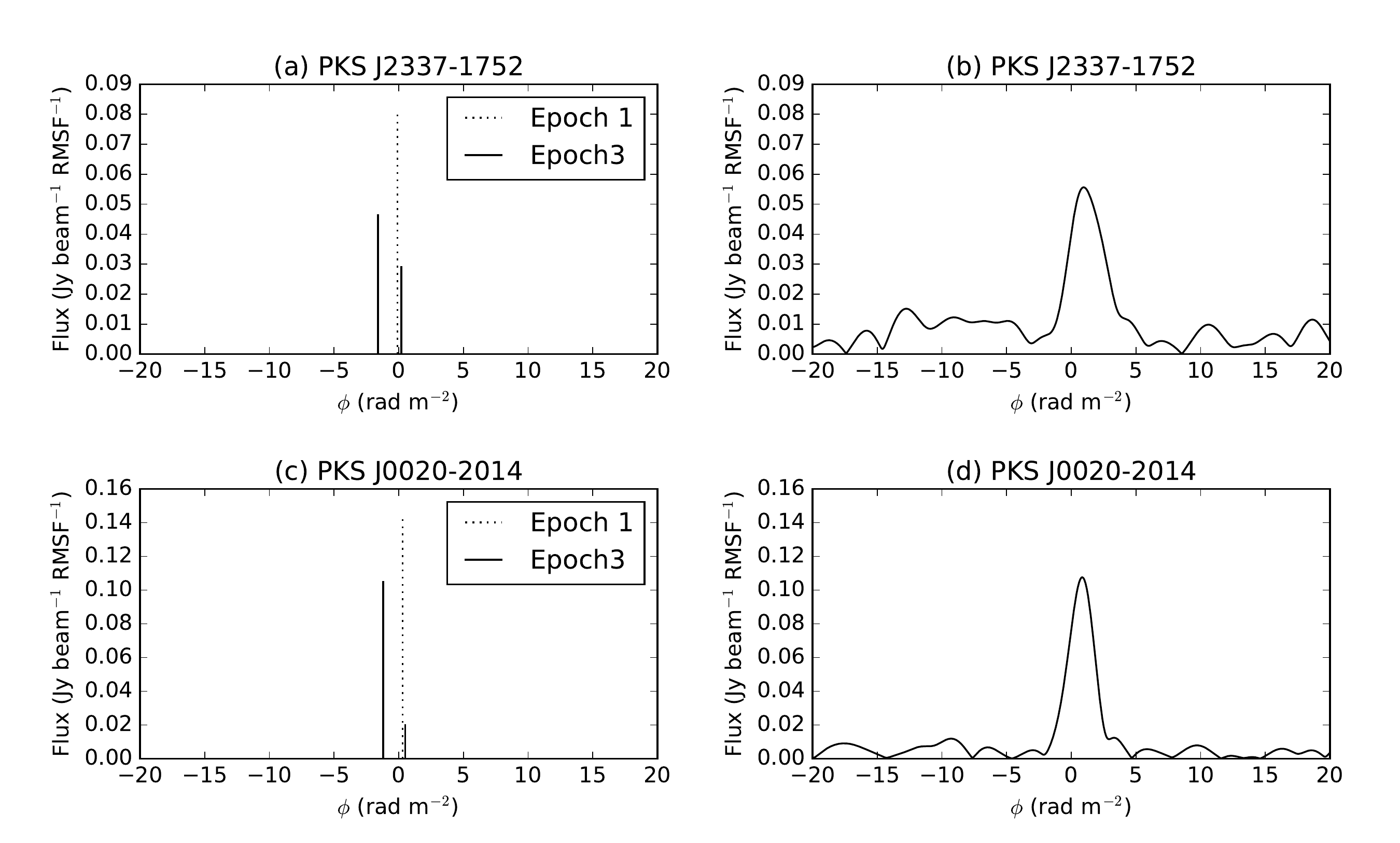}
\caption{Sources found in a multi-epoch analysis that distinguishes real sources from instrumental effects by searching for peaks shifted in Faraday space as a result of ionospheric Faraday rotation. (a) Peaks detected in RM synthesis in epochs 1 and 3 for PKS J2337$-$1752 before correcting for the effects of ionospheric Faraday rotation. (b) Faraday dispersion function for PKS J2337$-$1752 in epoch 3 after correcting for the effects of ionospheric Faraday rotation. RM$=+0.8$ rad\,m$^{-2}$. Noise is 3.9\,mJy\,beam$^{-1}$ RMSF$^{-1}$ rms. (c) Peaks detected in RM synthesis in epochs 1 and 3 for PKS J0020$-$2014 before correcting for the effects of ionospheric Faraday rotation. (d) Faraday dispersion function for PKS J0020$-$2014 in epoch 3 after correcting for the effects of ionospheric Faraday rotation. RM$=+1.0$ rad\,m$^{-2}$. Noise is 3.4\,mJy\,beam$^{-1}$ RMSF$^{-1}$ rms. }
\label{fig:fdfpks2}
\end{figure*}

Since deep observations were made at two epochs, a further test of the sources was made by checking whether their RMs were shifted in Faraday space by an amount that was consistent with the expected shift caused by the different ionospheric conditions between the two epochs. The advantage of this method is that it can identify real sources that were confused with instrumental leakage, because in at least one of the epochs, such a source would be shifted sufficiently away from RM$=0.0$ rad\,m$^{-2}$ to allow a positive identification. All of the previously detected sources were verified using this method and two additional sources were also identified: PKS J2337$-$1752 (PKS B2335$-$181) and PKS J0020$-$2014 (PKS B0017$-$205). Figures \ref{fig:fdfpks2}(a) and \ref{fig:fdfpks2}(c) show the RM synthesis components detected in epochs 1 and 3. In epoch 1 the instrumental component and the source components are confused near RM=0.0 rad\,m$^{-2}$ for both PKS J0020$-$2014 and PKS J2337$-$1752. In epoch 3, the ionosphere clearly shifts the source RM away from the instrumental component, thus enabling a positive identification. The Faraday dispersion functions for these two sources are shown in Figure \ref{fig:fdfpks2}.

The parameters associated with all detected point sources are summarized in Table \ref{table:pointSummary}. All but PKS J2337$-$1752 have RMs consistent with those measured by \citet{Taylor:2009v702v1230}. Not all of the sources appear to have been significantly depolarized at MWA wavelengths compared to observations at 1.4\,GHz, which suggests that there is not a systematic reason to explain the overall small number of detections. The two most highly polarized extragalactic sources, PKS J0020$-$2014 and PKS J0021$-$1910, are also the two largest sources in spatial extent amongst our detected sources. PKS J0020$-$2014 is a giant radio galaxy with a redshift of $z=0.197$ \citep{Ishwara-Chandra:1999v309p100} and an extent of 1.22 Mpc, while PKS J0021$-$1910 is a known double radio source with a redshift of $z=0.0952$ and an extent of 270\,kpc \citep{Reid:1999v124p285}\footnote{Assuming a spatially flat $\Lambda$CDM cosmology with matter density $\Omega_{M}=0.286$, vacuum energy density $\Omega_{\Lambda}=0.714$, and Hubble constant $H_{0} = 69.6$\,km\,s$^{-1}$\,Mpc$^{-1}$ \citep{Wright:2006v118p1711}.}. The remaining sources are all relatively compact.

\begin{figure}[t]
\epsscale{1.1}
\plotone{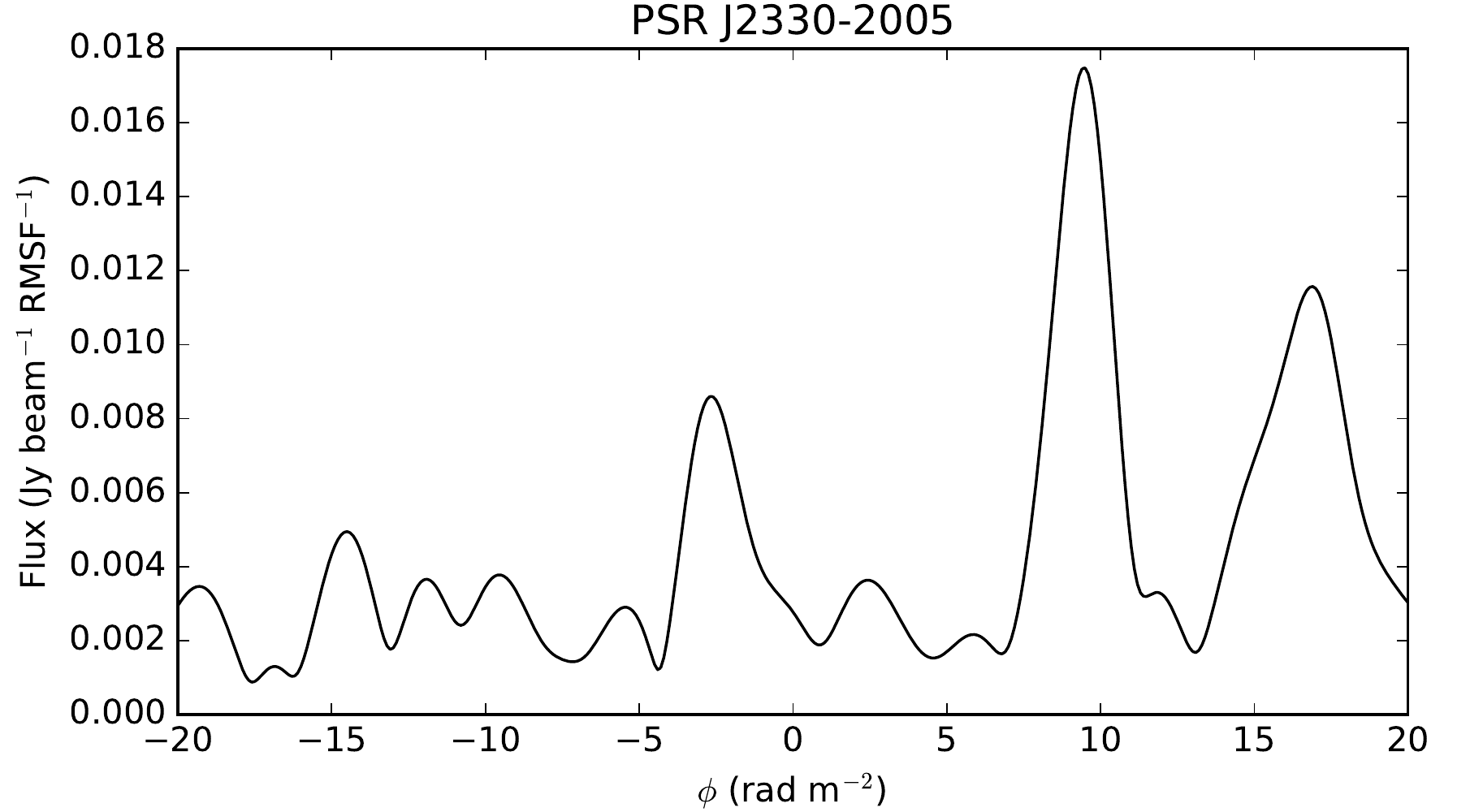}
\caption{Faraday dispersion function for PSR J2330$-$2005 from epoch 1 after correcting for the effects of ionospheric Faraday rotation. RM$=+9.6$ rad\,m$^{-2}$. Noise is 1\,mJy\,beam$^{-1}$ RMSF$^{-1}$ rms.}
\label{fig:fdfpsr}
\end{figure}

The pulsar PSR J2330$-$2005 does not appear in the \citet{Taylor:2009v702v1230} catalogue and is not detected in the uniformly weighted MWA data. It is, however, detected in both circular and linear polarization and in both epochs in the targeted search, using naturally-weighted data with short baselines removed (to avoid confusion from the diffuse structures). The parameters associated with the pulsar are summarized in Table \ref{table:pointSummary} and the Faraday dispersion function is shown in Figure \ref{fig:fdfpsr}. The secondary peaks in the Faraday dispersion function for the pulsar are unlikely to be real - most likely they are due to a combination of thermal noise and sidelobe noise as a result of the complicated PSF beam shape that results from natural weighting. In circular polarization, we measure a total flux \edit1{density} for the pulsar at 154\,MHz of $8.9\pm 1.1$\,mJy in epoch 1 and $9.6\pm 1.0$\,mJy in epoch 3. We estimate a fractional circular polarization of $\sim$$7\%$ based on the total intensity of 140 mJy measured at the pulsar position; however, the field is highly confused in total intensity at this level. As such, the total intensity is most likely over-estimated and the fractional polarization is thus a lower limit. This would be consistent with the $22\%$ circular polarization observed in the integrated pulse profile of the pulsar at 648\,MHz \citep{McCulloch:1978v183p645} .

\citet{Offringa:2016v458p1057} performed a deep point-source survey of the EoR-0 field using 45 hours of MWA data and achieved a sensitivity of 0.6\,mJy\,beam$^{-1}$ in polarization. Using a novel peeling algorithm, spectra for the 586 brightest sources in the field were presented. Unfortunately, PKS J0020$-$0014 is a resolved source and so was discarded from the catalogue and PSR J2330$-$2005 fell outside of the restricted field-of-view of the survey. PKS J0021$-$1910 appears in the catalogue but is not detected in polarization. The \citet{Offringa:2016v458p1057} survey did not consider Faraday rotation of the source and so linearly polarized sources with non-zero RM are depolarized. In addition, when combining results from multiple epochs, ionospheric Faraday rotation was also not considered; this too would lead to depolarization of linearly polarized sources (a similar issue was encountered by \citealt{Moore:2015} in PAPER observations). It is thus not surprising that linearly polarized sources were not detected by \citet{Offringa:2016v458p1057}, but circularly polarized sources should not be as greatly affected when combining data from multiple epochs. Indeed, on closer inspection, PSR J2330$-$2005 is detected in the \citet{Offringa:2016v458p1057} data at the $6.3\%$ level with a circularly polarized flux density of $6.5$ mJy (Offringa, private communication).

\begin{table*}[t]
\centering
\caption{Details of polarized point sources detected in the EoR-0 field. RM$_{\text{MWA}}$ is the rotation measure determined from MWA observations (corrected for ionospheric Faraday rotation). RM$_{\text{lit}}$ is the rotation measure in literature. P$_{\text{MWA}}$ and p$_{MWA}$ are the polarized flux density and fraction polarization derived from MWA observations, respectively. p$_{\text{lit}}$ is the fractional polarization in literature, at an observing frequency of $\nu_{\text{obs}}$. $\text{DP}(1400, 154)$ is the depolarization ratio from 1.4 GHz to 154 MHz.}
\label{table:pointSummary}
\begin{tabular}{l c c c c c c c c}
\hline\hline
Source         & RM$_{\text{MWA}}$ & RM$_{\text{lit}}$ & P$_{\text{MWA}}$ &  p$_{MWA}$ & p$_{\text{lit}}$ & $\nu_{\text{obs}}$ & $\text{DP}(1400, 154)$ \\
               & (rad\,m$^{-2}$) & (rad\,m$^{-2}$) & (mJy\,beam$^{-1}$) & & & (MHz) & \\ [0.5ex]
\hline
PSR J2330$-$2005 & $+9.6\pm0.1$ & $+16\pm3$ \textsuperscript{a} & 19 & 14\% & 16\% & 648 \textsuperscript{b}  & \\
  &  & $+9.5\pm0.2$ \textsuperscript{a} &  &  &  & & & \\
  &  & $+9.5\pm0.6$ \textsuperscript{c} &  &  &  & & & \\
PKS J2337$-$1752 & $+0.6\pm0.1$ & $+12.2\pm1.3$ \textsuperscript{d}   & 46 & 1.3\% & $3.67\pm0.07$\% & 1400 \textsuperscript{d} & 0.26 \\
PKS J0002$-$2153 & $+5.8\pm0.1$ & $+6.0\pm4.9$ \textsuperscript{d}   & 33 & 2.1\% & $6.0\pm4.9$\% & 1400 \textsuperscript{d} & 0.32 \\
PKS J0020$-$2014 & $+1.6\pm0.1$ & $+1.5\pm2.7$ \textsuperscript{d}   & 105 & 3.5\% & $12\pm1.3$\% & 1400 \textsuperscript{d} & 0.37 \\
PKS J0021$-$1910 & $+7.9\pm0.1$ & $+3.6\pm5.2$ \textsuperscript{d}   & 140 & 3.0\% & $3.6\pm5.2$\% & 1400 \textsuperscript{d} & 0.91 \\ [1ex]
\hline
\multicolumn{9}{l}{\textsuperscript{a} \footnotesize{\citet{Hamilton:1987v224p1073}};\textsuperscript{b} \footnotesize{\citet{McCulloch:1978v183p645}}; \textsuperscript{c} \footnotesize{\citet{Johnston:2007v381p1625}}; \textsuperscript{d} \footnotesize{\citet{Taylor:2009v702v1230}}} \\
%\multicolumn{8}{l}{\textsuperscript{h} \footnotesize{\citet{Hamilton:1987v224p1073}}} \\
%\multicolumn{8}{l}{\textsuperscript{i} \footnotesize{\citet{McCulloch:1978v183p645}}} \\
%\multicolumn{8}{l}{\textsuperscript{j} \footnotesize{\citet{Johnston:2007v381p1625}}} \\
\end{tabular}
\end{table*}

% ***********************************************************
\section{Discussion}
\label{sec:discussion}

\subsection{Size-scale of Structures in Linearly Polarized Emission}

The linearly polarized features seen in Figure \ref{fig:eoriqu} are highly prominent even in single 112\,s snapshot images. \edit1{LOFAR observes similar features at 160\,MHz in long ($\sim$$6$ hr), high resolution ($\sim$$3\arcmin-4\arcmin$) and sensitive ($\sim$$70-300$ $\mu$Jy PSF$^{-1}$) observations \citep{Jelic:2014v1407p2093} of the ELAIS-N1 field ($l=84\arcdeg$, $b=+45\arcdeg$) and \citet{Jelic:2015v583p137} observations of the 3C196 field ($l=171\arcdeg$, $b=+33\arcdeg$) but at significantly lower signal-to-noise.} The LOFAR observations differ from the MWA observations in that they incorporate longer baselines, are at lower Galactic latitudes and their imaging utilizes a robust image weighting of 0 which results in higher resolution ($\sim$$3\arcmin$--$4\arcmin$) images compared to the $\sim$$50\arcmin$ naturally-weighted and $(u,v)$-tapered images of the MWA presented here. Factoring in the beam size, the \citet{Jelic:2014v1407p2093} LOFAR observations have a sensitivity of $\sim$$650$ mK at the $\sim$$4\arcmin$ scale, whereas the MWA epoch 1 observations have a sensitivity of $\sim$$33$ mK at the $\sim$$50\arcmin$ scale.

\begin{figure*}[t]
\centering
\epsscale{1.15}
\plotone{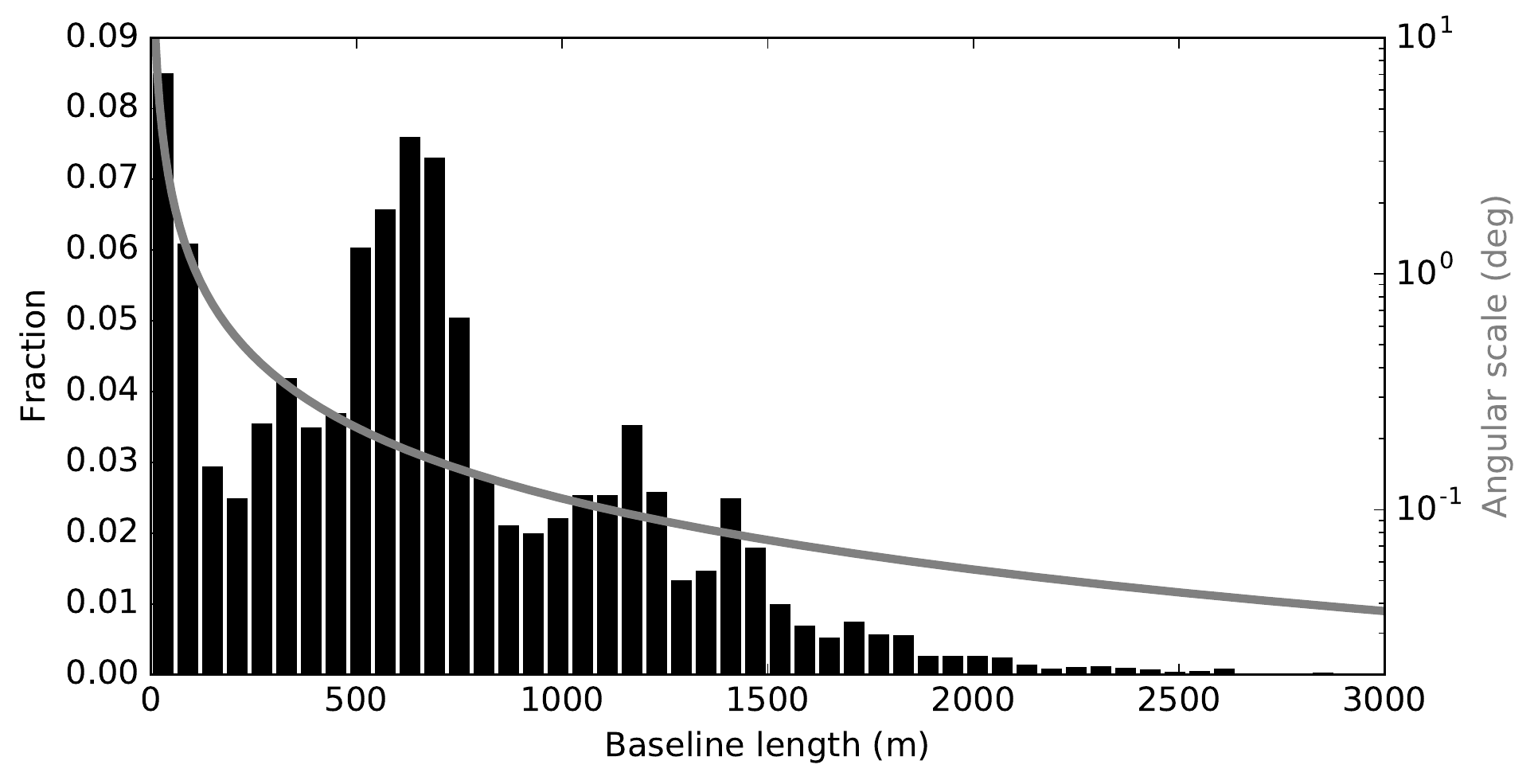}
\caption{Snapshot imaging baseline distribution for MWA 128T when observing at zenith ($\delta=-26.7\arcdeg$). The grey curve plots the angular scale probed as a function of baseline length.}
\label{fig:blsit}
\end{figure*}

The unique baseline distribution and radio quiet location of the MRO \citep{Offringa:2015v32p8O} give excellent sensitivity to structures $1\arcdeg-10\arcdeg$ in extent and sample a region not accessible to other low frequency instruments such as LOFAR \citep{vanHaarlem:2013v556p2}, for example. While the 128 tiles of the MWA provide a total of 8128 baselines out to almost 3\,km, the peak sensitivity of the array is derived from its dense inner core which was specifically designed for EoR science. Figure \ref{fig:blsit} shows the fraction of baselines as a function of baseline length. Approximately 8.5\% of the available baselines (689) are shorter than 60\,m and nearly 15\% (1183) are shorter than 120 m. The combination of sensitivity to large-scale structure and the relatively large $(u,v)$-tapered naturally weighted beam explain why the observed polarized features are so much brighter in the MWA images. In addition to providing increased sensitivity, the large number of short baselines also provide excellent snapshot imaging capabilities. This enables variations in ionospheric Faraday rotation to be monitored and calibrated for on short time-scales. Although not utilized here, it is possible that observations of the linearly polarized emission may also be used to constrain XY-phase during the calibration phase; thus reducing the effect of leakage from Stokes U into Stokes V.

Ultimately, the Amsterdam-ASTRON Radio Transients Facility and Analysis Centre \citep{Cendes:2012} should provide LOFAR with short baseline imaging capability. Similarly, the LWA (Long Wavelength Array, \citealt{Ellingson:2009v97p1421}) has a compact baseline configuration, however, it operates at longer wavelengths compared to the MWA and will be more greatly affected by the ionosphere.

The single-dish observations of \citet{Vinyaikin:2015v59p672} at low frequencies ($151.5$--$290$\,MHz) over a number of selected regions of the sky also detected the presence of $5\arcdeg-10\arcdeg$ features in linear polarization. They suggested that these features would be undetectable by interferometric observations because of the lack of short-spacings. However, despite being an interferometer, the MWA provides sensitivity to these scale sizes and thus bridges the gap between traditional single-dish and interferometric observations.

\subsection{The Nature of Diffuse Polarization}
\label{sec:dintro}

A feature of the observed linear polarization is the lack of correlation with the Stokes I map at 154\,MHz (see Figure \ref{fig:eoriqu}). The stark difference between features in linear polarization and total intensity has been noted at other wavelengths e.g. \citet{Wieringa:1993v268p215} at 325\,MHz; \citet{Haverkorn:2003v403p1045} at 350\,MHz; \citet{Bernardi:2013v771p105} at 189\,MHz; \citet{Gray:1998v393p660} and \citet{Gaensler:2011v478p214} at 1.4\,GHz; and \citet{Sun:2014v437p2936} at 2.3\,GHz and 4.8\,GHz. The prevailing interpretation is that the ionized foreground gas modulates small-scale RM structures onto an intrinsically highly polarized smooth synchrotron background via Faraday rotation.

The observed linearly polarized emission at 154\,MHz is restricted to low Faraday depths ranging from $\phi=-2$ to 9\,rad\,m$^{-2}$ (see Figure \ref{fig:rmdist}), with no other significant emission seen out to $\lvert \phi_{max} \rvert=160$\,rad\,m$^{-2}$. The distribution in RM is similar to that observed with the 32-tile MWA prototype \citep{Bernardi:2013v771p105} at 189\,MHz in a larger region that includes the EoR-0 field, and to LOFAR observations at 150\,MHz in the ELAIS-N1 \citep{Jelic:2014v1407p2093} and 3C196 fields \citep{Jelic:2015v583p137}.

The mean brightness temperature of the linearly polarized emission observed with the MWA at 154\,MHz is $\sim$$1.8$\,K in the east (towards the SGP) for the band-averaged data, but increases towards the centre of the EoR-0 field to a mean brightness temperature of $\sim$$4$\,K and a peak of 11\,K. Increased levels of polarized emission were observed in this region by \citet{Bernardi:2013v771p105} at 189\,MHz with $15.6\arcmin$ resolution using the 32-tile prototype of the MWA, with peaks up to 13\,K. Similarly, observations by \citet{Mathewson:1965v18p635} at 408\,MHz with $48\arcmin$ resolution and \citet{Wolleben:2006v448p411} at 1.4\,GHz with $30\arcmin$ resolution also suggest higher than ambient levels of polarized emission in this region. The emission appears to be coincident with part of a polarized structure identified by \citet{Wolleben:2007v664p349}, which may be associated with the southern extension of the North Polar Spur. However, as noted by \citet{Bernardi:2013v771p105}, there is little correspondence in detailed structure of the linearly polarized maps seen at MWA wavelengths and those at either 408\,MHz and 1.4\,GHz.

Diffuse emission in total intensity is weak in the EoR-0 field and difficult to separate from bright confusing sources that are within the field; a significant fraction of the diffuse emission may exist at spatial-scales to which the MWA is not sensitive. However, an estimate of fractional polarization can be obtained by extrapolating total intensity measurements from higher frequency observations. At 408\,MHz \citep{Mathewson:1965v18p635} the total intensity flux is 19.5$\pm$2\,K, the measured polarized flux is $\sim$$3$--$4$\,K ($15$--$21$\% polarized) and the temperature spectral index\footnote{$T\propto\nu^{-\beta}$ and $\alpha=2-\beta$; where $T$ is the brightness temperature, $\nu$ is the observing frequency, $\beta$ is the temperature spectral index and $\alpha$ is the spectral index.} for the total intensity emission around the SGP \citep{Guzman:2011v525p138} is $\beta=2.55$ ($\alpha=-0.55$). Extrapolating to 154\,MHz, the total intensity is estimated to be $\sim$$246$\,K. We observe $\sim$$4$\,K polarization at 154\,MHz, which corresponds to 1.6\% fractional polarization and a peak of 11\,K ($\sim$$4.5\%$ fractional polarization). This is significantly lower than the fractional polarization at 408\,MHz, but is in line with the $1.5\%$ fractional polarization seen with LOFAR at 150\,MHz in the ELAIS-N1 field \citep{Jelic:2014v1407p2093}.

\subsection{Localization of Polarized Emission}
\label{sec:horizon}

An estimate of the distance to the polarized emission can be determined by comparing the rotation measure of the emission against the overall contribution to RM of the Galaxy in the direction of the field. In the direction of the brightest features in the polarized intensity map, we measure a typical ionosphere-corrected RM of $+1.0\pm 0.3$\,rad\,m$^{-2}$ (see Figure \ref{fig:rmdist}). Estimates of the full Galactic contribution to RM in this same region, based on measurements of extragalactic Faraday rotation \citep{Oppermann:2015v575p118O}, result in an RM of $+9.0\pm 1.6$\,rad\,m$^{-2}$. If the thermal electron density in the Milky Way is assumed to be an exponential disk with a mid-plane free electron density $n_{e,0}$ (cm$^{-3}$) with a scale height $H$ and a uniform vertical magnetic field $B_{z}$ ($\mu G$), the expected RM (rad\,m$^{-2}$) out to a distance $z$ (pc) is \citep{Mao:2010v717p1170}:

\begin{equation}
\label{eq:rmdist}
\text{RM}=0.812B_{z}n_{e,0}H(1-e^{-z/H})\,.
\end{equation}

Using the measured RM at extragalactic distances, $+9.0\pm 1.6$\,rad\,m$^{-2}$, we can estimate the conditions of the magnetized plasma in the direction of the EoR-0 field as a function of the scale height $H$. Solving for $z$, using the measured RM of the observed diffuse emission ($+1.0\pm 0.3$\,rad\,m$^{-2}$), we estimate the distance to this emission is $z\sim(0.12\pm 0.04) H$. Estimates of the scale height toward the SGP, which is effectively the thick-disk component of the Milky Way, range between 930\,pc \citep{Berkhuijsen:2006v327p82} and 1830\,pc \citep{Gaensler:2008v25p184}; this corresponds to a distance of $\sim$$110$--$220$\,pc to the polarized features. There are a significant number of assumptions and uncertainties associated with this estimate, but it is sufficient to determine that the source of the polarized emission is in the local region of the Galaxy. The structures may even be constrained to lie within the local bubble, which extends out to $50$--$200$\,pc from the Sun but is elongated toward high Southern Galactic latitudes \citep{Lallement:2003v411p447}.

A more significant effect that may be used to localize the features with improved precision is that of depolarization. There are three prominent effects that can cause depolarization at long wavelengths: bandwidth, beam and depth depolarization. Bandwidth depolarization occurs when there is a significant rotation of the polarization angle across a single spectral channel. Beam depolarization is caused by fluctuations in polarization angle across the synthesized beam. Depth depolarization is caused by fluctuations in polarization angle along the line of sight. For the MWA, bandwidth depolarization is negligible for Faraday depths out to $\lvert\phi_{max}\rvert=160$ rad\,m$^{-2}$ (see Section \ref{sec:rmsynthesis}). The combined effects of depth depolarization and beam depolarization limit our ability to detect polarized emission beyond a certain distance, known as the polarization horizon \citep{Landecker:2002v609p9}. The polarization horizon depends on frequency, synthesized beam width, and \edit1{physical properties of the medium} in the observing direction \edit1{(due to variations in the magnetic field and length-scale density along different lines of sight)}.

At 1.4\,GHz, the polarization horizon is typically of the order of thousands of parsecs, e.g. \citet{Gaensler:2001v549p959}, whereas at 408\,MHz this reduces to $\sim$$150$\,pc \citep{Mathewson:1965v18p635}. \edit1{At 154\,MHz with the MWA, beam depolarization is a major concern owing to the large PSF of the observations presented here. However, LOFAR observations at significantly higher resolution observe levels of fractional polarization \citep{Jelic:2014v1407p2093} similar to that seen with the MWA. Depth depolarization is also a significant effect that will limit our long wavelength observations to structures that are relatively local compared to higher frequency observations, because more distant structures are significantly depolarized by the foreground ISM.} Assuming depth depolarization is the dominant factor, and assuming uniform synchrotron emissivity, electron density, and magnetic field in a volume of ISM, the path length $L$ \edit1{at} which integrated emission is totally depolarized is defined as \citep{Uyaniker:2003v585p785}:

\begin{equation}
\label{eq:ddpol}
L\sim\frac{\pi} {(0.81\lambda^{2}n_{e}B_{\parallel})}\,.
\end{equation}

Here $n_{e}$ is the electron density (cm$^{-3}$), $\lambda$ is the wavelength (m) and $B_{\parallel}$ is the magnetic field parallel to the line of sight ($\mu$G). We assume an electron density of $n_{e}=0.015$\,cm$^{-3}$, which is consistent with most estimates of the volume-average electron density in the thick-disk component of the warm ionized medium \citep{Gaensler:2008v25p184}. We can estimate $B_{\parallel}$ using the local horizontal field of $\sim$$2.0$\,$\mu$G \citep{Beck:1996v34p155} projected for the direction of the observation; this gives $B_{\parallel}=0.6$\,$\mu$G. At the center of the MWA band, total depolarization occurs at a distance of $L\sim$$125$\,pc and so \edit1{most} polarized features observed \edit1{can be assumed to be} at $\lesssim 125$\,pc. This estimate contains uncertainties with respect to the value of $n_{e}$ and $B_{\parallel}$ used in the direction of the the EoR field. \edit1{We also note that at path-lengths beyond the polarization horizon, the radiation is partially repolarized again \citep{Burn:1966v133p67B,Sokoloff:1998v299p189S} and so some fraction of emission will exist from beyond the horizon.} An even greater uncertainty exists with respect to beam depolarization, which will be significant at MWA wavelengths. However, as the observed polarized structures are larger in extent than the MWA beam and exhibit smooth features in Faraday space (see Figure \ref{fig:rmmap}) this effect may not be as great in this instance.

A third estimate of the distance to the emission can be derived from known pulsars within the field. This approach is similar to the first approach, which used the RM contribution of the Galaxy, but relies on the RM towards a nearby pulsar to reduce the uncertainty associated with current models of the Galaxy. Using this approach, the distance $L$ to the polarized emission can be estimated as

\begin{equation}
\label{eq:pulsardist}
L=\frac{d_{\text{pulsar}} \times \text{RM}} { \text{RM}_{\text{pulsar}}}\,.
\end{equation}

Here $d_{\text{pulsar}}$ is the distance to the pulsar, $RM$ is the measured RM from the polarized emission and $RM_{\text{pulsar}}$ is the RM of the pulsar.

% Gl=49.39 Gb=-70.19  (23:30:26.885, -20:05:29.63)
% B2327-20: DM=8.456(2)  RM=16(3)    dist_DM=0.49 kpc
% B2327-20: DM=8.456(2)  RM=9.5(0.2) dist_DM=0.49 kpc
% B2327-20: DM=8.456(2)  RM=9.5(0.6) dist_DM=0.5 kpc
% B = 9.6*sin(-70.19 * pi/180) / (0.812 * 8.456) = 1.31 uG
% n = 9.6 / (490 * 0.812 * 1.31) = 0.0183
The diffuse polarized emission in the EoR-0 field has an RM distribution that peaks at $+1.0\pm 0.3$\,rad\,m$^{-2}$. A search through the ATNF Pulsar Catalogue\footnote{\url{http://www.atnf.csiro.au/research/pulsar/psrcat}} v1.54 \citep{Manchester:2005v129p1993} revealed a single known pulsar within the EoR-0 field, PSR J2330$-$2005 (see Figure \ref{fig:rmfeatures}). The pulsar has a dispersion measure (DM) of $8.456\pm0.002$\,pc\,cm$^{-3}$ \citep{Stovall:2015v808p156}, an estimated DM-based distance of $490\pm100$\,pc \citep{Taylor:1993v411p674} and an RM of $+16\pm 3$\,rad\,m$^{-2}$ \citep{Hamilton:1987v224p1073}. Based on these parameters, this would place the polarization emission at a distance of $31\pm 12$\,pc.

In our targeted search of the pulsar field we detect PSR J2330$-$2005 in both linear and circular polarization. In linear polarization we consistently find a weak 19\,mJy\,beam$^{-1}$ peak ($14\%$ fractional polarization) and measure an RM of $+9.6\pm0.1$\,rad\,m$^{-2}$ in both epoch 1 and 3. The measured RM is lower than that reported in the ATNF Pulsar Catalogue. However, \citet{Hamilton:1987v224p1073} report an RM of $+9.5\pm0.2$\,rad\,m$^{-2}$ from unpublished observations and \citet{Johnston:2007v381p1625} measure an RM of $+9.5\pm0.6$\,rad\,m$^{-2}$, both of which are highly consistent with our measurement. If we take our measured RM instead of the RM from the ATNF Pulsar Catalogue, we place the distance to the polarized emission at $51\pm20$\,pc. Based on the measured RM to the pulsar we estimate the average electron density on the line of sight of this pulsar to be $0.0183\pm0.002$\,cm$^{-2}$ and the magnetic field $B_{\parallel}=1.31\pm0.01$\,$\mu$G; these are consistent with those expected in the solar neighbourhood.

We recognize that the estimate of the distance towards the polarized emission of $\sim$$110$--$220$\,pc, based on the estimated contribution of Galactic RM towards extragalactic sources, and the $L\lesssim 125$\,pc distance based on depth depolarization contain significant uncertainties. The estimate based on a relatively nearby pulsar within the observed field, however, is only limited by uncertainties in the measured distance of the pulsar and any inhomogeneities that may exist in the magnetic field and electron density towards the pulsar. As such, we adopt $51\pm 20$\,pc as our estimate of the distance towards the observed polarized emission.

\subsection{Polarized Point Source Population}
\label{sec:pointpop}

Based on \citet{Taylor:2009v702v1230} observations at $1.4$\,GHz, there are 399 known polarized sources within the 400 sq. degree region of the EoR-0 field. The $154$\,MHz flux density of these sources, in total intensity, cannot be accurately determined from the high resolution MWA maps shown in Figure \ref{fig:eorallI} because the field suffers greatly from sidelobe confusion. Instead, the spectral slope of each source can be determined by comparing the $1.4$\,GHz \citet{Taylor:2009v702v1230} observations with $158$\,MHz GLEAM observations of the field (Hurley-Walker et al., in prep.). Based on the measured spectral slopes and assuming no depolarization, one would expect $\sim$$191$ of the \citet{Taylor:2009v702v1230} $1.4$\,GHz sources to be detected in polarization at $154$\,MHz. In all, only four of these sources were detected at 154 MHz with the MWA.

It is useful to consider the depolarization ratio for the sources within the MWA field. We determine the depolarization ratio ($\text{DP}$) between 1.4\,GHz and 154\,MHz using \citep{Beck:2007v470p539}:

\begin{equation}
\label{eq:dpratio}
\text{DP}(1400, 154)=(P_{154}/P_{1400})(1400/154)^{\alpha}\end{equation}

where $\alpha$ is the measured spectral index of the source. $\text{DP}(1400, 154)=1$ when there is no change in fractional linear polarization from 1.4 GHz to 154 MHz, i.e. no depolarization. $\text{DP}(1400, 154)=0.5$ when the fractional linear polarization at 154 MHz is half that at 1.4 GHz. In order to depolarize all remaining $187$ \citet{Taylor:2009v702v1230} sources at 154\,MHz to below the sensitivity limits of the MWA observations, a $\text{DP}(1400, 154)$ upper limit of $<0.08$ would be required.

\citet{Mulcahy:2014v568p74} searched for polarized sources in a deep 8 hour 151\,MHz LOFAR observation around M51 with significantly higher resolution (20$\arcsec$) and sensitivity (100\,$\mu$Jy\,beam$^{-1}$), albeit over a much smaller field-of-view (17 square degrees). In all, a total of 6 polarized sources were detected. Three of the sources have \citet{Taylor:2009v702v1230} counterparts and have a measured $\text{DP}(1400, 151)$ of 0.196, 0.038, and 0.029. These depolarization ratios would be consistent with those required to depolarize all known 1.4\,GHz polarized sources in our MWA field-of-view even without taking into consideration the additional beam depolarization that would be expected with the larger MWA beam.

While it is beyond the scope of this paper to analyze further, it is interesting to note that the four extragalactic sources detected with the MWA are not significantly more depolarized at 154\,MHz compared to 1.4\,GHz, i.e. $\text{DP}(1400, 154)$ ranges between 0.26 and 0.91 (see Table \ref{table:pointSummary}). As such, they are characteristically different to the sources detected by \citet{Mulcahy:2014v568p74} with LOFAR and the MWA field sources that have clearly depolarized below our detection threshold. This could hint at a very small population of sources (one per 100 sq. deg) that do not show significant changes in depolarization with wavelength. The population is small enough that LOFAR may not yet have observed such sources with the limited number of fields observed in full polarization with its smaller field-of-view. We do note, however, that two of the least depolarized sources detected with the MWA are associated with unresolved polarized hot spots of relatively large radio galaxies (0.27\,Mpc and 1.22\,Mpc in extent). If these hot spots lie outside the local environment of the host galaxy, they may not suffer as greatly from the effects of depolarization as ones that are embedded within a magnetized plasma. 

\subsection{Turbulence in the ISM}
\label{sec:structure}

The structures seen in polarization gradient maps are generally caused by: Differential Faraday rotation \citep{Shukurov:2003v342p496S, Fletcher:2007v23p109}, a foreground Faraday screen \citep{Haverkorn:2004v421p1011, Fletcher:2007v23p109, Gaensler:2011v478p214} or intrinsic emission \citep{Sokoloff:1998v299p189S, Sun:2014v437p2936}.

Differential Faraday rotation causes depolarization contours that may manifest themselves as polarization gradients. They arise where synchrotron emission and Faraday rotation occur in the same region. For a given wavelength ($\lambda$), depolarization occurs at position $x$ in the plane of the sky under the condition where $2\lvert$RM$(x) \rvert\lambda^{2}=\pi n$ \citep{Shukurov:2003v342p496S} and $n=1,2,3,\ldots$\ . The resulting depolarization contours are infinitely thin, i.e. unresolved by the beam, and will shift as a function of frequency. As such, the contours in this interpretation are not directly related to any real structures in the ISM - hence the alternate name of ``Faraday ghosts''.

A second interpretation of depolarization canals is that they are caused by gradients in a foreground Faraday screen \citep{Fletcher:2007v23p109}. In this interpretation, features in the radio polarization gradient map are physically associated with specific structures in the ISM. These structures are caused by sudden increases or decreases of the electron density or magnetic field. As such, these features remain spatially fixed but appear and disappear as a function of frequency as different depths are probed. Features exhibiting such behaviour have been observed at centimeter wavelengths, see \cite{Gaensler:2011v478p214}, but the evolution of these features has not yet been explored over large fractional bandwidths.

A third interpretation is that the features are intrinsically polarized and caused by random anisotropic magnetic fields \citep{Shukurov:2003v342p496S}. This results in smooth synchrotron emission in total intensity, which is not observable since it is spatially filtered by the instrument, but with intrinsically polarized structures on smaller scales that are observable. In general, the structures will not shift or evolve as a function of frequency, however, more distant features will only be seen at shorter wavelengths as a result of the polarization horizon (see Section \ref{sec:horizon}). As such, there will be an increase in the number of observed features as a function of increasing frequency. 

The three different interpretations can be easily distinguished with a multi-frequency analysis of the gradient maps. We have shown that significant features are observed in the EoR gradient maps when full-band 154\,MHz data are utilized (Figure \ref{fig:eorgt}). These features are of order $1\arcdeg$ in extent and are consistent with the beam size i.e. unresolved. Assuming a distance of $51\pm20$\,pc, they have a spatial extent of $0.9\pm0.3$\,pc, which is consistent with the $\sim$$0.5$\,pc spatial extent observed in similar unresolved features in the Galactic plane at 1.4\,GHz \citep{Gaensler:2011v478p214}. These gradient map features are also present in multi-band GLEAM snapshot data (Figure \ref{fig:gleamgv}) but the maps are limited by poor sensitivity and instrumental leakage that affect the higher frequency end of the band.

The deep epoch 1 observations, however, provide sufficient sensitivity on a per-160\,kHz channel basis to examine the frequency-dependent behaviour of the gradient function, see Figure \ref{fig:eorgv2}. If we consider the polarization horizon, as described in Equation \ref{eq:ddpol}, the gradient function cube explores a square frustum centered on the EoR-0 field. In this instance, the back of the frustum (upper end of the observing band) probes more deeply (polarization horizon of 150\,pc) and front of the frustum (lower end of the observing band) probes nearby features (polarization horizon of 100\,pc). 

When the gradient function cube is explored, the gradient features vary smoothly as a function of frequency but show no significant spatial movement. The lack of spatial translation of the features rules out the differential Faraday rotation interpretation. Furthermore, most features that peak at the lower end of the band tend to persist towards the higher end of the band, with an accumulation of features with increasing frequency. This observation tends to support an intrinsic polarization interpretation but does not completely rule out a foreground Faraday screen (which generally results in features that do not vary as a function of wavelength).

A limitation of the polarization gradient method is that it is only sensitive to structures that have a scale-size similar to that of the synthesized beam of the instrument \citep{Robitaille:2015v451p372}. Gradient features larger than the beam size are resolved spatially in the plane of the sky. The same is not true for features that extend spatially in a direction perpendicular to the plane of the sky since the gradient function is only performed over the two spatial dimensions and not in the frequency direction (which as described in the previous paragraph can act as a proxy for depth). This may result in features appearing larger in depth than in spatial extent and confuse the distinction between features caused by interpretations 2 and 3 above.

To distinguish between interpretation 2 and 3, we can determine whether the observed RM gradient in the field is sufficiently large enough to result in the polarization gradients we observe. If an RM gradient results in a polarization gradient then this is evidence of a foreground Faraday screen (interpretation 2). However, if a polarization gradient appears where there is no clear RM gradient then this is evidence of intrinsic polarization (interpretation 3). To test this, we consider a uniform linearly polarized background:

\begin{equation}
\label{eq:lpol}
\vec{P}=P_{0} e^{2 i \left(\psi+\phi\lambda^{2}\right)}
\end{equation}

where $\psi$ is the intrinsic polarization angle, $\phi$ is the Faraday depth and $P_{0}$ is magnitude of the polarized intensity. From this equation, assuming $P_{0}$ is constant, the relationship between the polarization gradient $(\lvert \vec{\nabla P} \rvert$) and RM gradient ($\lvert \nabla \phi \rvert$) for Faraday-thin polarized emission can be derived as \citep{Burkhart:2012v749p145}:

\begin{equation}
\label{eq:pgvsrmg}
\lvert \vec{\nabla P} \rvert = 2\lambda^{2}P_{0} \lvert \nabla \phi \rvert\,.
\end{equation}

The intrinsic polarized intensity, $P_{0}$, cannot be obtained from our MWA observations directly because of depolarization. Instead, we can estimate it based on the intensity of synchrotron emission out to our adopted distance of the polarized emission. \citet{Nord:2006v132p242} estimate a local emissivity $\epsilon_{74}=0.36\pm 0.17$\,K\,pc$^{-1}$ at 74\,MHz. Taking a distance of $51\pm 20$\,pc, the estimated total intensity is $500\pm 300$\,mJy\,beam$^{-1}$ assuming a temperature spectral index of $\beta=-2.55$ (see Section \ref{sec:dintro}). If we assume approximately 30\% intrinsic polarization \citep{Sun:2008v477p573} this equates to $P_{0}=150 \pm 90$\,mJy\,beam$^{-1}$ at 154 MHz. Similarly, \citet{Peterson:2002v575p217} estimate $\epsilon_{10}=3.0\pm 0.7\times10^{-40}$\,W\,m$^{-3}$\,Hz$^{-1}$\,sr$^{-1}$ at 10\,MHz. Using the same assumptions as above, this results in $P_{0}=530 \pm 230$\,mJy\,beam$^{-1}$. At 154\,MHz, these estimates suggest that a 0.01 rad\,m$^{-2}$\,beam$^{-0.5}$ RM gradient would result in a $0.011\pm0.007$\,Jy\,beam$^{-1.5}$ gradient in polarization for the \citet{Nord:2006v132p242} estimate and $0.040\pm0.017$\,Jy\,beam$^{-1.5}$ gradient for the \citet{Peterson:2002v575p217} estimate.

\begin{figure*}[t]
\epsscale{1.25}
\plotone{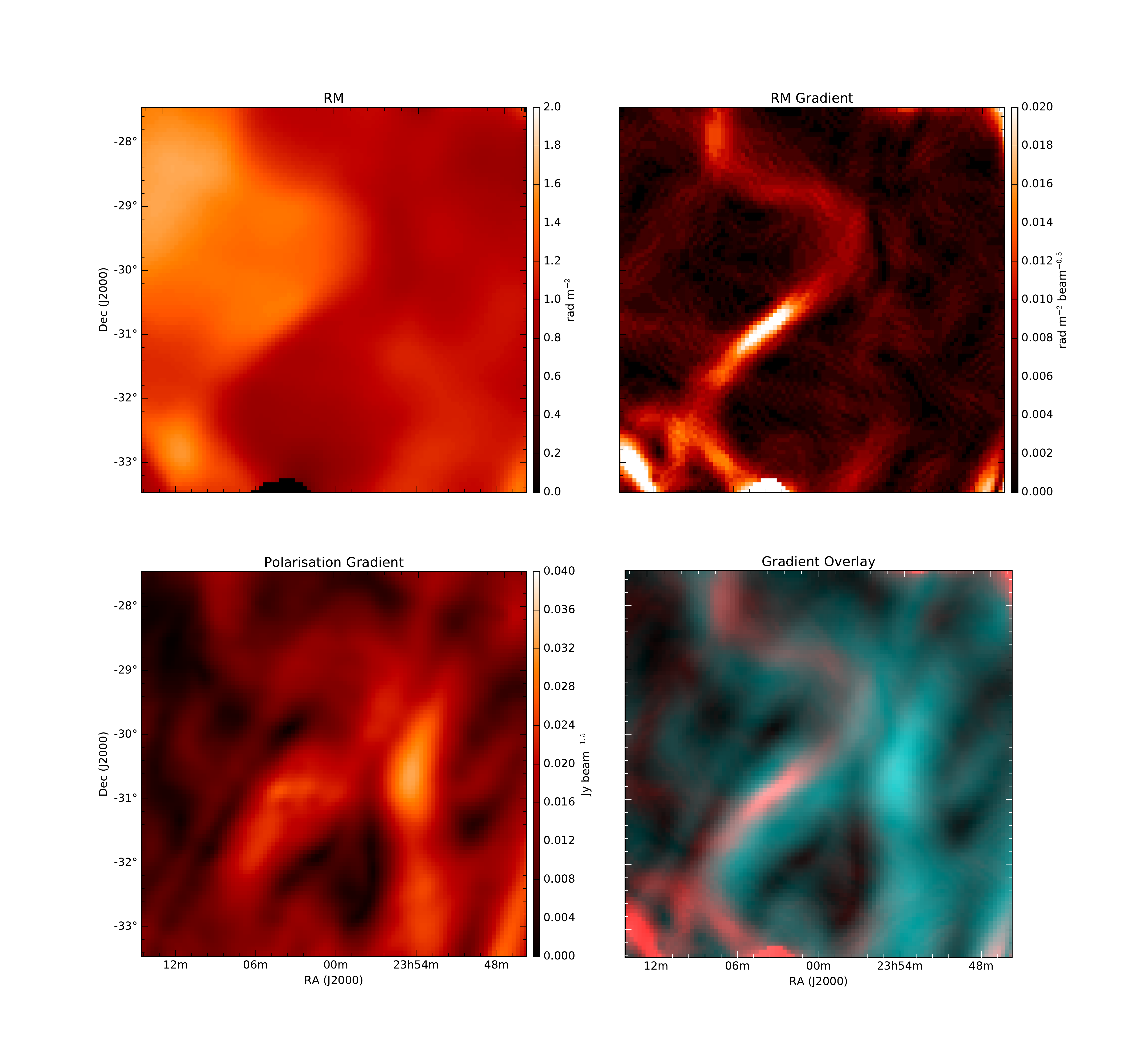}
\caption{A zoomed in view of a region containing both significant polarized flux and smooth RM variations. [Top Left] Map showing RM (rad\,m$^{-2}$) at peak polarized intensity for each pixel in the Faraday depth cube. The map is the same as that highlighted by the dashed square in Figure \ref{fig:rmmap} but with a colour scale that highlights finer variations in RM. [Top Right] The rotation measure gradient map (rad\,m$^{-2}$\,beam$^{-0.5}$) - $\lvert \nabla \phi \rvert$. This is the spatial gradient of the RM map (shown top left). [Bottom Left] The polarization gradient map (Jy\,beam$^{-1.5}$) - $\lvert \vec{\nabla P} \rvert$. The map is the same as that highlighted by the dashed square in Figure \ref{fig:gleamgv} for the time-averaged and band-averaged epoch 1 data but with a different colour scale. [Bottom Right] Overlap map of RM gradient (red) and polarization gradient (cyan).}
\label{fig:gradrm}
\end{figure*}

We can compare this with our observed gradients. Figure \ref{fig:gradrm} shows the RM, RM gradient and polarization gradient in a region of the EoR-0 field where significant gradients are observed in polarization. There is one clear RM gradient feature, the S-shaped feature on the left of the RM gradient map running from top to bottom, that is associated with one of the filaments observed in the polarization gradient map. The feature peaks at $\sim$$0.02$\,rad\,m$^{-2}$\,beam$^{-0.5}$ in the RM gradient map and at $\sim$$0.024$\,Jy\,beam$^{-1.5}$ in the polarization map. This particular feature seems consistent with a polarization gradient resulting from a Faraday screen in which $P_{0}=160$\,mJy\,beam$^{-1}$, a value that is within the \citet{Nord:2006v132p242} and \citet{Peterson:2002v575p217} estimates. The structure is reminiscent of the \citet{Burkhart:2012v749p145} ``Case 2'' scenario associated with supersonic- and subsonic-type turbulence. In this scenario there is a jump in RM spatially as a result of strong turbulent fluctuations in the magnetic field or electron density along the line-of-sight, weak shocks, or edges of a foreground cloud. 

The brightest polarization gradient feature, just west of the S-shaped feature in Figure \ref{fig:gradrm}, has no obvious counterpart in the RM gradient map. It is likely that this, and similar features throughout the wider field, are polarization gradients resulting from intrinsically polarized structures and are not caused by foreground Faraday screens. The presence of polarization gradients with RM gradient counterparts and also those without counterparts hint that the observed polarized structure results from a combination of both instrinsically polarized structures and a foreground Faraday screen.

\subsection{Structure function}
\label{sec:structureFunction}

In Section \ref{sec:structure} we investigated possible causes for the structures seen in the polarization gradient maps and concluded that they could result from a combination of both instrinsically polarized structures and a foreground Faraday screen. An alternate method of distinguishing between these two causes is through a structure function analysis \citep{Sun:2014v437p2936}. The structure function of complex polarization ($P=Q+iU$) and that of polarized intensity ($p=\lvert P\rvert$) are defined respectively as:
\begin{equation}
\label{eq:sfP}
SF_{P}\left(\delta\theta\right)\equiv\left<\left\lvert P\left(\theta\right)-P\left(\theta+\delta\theta\right)\right\rvert^{2}\right>
\end{equation}
\begin{equation}
\label{eq:sfp}
SF_{p}\left(\delta\theta\right)\equiv\left<\left[ p\left(\theta\right)-p\left(\theta+\delta\theta\right)\right]^{2}\right>
\end{equation}

Here $\delta\theta$ is the angular separation between two lines of sight. A comparison of the two structure functions can indicate whether the observed polarized structures are intrinsic or caused by Faraday screens \citep{Sun:2014v437p2936}. For intrinsic polarization there will be no correlation between polarized intensity and polarization angle, so the slope of $SF_{p}$ will be similar to that of $SF_{P}$. Alternatively, if the polarized structures are caused by Faraday screens with beam depolarization, then the slope of $SF_{P}$ will be flatter than $SF_{p}$ since much of the intrinsic structure will be smeared out by Faraday screens.

\begin{figure*}[t]
\epsscale{1.2}
\plotone{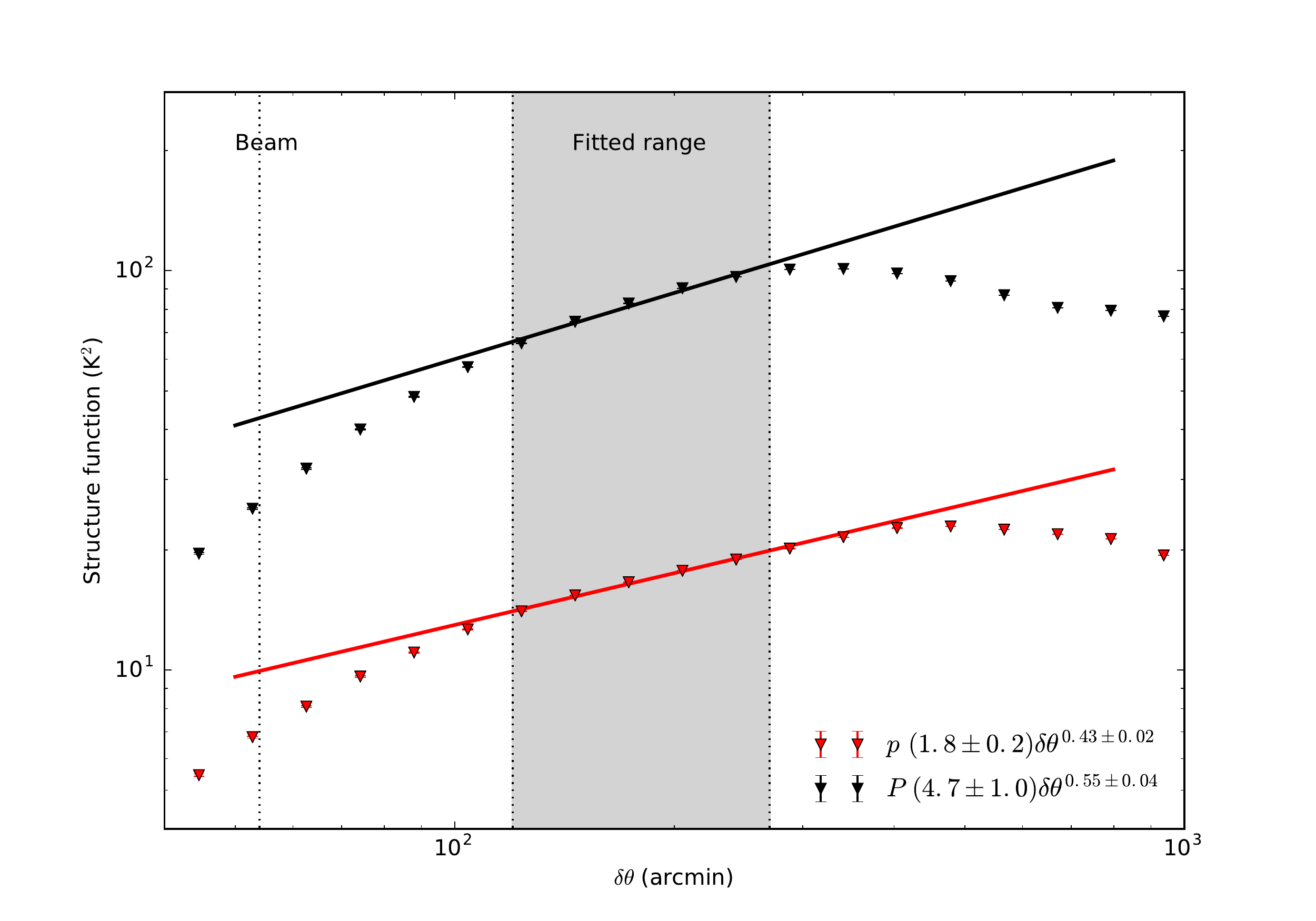}
\caption{Structure function for epoch 1 data in the EoR-0 field using ionosphere-corrected continuum images. The fit to the curve in the region  marked in grey ($120\arcmin<\delta\theta<270\arcmin$) is shown for both $p$ and $P$. The beam is marked by a dotted line at $54\arcmin$. The noise floor is not shown here but it is $\sim$$4$ orders of magnitude below each of the fitted curves.}
\label{fig:eorsf}
\end{figure*}

The resulting structure functions for $SF_{p}$ and $SF_{P}$ are shown in Figure \ref{fig:eorsf} for epoch 1 observations of the EoR-0 field. At angular separations less than about $90\arcmin$, the slope of the structure function effectively represents the smoothing effect of the MWA PSF ($54\arcmin$). At very large angular separations, the structure function is less constrained because of limited sensitivity of the MWA to structures $\gtrsim10\arcdeg$ and the available field-of-view ($25\arcdeg\times25\arcdeg$). For this analysis, we focus on the region between $120\arcmin<\delta\theta<270\arcmin$ to avoid data affected by instrumental constraints. \edit1{We note that there is observed curvature in the structure function for $SF_{P}$ in the fitting region, possibly associated with contaminating sidelobe structure. While this results in a poorer fit for $SF_{P}$ compared to that of $SF_{p}$, the slope of $SF_{P}$ is consistent to, or at most steeper, than that of $SF_{p}$.} If the slope of $SF_{P}$ were flatter than that of $SF_{p}$ this would be suggestive of polarization caused by Faraday screens \citep{Sun:2014v437p2936}. \edit1{However, the similarity in slope of $SF_{p}$ and $SF_{P}$, within $3\sigma$ fitting errors, suggests that the structures are dominated by intrinsic polarization.}

In observations of the Galactic plane at 4.8\,GHz, \citet{Sun:2014v437p2936} also find evidence of intrinsic polarization whereas at 2.3\,GHz a structure function analysis suggests the presence of a Faraday screen. The reasoning for the different behaviour is that higher frequency observations (4.8\,GHz) probe more deeply as they are less affected by the presence of a Faraday screen. One would expect that MWA observations would be particularly sensitive to the effects of a Faraday screen because they are observed at long wavelengths. However, the structure function analysis suggests that the observed diffuse polarization is intrinsic in nature and must therefore be associated with structures that are very local.

\edit1{The structure function analysis performed here suggests that intrinsic polarization is dominating in this field.} This is consistent with the multi-frequency polarization gradient analysis performed in Section \ref{sec:structure}, which found evidence of both intrinsic and Faraday screen causes for the observed polarization gradients. Currently, the structure function analysis is limited by the effective range of angular separations that could be used. Deconvolution and imaging of even larger fields would aid in widening this range and improving the structure function analysis; however, this will be left for future work.

\subsection{Faraday Depth Structure}
\label{sec:FaradayStructure}

As described in Section \ref{sec:rmsynthesis}, the 154\,MHz MWA observations provide a narrow RMSF of $\delta\phi=2.3$\,rad\,m$^{-2}$. Combined with the maximum-scale size sensitivity of 1.0\,rad\,m$^{-2}$ the MWA observations are only able to detect simple components even in the presence of Faraday complex structure. \edit1{The presence of a Faraday-thick structure would be entirely resolved by the MWA RMSF resulting in a two-peak FDF \citep[see Figure 3, ][]{Li:2011v531p126}.} A Faraday-thin structure, however, would result in a simple FDF with a single component; see Appendix B of \citet{Brentjens:2005v441p1217}.

A mix of structures has been observed in Faraday spectra observed in the Galactic anti-center with the Westerbork Synthesis Radio Telescope (WSRT) at 324-387\,MHz with $\sim$$3\arcmin$ resolution \citep{Schnitzeler:2007v471p21S, Schnitzeler:2009v494p611S}. The vast majority of lines of sight observed in a $9\arcdeg\times 9\arcdeg$ field have Faraday spectra that are reasonably well fit by a simple model dominated by a single bright peak. Only a small fraction of lines of sight exhibit Faraday complexity.

Looking at the Faraday spectra from the 154\,MHz MWA EoR-0 observations, the vast majority of Faraday structure is simple and dominated by peaks at low RM; see Figures \ref{fig:fdfsample} and \ref{fig:rmdist}. \edit1{This indicates that the extent of the polarized emission in Faraday depth is less than the MWA RMSF.} The Faraday spectra are similar to those observed in diffuse polarization at similar wavelengths with LOFAR in the ELAIS-N1 \citep{Jelic:2014v1407p2093} and 3C196 fields \citep{Jelic:2015v583p137} \edit1{with a narrower RMSF ($0.9$\,rad\,m$^{-2}$)}. It is unlikely that these are unresolved Faraday-thick structures because of the excellent resolution available in Faraday space with the MWA\edit1{ and LOFAR}. Without introducing a more involved scenario, in which a secondary peak in Faraday space is weakened to a level below our detection threshold, a simple explanation of the peak is that \edit1{the polarized emission is Faraday thin, meaning the polarized structure is intrinsic.}. This would be consistent with findings of the polarization gradient function analysis and the structure function analysis discussed in Sections \ref{sec:structure} and \ref{sec:structureFunction}.

% ***********************************************************
\section{Conclusions}
\label{sec:conclusion}

We have presented a 625 square degree survey of diffuse linear polarization at 154\,MHz carried out with the MWA. The survey, centered on the MWA EoR-0 field (0\rah, -27\arcdeg), achieved a sensitivity of 5.9\,mJy\,beam$^{-1}$ at $\sim$$54\arcmin$ resolution. The compact baselines of the MWA have been shown to be particularly sensitive to diffuse structures spanning $1\arcdeg-10\arcdeg$, something that has traditionally only been within reach of single-dish instruments.

Our MWA observations reveal smooth large-scale diffuse structures that are $\sim$$1\arcdeg-8\arcdeg$ in extent in linear polarization and clearly detected even in 2 minute snapshots. The brightness temperature of these structures is on average 4\,K in polarized intensity, peaking at 11\,K. We estimate a distance of $51\pm 20$\,pc to the polarized emission based on RM measurements of the in-field pulsar PSR J2330$-$2005.

Rotation measure synthesis reveals that the structures have Faraday depths ranging from $-2$\,rad\,m$^{-2}$ to 10\,rad\,m$^{-2}$. A large fraction of these peak at $+1.0$\,rad\,m$^{-2}$ but smaller structures are also observed to peak at $+3.0$, $+7.1$ and $\sim$$+9$\,rad\,m$^{-2}$. The observed RM structure is smooth, particularly around the region where polarized intensity peaks, with a peak RM gradient of $\sim$$0.02$\,rad\,m$^{-2}$\,beam$^{-0.5}$.

The sensitivity available in our deep observations allowed a frequency-dependent analysis of the polarized structure to be performed for the first time at long wavelengths. The results of the analysis suggested that the polarized structures are dominated by intrinsic emission but may also have a component that is due to a foreground Faraday screen. A structure function analysis of our linearly polarized images and an analysis of Faraday structure also suggest that intrinsic polarized emission tends to dominate.

A 400 square degree subset of the field was re-imaged at full resolution ($\sim$$2.4\arcmin$) and 2.3\,mJy\,beam$^{-1}$ sensitivity to search for polarized point sources.  We detect 4 extragalactic linearly polarized point sources within the EoR-0 field and have confirmed these by observing the shift in their RM over two epochs as a result of observably different ionospheric conditions in those epochs. Based on known polarized field sources at 1.4\,GHz and non-detections at 154\,MHz, we estimate an upper limit on the depolarization ratio of 0.08 from 1.4\,GHz to 154\,MHz. Such levels of depolarization are not surprising at long wavelengths, however, we note that the four detected sources did not exhibit significantly increased levels of depolarization compared to 1.4\,GHz. This may hint towards a small population of sources (one per 100 sq. deg) with this behaviour. We also note that these may be associated with relatively large radio galaxies where unresolved polarized hot spots lie outside of the local environment and are less likely to suffer the effects of depolarization.

With its high sensitivity to large-scale structure, the MWA has proven itself to be a formidable instrument for the study of diffuse polarization in the local ISM. In combination with RM synthesis, it also provides a unique ability to measure the effect of ionospheric Faraday rotation on the diffuse polarized background both spatially ($\sim$$1\arcdeg$ resolution) and temporally ($\sim$$2$ minute resolution). This not only allows ionospheric effects to be calibrated without the need for ionospheric models but also provides an opportunity to observe local solar events by measuring their effect on the background RM.

The survey presented in this paper utilized only a very small fraction of the data currently available in the EoR and GLEAM projects. There is great potential to extend the survey to look more deeply within the EoR fields, to look over an increased field-of-view with the GLEAM data, and also to explore the full $80$--$230$\,MHz range of frequencies available in the GLEAM data. Additional epochs of GLEAM data will aid in improving sensitivity and mitigating the sidelobe confusion that affects some snapshots. An implementation of an improved beam model will drastically reduce the apparent leakage observed in the highest frequency bands and so increase the range of reliable data available for subsequent frequency-dependent analysis. Furthermore, deconvolution of the diffuse structure and a multi-scale analysis of the gradient of linear polarization, e.g. \citet{Robitaille:2015v451p372}, would allow the observations to probe more deeply.

While the sensitivity of the observations presented in this paper prevented such an analysis, a measurement of instrumental leakage from linear polarization into total intensity would provide valuable information to the EoR community. Understanding leakage of this form is of particular importance for EoR science because it can act as a possible contamination source for EoR measurements \citep{Jelic:2008v389p1319, Geil2011v418p516, Moore:2013v769p154, Asad:2015v451p3709}. This has been left for future work and will be performed once the new MWA beam model has been implemented. An improved analysis will also be possible with an extension to the MWA that is currently underway. The extension provides additional compact configurations and redundant baselines that will aid in calibration.

% ***********************************************************
\section*{Acknowledgments}

\edit1{This research was conducted by the Australian Research Council Centre of Excellence for All-sky Astrophysics (CAASTRO), through project number CE110001020. DCJ is supported by the US National Science Foundation under grant number 1401708. This scientific work makes use of the Murchison Radio-astronomy Observatory, operated by CSIRO. We acknowledge the Wajarri Yamatji people as the traditional owners of the Observatory site. Support for the operation of the MWA is provided by the Australian Government (NCRIS), under a contract to Curtin University administered by Astronomy Australia Limited. We acknowledge the Pawsey Supercomputing Centre which is supported by the Western Australian and Australian Governments. The Dunlap Institute is funded through an endowment established by the David Dunlap family and the University of Toronto. We are also grateful to the anonymous referee for useful comments on the original version of this paper.}

%\allauthors


\begin{thebibliography}{}

% Ionospheric Modelling using GPS to Calibrate the MWA. I: Comparison of First Order Ionospheric Effects between GPS Models and MWA Observations
\bibitem[Arora et al.(2015)]{Arora:2015v32p29}
Arora, B.~S., Morgan, J., Ord, S.~M., et al.\ 2015, \pasa, 32, e029 

% Polarization leakage in epoch of reionization windows - I. Low Frequency Array observations of the 3C196 field
\bibitem[Asad et al.(2015)]{Asad:2015v451p3709}
Asad, K.~M.~B., Koopmans, L.~V.~E., Jeli{\'c}, V., et al.\ 2015, \mnras, 451, 3709

% Baars flux scale
\bibitem[Baars et al.(1977)]{Baars:1977v61p99B}
Baars, J.~W.~M., Genzel, R., Pauliny-Toth, I.~I.~K., \& Witzel, A.\ 1977, \aap, 61, 99 

% Galactic Magnetism: Recent Developments and Perspectives
\bibitem[Beck et al.(1996)]{Beck:1996v34p155}
Beck, R., Brandenburg, A., Moss, D., et al.\ 1996, \araa, 34, 155

% Magnetism in the spiral galaxy NGC 6946: magnetic arms, depolarization rings, dynamo modes, and helical fields
\bibitem[Beck(2007)]{Beck:2007v470p539}
Beck, R.\ 2007, \aap, 470, 539 

% Interstellar polarization at high galactic latitudes from distant stars. V. First results for the South Galactic Pole
\bibitem[Berdyugin \& Teerikorpi(2001)]{Berdyugin2001v368p635}
Berdyugin, A., \& Teerikorpi, P.\ 2001, \aap, 368, 635 

% Interstellar polarization at high galactic latitudes from distant stars. VII. A complete map for southern latitudes b < -70°
\bibitem[Berdyugin et al.(2004)]{Berdyugin2004v424p873}
Berdyugin, A., Piirola, V., \& Teerikorpi, P.\ 2004, \aap, 424, 873 

% Interstellar polarization at high galactic latitudes from distant stars. VIII. Patterns related to the local dust and gas shells from observations of ~3600 stars
\bibitem[Berdyugin et al.(2014)]{Berdyugin:2014v561p24}
Berdyugin, A., Piirola, V., \& Teerikorpi, P.\ 2014, \aap, 561, A24 

% Filling factors and scale heights of the diffuse ionized gas in the Milky Way
\bibitem[Berkhuijsen et al.(2006)]{Berkhuijsen:2006v327p82}
Berkhuijsen, E.~M., Mitra, D., \& Mueller, P.\ 2006, Astronomische Nachrichten, 327, 82 

% Foregrounds for observations of the cosmological 21 cm line.
% I. First Westerbork measurements of Galactic emission at 150 MHz in a low latitude field
\bibitem[Bernardi et al.(2009)]{Bernardi:2009v500p965}
Bernardi, G., de Bruyn, A.~G., Brentjens, M.~A., et al.\ 2009, \aap, 500, 965 

% Foregrounds for observations of the cosmological 21 cm line.
% II. Westerbork observations of the fields around 3C 196 and the North Celestial Pole
\bibitem[Bernardi et al.(2010)]{Bernardi:2010v522p67}
Bernardi, G., de Bruyn, A.~G., Harker, G., et al.\ 2010, \aap, 522, A67 

% A 189 MHz, 2400 deg2 Polarization Survey with the MWA 32-element Prototype.
\bibitem[Bernardi et al.(2013)]{Bernardi:2013v771p105}
Bernardi, G., Greenhill, L.~J.; Mitchell, D.~A. et al.\ 2013, \apj, 771, 105 

% Science with the Murchison Widefield Array
\bibitem[Bowman et al.(2013)]{Bowman:2013v30p31B}
Bowman, J.~D., Cairns, I., Kaplan, D.~L., et al.\ 2013, \pasa, 30, 31 

% RM synthesis
\bibitem[Brentjens \& de Bruyn(2005)]{Brentjens:2005v441p1217}
Brentjens, M.~A., \& de Bruyn, A.~G.\ 2005, \aap, 441, 1217 

% Properties of Interstellar Turbulence from Gradients of Linear Polarization Maps
\bibitem[Burkhart et al.(2012)]{Burkhart:2012v749p145}
Burkhart, B., Lazarian, A., \& Gaensler, B.~M.\ 2012, \apj, 749, 145 

% On the depolarization of discrete radio sources by Faraday dispersion
\bibitem[Burn(1966)]{Burn:1966v133p67B}
Burn, B.~J.\ 1966, \mnras, 133, 67

% A New 1.4 GHz Radio Continuum Map of the Sky South of Declination +25deg
\bibitem[Calabretta et al.(2014)]{Calabretta:2014v31p7}
Calabretta, M.~R., Staveley-Smith, L., \& Barnes, D.~G.\ 2014, \pasa, 31, e007

% Galactic Foregrounds and CMB Polarization / S-PASS
\bibitem[Carretti(2010)]{Carretti:2010v438p276}
Carretti, E.\ 2010, in ASP Conf. Ser. 438, The Dynamic Interstellar Medium: A Celebration of the Canadian Galactic Plane Survey, ed. R. Kothes, T. L. Landecker, \& A. G. Willis (San Francisco, CA: ASP), 276 

% AARTFAAC
\bibitem[Cendes \& AARTFAAC Project Team(2012)]{Cendes:2012}
Cendes, Y., \& AARTFAAC Project Team 2012, American Astronomical Society Meeting Abstracts \#219, 219, 422.36

% The long wavelength array
\bibitem[Ellingson et al.(2009)]{Ellingson:2009v97p1421}
Ellingson, S.~W., Clarke, T.~E., Cohen, A., et al.\ 2009, IEEE Proceedings, 97, 1421 

% Polarization gradient
\bibitem[Fletcher \& Shukurov(2007)]{Fletcher:2007v23p109}
Fletcher, A., \& Shukurov, A.\ 2007, EAS Publications Series, 23, 109 

% Limits of noise and confusion in the MWA GLEAM year 1 survey
\bibitem[Franzen et al.(2015)]{Franzen:2015}
Franzen, T.~M.~O., Jackson, C.~A., Callingham, J.~R., et al.\ 2015, The many facets of extragalactic radio surveys: towards new scientific challenges (EXTRA-RADSUR2015).~20-23 October 2015, Bologna, Italy. \url{http://pos.sissa.it/archive/conferences/267/003/EXTRA-RADSUR2015\_053.pdf}

% Low-Mach-number turbulence in interstellar gas revealed by radio polarization gradients
\bibitem[Gaensler et al.(2011)]{Gaensler:2011v478p214}
Gaensler, B.~M., Haverkorn, M., Burkhart, B., et al.\ 2011, \nat, 478, 214

% Radio Polarization from the Inner Galaxy at Arcminute Resolution
\bibitem[Gaensler et al.(2001)]{Gaensler:2001v549p959}
Gaensler, B.~M., Dickey, J.~M., McClure-Griffiths, N.~M., et al.\ 2001, \apj, 549, 959 

% The Vertical Structure of Warm Ionised Gas in the Milky Way
\bibitem[Gaensler et al.(2008)]{Gaensler:2008v25p184}
Gaensler, B.~M., Madsen, G.~J., Chatterjee, S., \& Mao, S.~A.\ 2008, \pasa, 25, 184

% Polarized foreground removal at low radio frequencies using
% rotation measure synthesis: uncovering the signature of hydrogen reionization
\bibitem[Geil et al.(2011)]{Geil2011v418p516}
Geil, P.~M., Gaensler, B.~M., \& Wyithe, J.~S.~B.\ 2011, \mnras, 418, 516 

% A large-scale, interstellar Faraday-rotation feature of unknown origin
\bibitem[Gray et al.(1998)]{Gray:1998v393p660}
Gray, A.~D., Landecker, T.~L., Dewdney, P.~E., \& Taylor, A.~R.\ 1998, \nat, 393, 660 

% All-sky Galactic radiation at 45 MHz and spectral index between 45 and 408 MHz
\bibitem[Guzm{\'a}n et al.(2011)]{Guzman:2011v525p138}
Guzm{\'a}n, A.~E., May, J., Alvarez, H., \& Maeda, K.\ 2011, \aap, 525, A138 

% Faraday rotation measurements on 163 pulsars
\bibitem[Hamilton \& Lyne(1987)]{Hamilton:1987v224p1073}
Hamilton, P.~A., \& Lyne, A.~G.\ 1987, \mnras, 224, 1073 

% The Sino-German λ6 CM Polarization Survey of the Galactic Plane: a Summary
\bibitem[Han et al.(2013)]{Han:2013v23p82}
Han, J.~L., Reich, W., Sun, X.~H., et al.\ 2013, International Journal of Modern Physics Conference Series, 23, 82 

% A 408 MHz all-sky continuum survey. II - The atlas of contour maps
\bibitem[Haslam et al.(1982)]{Haslam:1982v47p1}
Haslam, C.~G.~T., Salter, C.~J., Stoffel, H., \& Wilson, W.~E.\ 1982, \aaps, 47, 1 

%
\bibitem[Haverkorn et al.(2000)]{Haverkorn:2000v356p13}
Haverkorn, M., Katgert, P., \& de Bruyn, A.~G.\ 2000, \aap, 356, L13

% Multi-frequency polarimetry of the Galactic radio background around 350 MHz
% I. A region in Auriga around l = 161, b = 16
\bibitem[Haverkorn et al.(2003a)]{Haverkorn:2003v403p1031}
Haverkorn, M., Katgert, P., \& de Bruyn, A.~G.\ 2003a, \aap, 403, 1031 

% Characteristics of the structure in the Galactic polarized radio background at 350 MHz
\bibitem[Haverkorn et al.(2003b)]{Haverkorn:2003v403p1045}
Haverkorn, M., Katgert, P., \& de Bruyn, A.~G.\ 2003b, \aap, 403, 1045 

% Multi-frequency polarimetry of the Galactic radio background around 350 MHz
% II. A region in Horologium around l = 137, b = 7
\bibitem[Haverkorn et al.(2003c)]{Haverkorn:2003v404p233}
Haverkorn, M., Katgert, P., \& de Bruyn, A.~G.\ 2003c, \aap, 404, 233 

% Canals beyond Mars: Beam depolarization in radio continuum maps of the warm ISM
\bibitem[Haverkorn \& Heitsch(2004)]{Haverkorn:2004v421p1011}
Haverkorn, M., \& Heitsch, F.\ 2004, \aap, 421, 1011 

% Measuring magnetism in the Milky Way with the Square Kilometre Array
\bibitem[Haverkorn et al.(2015)]{Haverkorn:2015}
Haverkorn, M., Akahori, T., Carretti, E., et al.\ 2015, Proceedings of Science, Advancing Astrophysics with the Square Kilometre Array (AASKA14), 96

% The Faraday rotation measure synthesis technique
\bibitem[Heald(2009)]{Heald:2009v259p591}
Heald, G.\ 2009, IAU Symposium, 259, 591 

% MWACS
\bibitem[Hurley-Walker et al.(2014)]{Hurley-Walker:2014v31p45}
Hurley-Walker, N., Morgan, J., Wayth, R.~B., et al.\ 2014, \pasa, 31, e045

% GLEAM catalogue
%\bibitem[Hurley-Walker et al.(2016)]{Hurley-Walker:2016}
%Hurley-Walker, N., et al.\ 2016, in prep.

% Rotation measure synthesis at the 2 m wavelength of the FAN region: unveiling screens and bubbles
\bibitem[Iacobelli et al.(2013)]{Iacobelli:2013v549p56}
Iacobelli, M., Haverkorn, M., \& Katgert, P.\ 2013, \aap, 549, A56

% Galactic interstellar turbulence across the southern sky seen through spatial gradients of the polarization vector
\bibitem[Iacobelli et al.(2014)]{Iacobelli:2014v566p5}
Iacobelli, M., Burkhart, B., Haverkorn, M., et al.\ 2014, \aap, 566, A5 

% Giant radio sources
\bibitem[Ishwara-Chandra \& Saikia(1999)]{Ishwara-Chandra:1999v309p100}
Ishwara-Chandra, C.~H., \& Saikia, D.~J.\ 1999, \mnras, 309, 100 

% EoR survey
\bibitem[Jacobs et al.(2016)]{Jacobs:2016}
Jacobs, D.~C., Hazelton, B.~J., Trott, C.~M., et al.\ 2016, \apj, accepted (arXiv:1605.06978)

% Foreground simulations for the LOFAR-epoch of reionization experiment
\bibitem[Jeli{\'c} et al.(2008)]{Jelic:2008v389p1319}
Jeli{\'c}, V., Zaroubi, S., Labropoulos, P., et al.\ 2008, \mnras, 389, 1319

% Initial LOFAR observations of epoch of reionization windows. II. Diffuse polarized emission in the ELAIS-N1 field
\bibitem[Jeli{\'c} et al.(2014)]{Jelic:2014v1407p2093}
Jeli{\'c}, V., de Bruyn, A.~G., Mevius, M., et al.\ 2014, \aap, 568, 101 

% Linear polarization structures in LOFAR observations of the interstellar medium in the 3C196 field
\bibitem[Jeli{\'c} et al.(2015)]{Jelic:2015v583p137}
Jeli{\'c}, V., de Bruyn, A.~G., Pandey, V.~N., et al.\ 2015, \aap, 583, 137

% Evidence for alignment of the rotation and velocity vectors in pulsars – II. Further data and emission heights
% 9.5(6) rad m^-2
\bibitem[Johnston et al.(2007)]{Johnston:2007v381p1625}
Johnston, S., Kramer, M., Karastergiou, A., et al.\ 2007, \mnras, 381, 1625

% Murchison Widefield Array Observations of Anomalous Variability:
% A Serendipitous Night-time Detection of Interplanetary Scintillation
\bibitem[Kaplan et al.(2015)]{Kaplan:2015v809p12}
Kaplan, D.~L., Tingay, S.~J., Manoharan, P.~K., et al.\ 2015, \apjl, 809, L12 

% 3D mapping of the dense interstellar gas around the Local Bubble
\bibitem[Lallement et al.(2003)]{Lallement:2003v411p447}
Lallement, R., Welsh, B.~Y., Vergely, J.~L., et al.\ 2003, \aap, 411, 447 

% Polarization imaging of the galactic emission at 1420 MHz-the Canadian Galactic Plane Survey
\bibitem[Landecker et al.(2002)]{Landecker:2002v609p9}
Landecker, T.~L., Uyan{\i}ker, B., \& Kothes, R.\ 2002, in Astrophysical Polarized Backgrounds, ed. S. Cecchini, S. Cortiglioni, R. Sault, \& C. Sbarra, Vol. 609 (Melville, NY: American Institute of Physics), 9-14

% The application of compressive sampling to radio astronomy.
% II. Faraday rotation measure synthesis
\bibitem[Li et al.(2011)]{Li:2011v531p126}
Li, F., Brown, S., Cornwell, T.~J., \& de Hoog, F.\ 2011, \aap, 531, A126 

% Real-time imaging of density ducts between the plasmasphere and ionosphere
\bibitem[Loi et al.(2015a)]{Loi:2015v42p3707}
Loi, S.~T., Murphy, T., Cairns, I.~H., et al.\ 2015a, \grl, 42, 3707 

% Power spectrum analysis of ionospheric fluctuations with the Murchison Widefield Array
\bibitem[Loi et al.(2015b)]{Loi:2015v50p574}
Loi, S.~T., Trott, C.~M., Murphy, T., et al.\ 2015b, Radio Science, 50, 574 

% Quantifying ionospheric effects on time-domain astrophysics with the Murchison Widefield Array
\bibitem[Loi et al.(2015c)]{Loi:2015v453p2731}
Loi, S.~T., Murphy, T., Bell, M.~E., et al.\ 2015c, \mnras, 453, 2731

% The ATNF Pulsar Catalogue
\bibitem[Manchester et al.(2005)]{Manchester:2005v129p1993}
Manchester, R.~N., Hobbs, G.~B., Teoh, A., \& Hobbs, M.\ 2005, \aj, 129, 1993 
 
% A Survey of Extragalactic Faraday Rotation at High Galactic Latitude.
% The Vertical Magnetic Field of the Milky Way Toward the Galactic Poles
\bibitem[Mao et al.(2010)]{Mao:2010v717p1170}
Mao, S.~A., Gaensler, B.~M., Haverkorn, M., et al.\ 2010, \apj, 717, 1170

% 408 MHz linear polarization survey with Parkes.
\bibitem[Mathewson \& Milne(1965)]{Mathewson:1965v18p635}
Mathewson, D.~S. \& Milne, M.~F.\ 1965, AuJPh, 18, 635 

% Polarization characteristics of southern pulsars. II - 640-MHz observations
\bibitem[McCulloch et al.(1978)]{McCulloch:1978v183p645}
McCulloch, P.~M., Hamilton, P.~A., Manchester, R.~N., \& Ables, J.~G.\ 1978, \mnras, 183, 645 

% Theory of Star Formation
\bibitem[McKee \& Ostriker(2007)]{McKee:2007v45p565}
McKee, C.~F., \& Ostriker, E.~C.\ 2007, \araa, 45, 565 

% RTS
\bibitem[Mitchell et al.(2008)]{Mitchell:2008v2p707}
Mitchell, D.~A., Greenhill, L.~J., Wayth, R.~B., et al.\ 2008, IEEE Journal of Selected Topics in Signal Processing, 2, 707 

% The Effects of Polarized Foregrounds on 21 cm Epoch of Reionization Power Spectrum Measurements
\bibitem[Moore et al.(2013)]{Moore:2013v769p154}
Moore, D.~F., Aguirre, J.~E., Parsons, A.~R., et al.\ 2013, \apj, 769, 154

% PAPER polarization
\bibitem[Moore et al.(2016)]{Moore:2015}
Moore, D., Aguirre, J.~E., Parsons, A., et al.\ 2016, \apj, submitted (arXiv:1502.05072)

% LOFAR M51 polarization
\bibitem[Mulcahy et al.(2014)]{Mulcahy:2014v568p74}
Mulcahy, D.~D., Horneffer, A., Beck, R., et al.\ 2014, \aap, 568, A74

% Emissivity
\bibitem[Nord et al.(2006)]{Nord:2006v132p242}
Nord, M.~E., Henning, P.~A., Rand, R.~J., et al.\ 2006, \aj, 132, 242 

% LOFAR: The potential for solar and space weather studies
\bibitem[Oberoi \& Kasper(2004)]{Oberoi:2004v52p1415}
Oberoi, D., \& Kasper, J.~C.\ 2004, \planss, 52, 1415 

% AOFLAGGER
\bibitem[Offringa et al.(2012)]{Offringa:2012v539p95O}
Offringa, A.~R., van de Gronde, J.~J., \& Roerdink, J.~B.~T.~M.\ 2012, \aap, 539, A95 

% wsclean
\bibitem[Offringa et al.(2014)]{Offringa:2014v444p606O}
Offringa, A.~R., McKinley, B., Hurley-Walker, N., et al.\ 2014, \mnras, 444, 606 

% RFI environment
\bibitem[Offringa et al.(2015)]{Offringa:2015v32p8O}
Offringa, A.~R., Wayth, R.~B., Hurley-Walker, N., et al.\ 2015, \pasa, 32, e008 

% Parametrising Epoch of Reionization foregrounds: A deep survey of low-frequency point-source spectra with the MWA
\bibitem[Offringa et al.(2016)]{Offringa:2016v458p1057}
Offringa, A.~R., Trott, C.~M., Hurley-Walker, N., et al.\ 2016, \mnras, 458, 1057 

% Estimating extragalactic Faraday rotation
\bibitem[Oppermann et al.(2015)]{Oppermann:2015v575p118O}
Oppermann, N., Junklewitz, H., Greiner, M., et al.\ 2015, \aap, 575, A118 

% Imaging with MWA
\bibitem[Ord et al.(2010)]{Ord:2010v122p1353}
Ord, S.~M., Mitchell, D.~A., Wayth, R.~B., et al.\ 2010, \pasp, 122, 1353 

% MWA Correlator
\bibitem[Ord et al.(2015)]{Ord:2015v32p6O}
Ord, S.~M., Crosse, B., Emrich, D., et al.\ 2015, \pasa, 32, e006

% Study of Redshifted H I from the Epoch of Reionization with Drift Scan
\bibitem[Paul et al.(2014)]{Paul:2014v793p28}
Paul, S., Sethi, S.~K., Subrahmanyan, R., et al.\ 2014, \apj, 793, 28 

% Interstellar Absorption of the Galactic Polar Low-Frequency Radio 
% Background Synchrotron Spectrum as an Indicator of Clumpiness in the Warm Ionized Medium
\bibitem[Peterson \& Webber(2002)]{Peterson:2002v575p217}
Peterson, J.~D., \& Webber, W.~R.\ 2002, \apj, 575, 217 

% Local ISM 3D distribution and soft X-ray background. Inferences on nearby hot gas and the North Polar Spur
\bibitem[Puspitarini et al.(2014)]{Puspitarini:2014v566p13}
Puspitarini, L., Lallement, R., Vergely, J.-L., \& Snowden, S.~L.\ 2014, \aap, 566, A13 

% VLA Images at 5 GHZ of 212 Southern Extragalactic Objects
\bibitem[Reid et al.(1999)]{Reid:1999v124p285}
Reid, R.~I., Kronberg, P.~P., \& Perley, R.~A.\ 1999, \apjs, 124, 285 

% An improved source-subtracted and destriped 408-MHz all-sky map
\bibitem[Remazeilles et al.(2015)]{Remazeilles:2015v451p4311}
Remazeilles, M., Dickinson, C., Banday, A.~J., et al.\ 2015, \mnras, 451, 4311

% Multiscale analysis of the gradient of linear polarization
\bibitem[Robitaille \& Scaife(2015)]{Robitaille:2015v451p372}
Robitaille, J.-F., \& Scaife, A.~M.~M.\ 2015, \mnras, 451, 372 

% Miriad
\bibitem[Sault et al.(1995)]{Sault:1995v77p433}
Sault, R.~J., Teuben, P.~J., \& Wright, M.~C.~H.\ 1995, in R. A. Shaw, H. E. Payne, \& J. J. E. Hayes ed., Astronomical Data Analysis Software and Systems IV Vol. 77 of ASP Conference Series. p. 433 

% Understanding radio polarimetry. II. Instrumental calibration of an interferometer array.
\bibitem[Sault et al.(1996)]{Sault:1996v117p149}
Sault, R.~J., Hamaker, J.~P., \& Bregman, J.~D.\ 1996, \aaps, 117, 149

% WSRT Faraday tomography of the Galactic ISM at λ ˜ 0.86 m. First results for a field at (l, b)~=~(181°,20°)
\bibitem[Schnitzeler et al.(2007)]{Schnitzeler:2007v471p21S}
Schnitzeler, D.~H.~F.~M., Katgert, P., \& de Bruyn, A.~G.\ 2007, \aap, 471, L21 

% WSRT Faraday tomography of the Galactic ISM at λ ~ 0.86 m. I. The GEMINI data set at (l, b) = (181°, 20°)
\bibitem[Schnitzeler et al.(2009)]{Schnitzeler:2009v494p611S}
Schnitzeler, D.~H.~F.~M., Katgert, P., \& de Bruyn, A.~G.\ 2009, \aap, 494, 611 

% Depolarization Contours
\bibitem[Shukurov \& Berkhuijsen(2003)]{Shukurov:2003v342p496S}
Shukurov, A., \& Berkhuijsen, E.~M.\ 2003, \mnras, 342, 496 

% Culgoora
\bibitem[Slee(1977)]{Slee:1977v43p1}
Slee, O.~B.\ 1977, Australian Journal of Physics Astrophysical Supplement, 43, 1 

\bibitem[Slee(1995)]{Slee:1995v48p143}
Slee, O.~B.\ 1995, Australian Journal of Physics, 48, 143

% Intrinsic cause of polarization gradients
\bibitem[Sokoloff et al.(1998)]{Sokoloff:1998v299p189S}
Sokoloff, D.~D., Bykov, A.~A., Shukurov, A., et al.\ 1998, \mnras, 299, 189 

% Calibrating high-precision Faraday rotation measurements for LOFAR and the next generation of low-frequency radio telescopes
\bibitem[Sotomayor-Beltran et al.(2013)]{SotomayorBeltran:2013v552p58S}
Sotomayor-Beltran, C., Sobey, C., Hessels, J.~W.~T., et al.\ 2013, \aap, 552, A58 

% Pulsar Observations Using the First Station of the Long Wavelength Array and the LWA Pulsar Data Archive
\bibitem[Stovall et al.(2015)]{Stovall:2015v808p156}
Stovall, K., Ray, P.~S., Blythe, J., et al.\ 2015, \apj, 808, 156 

% MWA beam
\bibitem[Sutinjo et al.(2015)]{Sutinjo:2015v50p52S}
Sutinjo, A., O'Sullivan, J., Lenc, E., et al.\ 2015, Radio Science, 50, 52 

% Radio observational constraints on Galactic 3D-emission models
\bibitem[Sun et al.(2008)]{Sun:2008v477p573}
Sun, X.~H., Reich, W., Waelkens, A., \& En{\ss}lin, T.~A.\ 2008, \aap, 477, 573 

% A Sino-German λ6 cm polarization survey of the Galactic plane. III. The region from 10° to 60° longitude
\bibitem[Sun et al.(2011)]{Sun:2011v527p74}
Sun, X.~H., Reich, W., Han, J.~L., et al.\ 2011, \aap, 527, A74 

% Absolutely calibrated radio polarimetry of the inner Galaxy at 2.3 and 4.8 GHz
\bibitem[Sun et al.(2014)]{Sun:2014v437p2936}
Sun, X.~H., Gaensler, B.~M., Carretti, E., et al.\ 2014, \mnras, 437, 2936 

% Pulsar distances and the galactic distribution of free electrons
\bibitem[Taylor \& Cordes(1993)]{Taylor:1993v411p674}
Taylor, J.~H., \& Cordes, J.~M.\ 1993, \apj, 411, 674

% The Canadian Galactic Plane Survey
\bibitem[Taylor et al.(2003)]{Taylor:2003v125p314}
Taylor, A.~R., Gibson, S.~J., Peracaula, M., et al.\ 2003, \aj, 125, 3145 


% A Rotation Measure Image of the Sky
\bibitem[Taylor et al.(2009)]{Taylor:2009v702v1230}
Taylor, A.~R., Stil, J.~M., \& Sunstrum, C.\ 2009, \apj, 702, 1230 

\bibitem[Thyagarajan et al.(2015)]{Thyagarajan:2015v804p14} 
Thyagarajan, N., Jacobs, D.~C., Bowman, J.~D., et al.\ 2015, \apj, 804, 14 

% The Murchison Widefield Array: 
% The Square Kilometre Array Precursor at Low Radio Frequencies
\bibitem[Tingay et al.(2013)]{Tingay:2013v30p7} 
Tingay, S.~J., Goeke, R., Bowman, J.~D., et al.\ 2013, \pasa, 30, 7 

% Comparison of Observing Modes for Statistical Estimation of the 21 cm Signal from the Epoch of Reionisation
\bibitem[Trott(2014)]{Trott:2014v31p26}
Trott, C.~M.\ 2014, \pasa, 31, e026

% Radio Polarization from the Galactic Plane in Cygnus
\bibitem[Uyan{\i}ker et al.(2003)]{Uyaniker:2003v585p785}
Uyan{\i}ker, B., Landecker, T.~L., Gray, A.~D., \& Kothes, R.\ 2003, \apj, 585, 785 

% LOFAR: The LOw-Frequency ARray
\bibitem[van Haarlem et al.(2013)]{vanHaarlem:2013v556p2}
van Haarlem, M.~P., Wise, M.~W., Gunst, A.~W., et al.\ 2013, \aap, 556, A2 

% Low-frequency polarization measurements of the diffuse radio emission of the galaxy
\bibitem[Vinyaikin \& Paseka(2015)]{Vinyaikin:2015v59p672}
Vinyaikin, E.~N., \& Paseka, A.~M.\ 2015, Astronomy Reports, 59, 672 

% GLEAM: The GaLactic and Extragalactic All-Sky MWA Survey
\bibitem[Wayth et al.(2015)]{Wayth:2015v32p25}
Wayth, R.~B., Lenc, E., Bell, M.~E., et al.\ 2015, \pasa, 32, e025 

% 325 MHz diffuse polarization
\bibitem[Wieringa et al.(1993)]{Wieringa:1993v268p215}
Wieringa, M.~H., de Bruyn, A.~G., Jansen, D., et al.\ 1993, \aap, 268, 215 

% ALBUS
\bibitem[Willis et al.(2016)]{Willis:2016}
Willis, A.~G., Mevius, M., Anderson, J.~M., et al.\ 2016, Astronomy and Computing, submitted

% 1.4 GHz polarization
\bibitem[Wolleben et al.(2006)]{Wolleben:2006v448p411}
Wolleben, M., Landecker, T.~L., Reich, W., \& Wielebinski, R.\ 2006, \aap, 448, 411 

% North polar spur extending to SGP region.
\bibitem[Wolleben(2007)]{Wolleben:2007v664p349}
Wolleben, M.\ 2007, \apj, 664, 349 

% A Cosmology Calculator for the World Wide Web
\bibitem[Wright(2006)]{Wright:2006v118p1711}
Wright, E.~L.\ 2006, \pasp, 118, 1711

\bibitem[Wrobel \& Walker(1999)]{Wrobel:1999v180p171}
Wrobel, J.~M., \& Walker, R.~C.\ 1999, Synthesis Imaging in Radio Astronomy II, 180, 171 

\end{thebibliography}
\end{document}